\documentclass[aps,prx,reprint,preprintnumbers,nofootinbib,floatfix]{revtex4-2}
\usepackage{xpatch}
\makeatletter
\patchcmd{\@ssect@ltx}
    {\addcontentsline{toc}{#1}{\protect\numberline{}#8}}
    {}
    {}
    {}
\makeatother
\pdfoutput=1
\usepackage[dvipsnames]{xcolor}

\definecolor{paperprimary}{HTML}{1155CC}
\definecolor{papersecondary}{HTML}{38761D}
\definecolor{ribbonfill}{HTML}{B8CCE8}
\colorlet{papertertiary}{paperprimary}
\usepackage{amssymb}
\usepackage{graphicx}
\usepackage{amsmath}
\usepackage{booktabs}
\usepackage{bm}
\usepackage{array}
\numberwithin{equation}{section}

\usepackage[english]{babel}
\usepackage{pict2e}

\colorlet{grownred}{paperprimary}
\colorlet{grownblue}{papertertiary}
\colorlet{actedgreen}{papersecondary}

\newcommand{\LatticeMidArrow}{%
  \raisebox{-.18ex}{\scalebox{.66}{$\blacktriangleright$}}%
}
\newcommand{\TensorMidArrow}{%
  \raisebox{-.18ex}{\scalebox{.60}{$\blacktriangleright$}}%
}

\newcommand{\NPLine}[4]{\Line(#1,#2)(#3,#4)}
\newcommand{\NPLabel}[3]{\put(#1,#2){\makebox(0,0){#3}}}
\newcommand{\NPLabelL}[3]{\put(#1,#2){\makebox(0,0)[l]{#3}}}
\newcommand{\NPLabelR}[3]{\put(#1,#2){\makebox(0,0)[r]{#3}}}
\newcommand{\NPArrow}[3]{%
  \put(#1,#2){\makebox(0,0){\rotatebox{#3}{\TensorMidArrow}}}%
}
\newcommand{\NPLatticeArrow}[3]{%
  \put(#1,#2){\makebox(0,0){\rotatebox{#3}{\LatticeMidArrow}}}%
}
\newcommand{\NPCopy}[2]{\put(#1,#2){\circle*{.22}}}
\newcommand{\NPMult}[2]{%
  \put(#1,#2){{\color{white}\circle*{.30}}}%
  \put(#1,#2){\circle{.32}}%
}
\newcommand{\NPCross}[2]{%
  \put(#1,#2){\Line(-.07,-.07)(.07,.07)\Line(-.07,.07)(.07,-.07)}%
}
\newcommand{\NPBlueDot}[2]{%
  \put(#1,#2){\color{grownblue}\circle*{.14}}%
}

\newcommand{\NativeUnfoldedAncilla}{%
  \begingroup\setlength{\unitlength}{.49cm}%
  \begin{picture}(6.6,7.7)(-2.1,-.2)
    \linethickness{.65pt}%
    \multiput(0,4)(0,1){4}{\line(1,0){4}}%
    \multiput(1,4)(1,0){3}{\line(0,1){3}}%
    \multiput(0,0)(0,1){4}{\line(1,0){4}}%
    \multiput(1,0)(1,0){3}{\line(0,1){3}}%
    {\color{grownblue}%
      \multiput(1,3)(0,.1){11}{\circle*{.04}}%
      \multiput(2,3)(0,.1){11}{\circle*{.04}}%
      \multiput(3,3)(0,.1){11}{\circle*{.04}}%
    }%
    \NPLabelR{-.35}{5.5}{\small$D(G')$}%
    \NPLabelR{-.35}{1.5}{\small$D(G)$}%
    \NPBlueDot{1}{3.5}\NPBlueDot{2}{3.5}\NPBlueDot{3}{3.5}%
    \NPLabelR{-.35}{3.5}{\color{grownblue}\scriptsize$\ket+_{G'}$}%
  \end{picture}\endgroup
}

\newcommand{\NativeFoldedAncilla}{%
  \begingroup\setlength{\unitlength}{.49cm}%
  \begin{picture}(6.6,8.7)(-2.1,-.2)
    \linethickness{.65pt}%
    \multiput(0,5)(0,1){4}{\line(1,0){4}}%
    \multiput(1,5)(1,0){3}{\line(0,1){3}}%
    \multiput(0,0)(0,1){4}{\line(1,0){4}}%
    \multiput(1,0)(1,0){3}{\line(0,1){3}}%
    {\color{grownblue}%
      \multiput(1,4)(0,.1){11}{\circle*{.04}}%
      \multiput(2,4)(0,.1){11}{\circle*{.04}}%
      \multiput(3,4)(0,.1){11}{\circle*{.04}}%
      \multiput(0,4)(.1,0){41}{\circle*{.04}}%
    }%
    \NPLabelR{-.35}{6.5}{\small$D(G')$}%
    \NPLabelR{-.35}{1.5}{\small$D(G)$}%
    \NPBlueDot{1}{4.5}\NPBlueDot{2}{4.5}\NPBlueDot{3}{4.5}%
    \NPBlueDot{.5}{4}\NPBlueDot{1.5}{4}%
    \NPBlueDot{2.5}{4}\NPBlueDot{3.5}{4}%
    \NPLabelR{-.35}{4.65}{\color{grownblue}\scriptsize$\ket+_{G'}$}%
    \NPLabelR{-.35}{3.75}{\color{grownblue}\scriptsize$\ket+_{N'}$}%
  \end{picture}\endgroup
}

\newcommand{\NativeCNOTLayout}{%
  \begingroup\setlength{\unitlength}{.64cm}%
  \begin{picture}(10.4,6.85)(-.9,0)
    \linethickness{.65pt}%
    \multiput(0,3.75)(0,.75){5}{\line(1,0){4}}%
    \multiput(1,3.75)(1,0){3}{\line(0,1){3}}%
    \multiput(5,3.75)(0,.75){5}{\line(1,0){4}}%
    \multiput(6,3.75)(1,0){3}{\line(0,1){3}}%
    \multiput(0,0)(0,.75){5}{\line(1,0){4}}%
    \multiput(1,0)(1,0){3}{\line(0,1){3}}%
    \NPLabelR{-.65}{6.37}{$\mathrm{INT}$}%
    \NPLabelL{9.65}{6.37}{$T$}%
    \NPLabelR{-.65}{2.62}{$C$}%
    \NPBlueDot{4.5}{4.125}\NPBlueDot{4.5}{4.875}%
    \NPBlueDot{4.5}{5.625}\NPBlueDot{4.5}{6.375}%
    \NPBlueDot{1}{3.375}\NPBlueDot{2}{3.375}\NPBlueDot{3}{3.375}%
  \end{picture}\endgroup
}

\newcommand{\NativeThreePatchLayout}{%
  \begingroup\setlength{\unitlength}{.66cm}%
  \begin{picture}(11.4,2.95)(-.1,0)
    \linethickness{.65pt}%
    \multiput(0,0)(.8,0){5}{\line(0,1){2.30}}%
    \multiput(0,.575)(0,.575){3}{\line(1,0){3.20}}%
    \multiput(4,0)(.8,0){5}{\line(0,1){2.30}}%
    \multiput(4,.575)(0,.575){3}{\line(1,0){3.20}}%
    \multiput(8,0)(.8,0){5}{\line(0,1){2.30}}%
    \multiput(8,.575)(0,.575){3}{\line(1,0){3.20}}%
    \NPLabel{1.6}{2.68}{$D(\mathbb Z_4)$}%
    \NPLabel{5.6}{2.68}{$D(D_4)$}%
    \NPLabel{9.6}{2.68}{$D(\mathbb Z_2)$}%
    \NPBlueDot{3.6}{.575}\NPBlueDot{3.6}{1.15}\NPBlueDot{3.6}{1.725}%
    \NPBlueDot{7.6}{.575}\NPBlueDot{7.6}{1.15}\NPBlueDot{7.6}{1.725}%
  \end{picture}\endgroup
}

\newcommand{\SMSAvInput}{%
  \begingroup\setlength{\unitlength}{.70cm}%
  \begin{picture}(5.6,3.7)(-2.8,-1.85)
    \linethickness{.75pt}\NPLine{-2.3}{0}{2.3}{0}\NPLine{0}{-1.35}{0}{1.35}%
    \NPLatticeArrow{-1.15}{0}{0}\NPLatticeArrow{1.15}{0}{0}%
    \NPLatticeArrow{0}{-.675}{90}\NPLatticeArrow{0}{.675}{90}%
    \NPLabel{-1.35}{.60}{\small$\ket{h_i}$}\NPLabel{1.35}{-.60}{\small$\ket{h_{i+1}}$}%
    \NPLabelR{-.40}{-1.10}{\small$\ket g$}\NPLabelL{.40}{1.10}{\small$\ket{q'}$}%
  \end{picture}\endgroup
}
\newcommand{\SMSAvOutput}{%
  \begingroup\setlength{\unitlength}{.62cm}%
  \begin{picture}(7.8,4.2)(-3.9,-2.1)
    \linethickness{.75pt}\NPLine{-3.6}{0}{3.6}{0}\NPLine{0}{-1.62}{0}{1.62}%
    \NPLatticeArrow{-1.8}{0}{0}\NPLatticeArrow{1.8}{0}{0}%
    \NPLatticeArrow{0}{-.81}{90}\NPLatticeArrow{0}{.81}{90}%
    \NPLabel{-2.12}{.65}{\small$\ket{h_ih^{-1}}$}%
    \NPLabel{2.12}{-.65}{\small$\ket{hh_{i+1}}$}%
    \NPLabelR{-.40}{-1.25}{\small$\ket{gh^{-1}}$}%
    \NPLabelL{.40}{1.25}{\small$\ket{h'q'}$}%
  \end{picture}\endgroup
}

\newcommand{\NativeAvUnfoldedInput}{%
  \begingroup\setlength{\unitlength}{.70cm}%
  \raisebox{-.65ex}{\begin{picture}(6.4,3.9)(-3.2,-1.95)
    \linethickness{.75pt}\NPLine{-2.85}{0}{2.85}{0}\NPLine{0}{-1.35}{0}{1.35}%
    \NPLatticeArrow{-1.42}{0}{0}\NPLatticeArrow{1.42}{0}{0}%
    \NPLatticeArrow{0}{-.61}{90}\NPLatticeArrow{0}{.74}{90}\put(0,0){\circle*{.11}}%
    \NPLabel{-1.65}{.62}{\small$\ket{h_i,h'_i}$}\NPLabel{1.65}{-.62}{\small$\ket{h_{i+1},h'_{i+1}}$}%
    \NPLabelR{-.42}{-1.14}{\small$\ket g$}\NPLabelL{.42}{1.14}{\small$\ket{q'}$}%
    \NPLabel{-.25}{-.32}{\scriptsize$v$}%
  \end{picture}}\endgroup
}
\newcommand{\NativeAvUnfoldedOutput}{%
  \begingroup\setlength{\unitlength}{.62cm}%
  \raisebox{-.65ex}{\begin{picture}(10.4,4.5)(-5.2,-2.25)
    \linethickness{.75pt}\NPLine{-4.8}{0}{4.8}{0}\NPLine{0}{-1.80}{0}{1.80}%
    \NPLatticeArrow{-2.40}{0}{0}\NPLatticeArrow{2.40}{0}{0}%
    \NPLatticeArrow{0}{-.90}{90}\NPLatticeArrow{0}{.90}{90}\put(0,0){\circle*{.12}}%
    \NPLabel{-2.70}{.68}{\scriptsize$\bigl(R^h\!\otimes R^{h'}\bigr)\ket{h_i,h'_i}$}%
    \NPLabel{2.70}{-.68}{\scriptsize$\bigl(L^h\!\otimes L^{h'}\bigr)\ket{h_{i+1},h'_{i+1}}$}%
    \NPLabelR{-.42}{-1.42}{\small$R^h\ket g$}\NPLabelL{.42}{1.42}{\small$L^{h'}\ket{q'}$}%
    \NPLabel{-.25}{-.32}{\scriptsize$v$}%
  \end{picture}}\endgroup
}

\newcommand{\NativeNLegMultTensor}{%
  \begingroup\setlength{\unitlength}{.72cm}%
  \raisebox{-.65ex}{\begin{picture}(4.1,2.9)(-2.05,-1.70)
    \linethickness{.82pt}%
    \NPLine{-1.35}{.82}{0}{0}\NPLine{-1.35}{-.82}{0}{0}%
    \NPLine{0}{0}{1.35}{-.82}\NPLine{0}{0}{1.35}{.82}%
    \NPArrow{-.68}{.41}{-31}\NPArrow{-.68}{-.41}{31}%
    \NPArrow{.68}{-.41}{-31}\NPArrow{.68}{.41}{31}%
    \NPMult{0}{0}%
    \NPLabelR{-1.50}{.82}{\scriptsize$g_1$}\NPLabelR{-1.50}{-.82}{\scriptsize$g_2$}%
    \NPLabelL{1.50}{-.82}{\scriptsize$g_{n-1}$}\NPLabelL{1.50}{.82}{\scriptsize$g_n$}%
    \NPLabel{0}{-1.20}{\scriptsize$\cdots$}%
  \end{picture}}\endgroup
}

\newcommand{\NativeCopyTensor}{%
  \begingroup\setlength{\unitlength}{.65cm}%
  \raisebox{-.65ex}{\begin{picture}(2.7,2.55)(-1.35,-1.22)
    \linethickness{.82pt}\NPCopy{0}{0}%
    \NPLine{0}{0}{-.91}{.525}\NPLine{0}{0}{-.742}{-.742}%
    \NPLine{0}{0}{.742}{-.742}\NPLine{0}{0}{.91}{.525}\NPLine{0}{0}{0}{1.05}%
    \NPLabelR{-1.02}{.59}{\scriptsize$g_1$}\NPLabelR{-.83}{-.83}{\scriptsize$g_2$}%
    \NPLabelL{.83}{-.83}{\scriptsize$g_3$}\NPLabelL{1.02}{.59}{\scriptsize$g_4$}%
    \NPLabel{0}{1.18}{\scriptsize$\cdots$}%
  \end{picture}}\endgroup
}

\newcommand{\NPPlaneSlice}[2]{%
  \NPLine{-2.15}{#1}{.95}{#1}%
  \NPLine{.95}{#1}{2.15}{#2}%
  \NPLine{2.15}{#2}{-.95}{#2}%
  \NPLine{-.95}{#2}{-2.15}{#1}%
}

\newcommand{\PlaquetteCompact}[4]{%
  \begingroup\setlength{\unitlength}{.80cm}%
  \begin{picture}(5.3,3.7)(-2.65,-1.85)
    {\color{paperprimary!65}\linethickness{.5pt}%
      \NPPlaneSlice{-.45}{.45}}%
    \linethickness{.8pt}%
    \NPLine{-1.55}{-.90}{-1.55}{.90}%
    \NPLine{-.60}{-1.35}{-.60}{.45}%
    \NPLine{.60}{-.45}{.60}{1.35}%
    \NPLine{1.55}{-.90}{1.55}{.90}%
    \NPLine{0}{0}{-.60}{-.45}\NPLine{-1.55}{0}{0}{0}%
    \NPLine{.60}{.45}{0}{0}\NPLine{0}{0}{1.55}{0}%
    \NPArrow{-.33}{-.2475}{217}\NPArrow{-.85}{0}{0}%
    \NPArrow{.33}{.2475}{217}\NPArrow{.85}{0}{0}%
    \NPCopy{-.60}{-.45}\NPCopy{-1.55}{0}%
    \NPCopy{.60}{.45}\NPCopy{1.55}{0}%
    \NPMult{0}{0}%
    \NPLabel{-.60}{-1.62}{\scriptsize$#1$}%
    \NPLabel{-1.55}{1.18}{\scriptsize$#2$}%
    \NPLabel{.60}{1.62}{\scriptsize$#3$}%
    \NPLabel{1.55}{-1.18}{\scriptsize$#4$}%
  \end{picture}\endgroup
}
\newcommand{\InterfacePlaquetteCompact}[4]{%
  \begingroup\setlength{\unitlength}{.80cm}%
  \begin{picture}(5.3,4.3)(-2.65,-2.15)
    {\color{paperprimary!65}\linethickness{.5pt}%
      \NPPlaneSlice{-.45}{.45}}%
    \linethickness{.8pt}%
    \NPLine{-1.55}{-.90}{-1.55}{.90}%
    \NPLine{-.60}{-1.35}{-.60}{.45}%
    \NPLine{.60}{-.45}{.60}{1.35}%
    \NPLine{1.55}{-.90}{1.55}{.90}%
    \NPLine{0}{0}{-.60}{-.45}\NPLine{-1.55}{0}{0}{0}%
    \NPLine{.60}{.45}{0}{0}\NPLine{0}{0}{1.55}{0}%
    \NPArrow{-.33}{-.2475}{217}\NPArrow{-.85}{0}{0}%
    \NPArrow{.33}{.2475}{217}\NPArrow{.85}{0}{0}%
    \NPCopy{-.60}{-.45}\NPCopy{-1.55}{0}%
    \NPCopy{.60}{.45}\NPCopy{1.55}{0}%
    \NPMult{0}{0}%
    \NPLabel{-.60}{-1.84}{\scriptsize$#1$}%
    \NPLabel{-1.55}{1.40}{\scriptsize$#2$}%
    \NPLabel{.60}{1.84}{\scriptsize$#3$}%
    \NPLabel{1.55}{-1.40}{\scriptsize$#4$}%
  \end{picture}\endgroup
}
\newcommand{\PlaquetteCircuitFrame}{%
  {\color{paperprimary!65}\linethickness{.45pt}%
    \NPPlaneSlice{-2.08}{-1.52}\NPPlaneSlice{-.88}{-.32}%
    \NPPlaneSlice{.32}{.88}\NPPlaneSlice{1.52}{2.08}}%
  \linethickness{.8pt}%
  \NPLine{-1.55}{-2.40}{-1.55}{2.40}%
  \NPLine{-.60}{-2.40}{-.60}{2.40}%
  \NPLine{.60}{-2.40}{.60}{2.40}%
  \NPLine{1.55}{-2.40}{1.55}{2.40}%
  \NPLine{0}{-2.88}{0}{2.88}%
  \NPArrow{0}{-2.48}{90}\NPArrow{0}{-1.20}{90}%
  \NPArrow{0}{0}{90}\NPArrow{0}{1.20}{90}\NPArrow{0}{2.48}{90}%
}
\newcommand{\RegularPlaquetteNetwork}[1]{%
  \begingroup\setlength{\unitlength}{.70cm}%
  \begin{picture}(5.8,7.9)(-2.9,-3.9)
    \PlaquetteCircuitFrame
    \NPCopy{-.60}{-2.08}\NPLine{0}{-1.80}{-.60}{-2.08}%
    \NPArrow{-.33}{-1.954}{205}\NPMult{0}{-1.80}\NPCross{0}{-1.80}%
    \NPCopy{-1.55}{-.60}\NPLine{-1.55}{-.60}{0}{-.60}%
    \NPArrow{-.85}{-.60}{0}\NPMult{0}{-.60}\NPCross{0}{-.60}%
    \NPCopy{.60}{.88}\NPLine{.60}{.88}{0}{.60}%
    \NPArrow{.33}{.754}{205}\NPMult{0}{.60}%
    \NPCopy{1.55}{1.80}\NPLine{0}{1.80}{1.55}{1.80}%
    \NPArrow{.85}{1.80}{0}\NPMult{0}{1.80}%
    \NPLabel{-1.55}{-2.68}{\scriptsize$g'_2$}%
    \NPLabel{-.60}{-2.68}{\scriptsize$g'_1$}%
    \NPLabel{.60}{-2.68}{\scriptsize$g'_3$}%
    \NPLabel{1.55}{-2.68}{\scriptsize$g'_4$}%
    \NPMult{0}{-2.88}\NPLabel{0}{-3.62}{\scriptsize$\ket{e}_{A}$}%
    \ifnum#1=0
      \NPMult{0}{2.88}\NPLabel{0}{3.52}{\scriptsize${}_{A}\bra e$}%
    \else
      {\linethickness{1.6pt}\NPLine{0}{2.88}{0}{3.50}}%
      \NPLabelL{.25}{3.52}{\scriptsize$m'_p$}%
    \fi
  \end{picture}\endgroup
}
\newcommand{\CosetPlaquetteNetwork}[1]{%
  \begingroup\setlength{\unitlength}{.74cm}%
  \begin{picture}(5.8,8.6)(-2.9,-4.5)
    \PlaquetteCircuitFrame
    \NPCopy{-.60}{-2.08}\NPLine{0}{-1.80}{-.60}{-2.08}%
    \NPArrow{-.33}{-1.954}{205}\NPMult{0}{-1.80}\NPCross{0}{-1.80}%
    \NPCopy{-1.55}{-.60}\NPLine{-1.55}{-.60}{0}{-.60}%
    \NPArrow{-.85}{-.60}{0}\NPMult{0}{-.60}\NPCross{0}{-.60}%
    \NPCopy{.60}{.88}\NPLine{.60}{.88}{0}{.60}%
    \NPArrow{.33}{.754}{205}\NPMult{0}{.60}%
    \NPCopy{1.55}{1.80}\NPLine{0}{1.80}{1.55}{1.80}%
    \NPArrow{.85}{1.80}{0}\NPMult{0}{1.80}%
    \NPLabel{-1.55}{-2.88}{\scriptsize$q'_{i-1}$}%
    \NPLabel{-.60}{-2.88}{\scriptsize$h_i$}%
    \NPLabel{.60}{-2.88}{\scriptsize$g'_i$}%
    \NPLabel{1.55}{-2.88}{\scriptsize$q'_i$}%
    \NPLine{0}{-2.88}{0}{-3.60}\NPMult{0}{-3.60}\NPLabel{0}{-4.16}{\scriptsize$\ket{N'}_A$}%
    \ifnum#1=0
      \NPMult{0}{2.88}\NPLabel{0}{3.62}{\scriptsize${}_A\bra{N'}$}%
    \else
      {\linethickness{1.6pt}\NPLine{0}{2.88}{0}{3.50}}%
      \NPLabelL{.42}{3.62}{\scriptsize$N'm'_i$}%
    \fi
  \end{picture}\endgroup
}

\newcommand{\NativeVertexCompact}[1]{%
  \begingroup\setlength{\unitlength}{.78cm}%
  \raisebox{-.55ex}{\begin{picture}(4.5,3.90)(-2.25,-1.95)
    {\color{grownblue}\linethickness{.55pt}%
      \NPLine{-2.05}{0}{2.05}{0}\NPLine{-1.45}{-.86}{1.45}{.86}}%
    \linethickness{.82pt}\NPCopy{0}{0}%
    \NPLine{-1.38}{-.94}{-1.38}{.94}\NPLine{1.38}{-.94}{1.38}{.94}%
    \NPLine{-.86}{-1.45}{-.86}{.43}\NPLine{.86}{-.43}{.86}{1.45}%
    \NPArrow{-1.38}{-.47}{90}\NPArrow{-1.38}{.47}{90}%
    \NPArrow{1.38}{-.47}{90}\NPArrow{1.38}{.47}{90}%
    \NPArrow{-.86}{-.98}{90}\NPArrow{-.86}{-.04}{90}%
    \NPArrow{.86}{.04}{90}\NPArrow{.86}{.98}{90}%
    \NPLine{-1.38}{0}{0}{0}\NPLine{-.86}{-.51}{0}{0}%
    \NPLine{0}{0}{1.38}{0}\NPLine{0}{0}{.86}{.51}%
    \NPArrow{-.69}{0}{0}\NPArrow{-.43}{-.255}{31}%
    \NPArrow{.69}{0}{0}\NPArrow{.43}{.255}{31}%
    \NPMult{-1.38}{0}\NPMult{1.38}{0}\NPMult{-.86}{-.51}\NPMult{.86}{.51}%
    #1%
  \end{picture}}\endgroup
}

\newcommand{\NativeVertexCircuit}[1]{%
  \begingroup\setlength{\unitlength}{.62cm}%
  \raisebox{-.55ex}{\begin{picture}(4.8,5.85)(-2.4,-2.72)
    {\color{grownblue}\linethickness{.55pt}%
      \multiput(-2.15,-2.25)(0,.90){6}{\line(1,0){4.30}}%
      \NPLine{-1.20}{-2.73}{1.20}{-1.77}\NPLine{-1.20}{-1.83}{1.20}{-.87}%
      \NPLine{-1.20}{-.93}{1.20}{.03}\NPLine{-1.20}{-.03}{1.20}{.93}%
      \NPLine{-1.20}{.87}{1.20}{1.83}\NPLine{-1.20}{1.77}{1.20}{2.73}}%
    \linethickness{.82pt}\NPLine{0}{-2.25}{0}{2.25}%
    \NPCopy{0}{-2.25}\NPCopy{0}{2.25}%
    \NPCopy{0}{-1.35}\NPLine{-1.90}{-1.35}{0}{-1.35}\NPArrow{-.95}{-1.35}{0}%
    \NPCopy{0}{-.45}\NPLine{-.95}{-.83}{0}{-.45}\NPArrow{-.48}{-.64}{22}%
    \NPCopy{0}{.45}\NPLine{0}{.45}{1.90}{.45}\NPArrow{.95}{.45}{0}%
    \NPCopy{0}{1.35}\NPLine{0}{1.35}{.95}{1.73}\NPArrow{.48}{1.54}{22}%
    \NPLine{-1.90}{-2.25}{-1.90}{2.25}\NPLine{-.95}{-2.25}{-.95}{2.25}%
    \NPLine{.95}{-2.25}{.95}{2.25}\NPLine{1.90}{-2.25}{1.90}{2.25}%
    \NPArrow{-1.90}{-1.80}{90}\NPArrow{-1.90}{.45}{90}%
    \NPArrow{-.95}{-1.54}{90}\NPArrow{-.95}{.71}{90}%
    \NPArrow{.95}{-.26}{90}\NPArrow{.95}{1.99}{90}%
    \NPArrow{1.90}{-1.35}{90}\NPArrow{1.90}{.90}{90}%
    \NPMult{-1.90}{-1.35}\NPMult{-.95}{-.83}%
    \NPMult{1.90}{.45}\NPMult{.95}{1.73}%
    #1%
  \end{picture}}\endgroup
}

\newcommand{\NativeOneLegKet}[1]{%
  \begingroup\setlength{\unitlength}{#1}%
  \raisebox{-.45ex}{\begin{picture}(.6,1.05)(-.3,-.10)
    \linethickness{.82pt}\NPLine{0}{0}{0}{.82}\NPCopy{0}{0}%
  \end{picture}}\endgroup
}
\newcommand{\NativeOneLegBra}[1]{%
  \begingroup\setlength{\unitlength}{#1}%
  \raisebox{-.45ex}{\begin{picture}(.6,1.05)(-.3,-.10)
    \linethickness{.82pt}\NPLine{0}{0}{0}{.82}\NPCopy{0}{.82}%
  \end{picture}}\endgroup
}

\newcommand{\NativeCLGate}{%
  \begingroup\setlength{\unitlength}{.78cm}%
  \raisebox{-1.05ex}{\begin{picture}(2.8,3.2)(-.45,-1.60)
    \linethickness{.82pt}\NPLine{0}{-1.10}{0}{1.10}\NPCopy{0}{0}%
    \NPLine{0}{0}{1.60}{0}\NPArrow{.80}{0}{0}%
    \NPLine{1.60}{-1.10}{1.60}{1.10}\NPArrow{1.60}{-.55}{90}\NPArrow{1.60}{.55}{90}%
    \NPMult{1.60}{0}%
    \NPLabel{0}{-1.48}{\scriptsize$g$}\NPLabel{1.60}{-1.48}{\scriptsize$h$}%
    \NPLabel{0}{1.48}{\scriptsize$g'$}\NPLabel{1.60}{1.48}{\scriptsize$h'$}%
  \end{picture}}\endgroup
}
\newcommand{\NativeControlled}[2]{%
  \begingroup\setlength{\unitlength}{.58cm}%
  \raisebox{-.55ex}{\begin{picture}(1.55,1.9)(-.25,-.95)
    \linethickness{.82pt}\NPLine{0}{-.82}{0}{.82}\NPCopy{0}{0}%
    \NPLine{0}{0}{1}{0}\ifnum#1=0\NPArrow{.50}{0}{180}\else\NPArrow{.50}{0}{0}\fi
    \NPLine{1}{-.82}{1}{.82}\NPArrow{1}{-.41}{90}\NPArrow{1}{.41}{90}%
    \NPMult{1}{0}%
    \ifnum#2=1\NPCross{1}{0}\fi
  \end{picture}}\endgroup
}

\newcommand{\bea}{\begin{eqnarray}}
\newcommand{\eea}{\end{eqnarray}}
\newcommand{\be}{\begin{equation}}
\newcommand{\ee}{\end{equation}}
\newcommand{\ba}{\begin{aligned}}
\newcommand{\ea}{\end{aligned}}
\newcommand{\bit}{\begin{itemize}}
\newcommand{\eit}{\end{itemize}}
\newcommand{\ben}{\begin{enumerate}}
\newcommand{\een}{\end{enumerate}}

\newcommand{\Z}{{\mathbb Z}}

\makeatother

\allowdisplaybreaks

\newcommand{\ket}[1]{\left\lvert{#1}\right\rangle}
\newcommand{\bra}[1]{\left\langle{#1}\right\rvert}
\newcommand{\braket}[2]{%
  \left\langle{#1}\middle\vert{#2}\right\rangle%
}

\usepackage{hyperref}
\hypersetup{
  pdfstartview={FitH},
  pdftitle={},
  colorlinks=true,
  linkcolor=paperprimary,
  citecolor=paperprimary,
  filecolor=paperprimary,
  urlcolor=paperprimary
}

\makeatletter
\newcommand\xlabel[2][]{\phantomsection\def\@currentlabelname{#1}\label{#2}}
\makeatother

\makeatletter
\def\l@subsubsection#1#2{}
\makeatother

\newcommand{\SMSWhite}[1]{{\setlength{\fboxsep}{.6pt}\colorbox{white}{#1}}}

\newcommand{\SMSCorrectionRow}[2]{%
\begingroup\setlength{\unitlength}{.59cm}%
\begin{picture}(13.4,4.15)(-.35,-1.10)
  \linethickness{.65pt}%
  \NPLine{0}{0}{10}{0}\NPLine{0}{2}{10}{2}%
  \multiput(2,-.52)(2,0){4}{\line(0,1){.52}}%
  \multiput(2,2)(2,0){4}{\line(0,1){.52}}%
  {\color{paperprimary}\linethickness{.85pt}%
    \multiput(2,0)(2,0){4}{\line(0,1){2}}}%
  \NPLabel{2}{1.20}{\SMSWhite{\scriptsize$q'_1$}}%
  \NPLabel{4}{1.20}{\SMSWhite{\scriptsize$q'_2$}}%
  \NPLabel{6}{1.20}{\SMSWhite{\scriptsize$q'_3$}}%
  \NPLabel{8}{1.20}{\SMSWhite{\scriptsize$q'_4$}}%
  \NPLabel{1}{2}{\SMSWhite{\scriptsize$g'_1$}}%
  \NPLabel{3}{2}{\SMSWhite{\scriptsize$g'_2$}}%
  \NPLabel{5}{2}{\SMSWhite{\scriptsize$g'_3$}}%
  \NPLabel{7}{2}{\SMSWhite{\scriptsize$g'_4$}}%
  \NPLabel{9}{2}{\SMSWhite{\scriptsize$g'_5$}}%
  \NPLabel{1}{0}{\SMSWhite{\scriptsize$#1_1$}}%
  \NPLabel{3}{0}{\SMSWhite{\scriptsize$#1_2$}}%
  \NPLabel{5}{0}{\SMSWhite{\scriptsize$#1_3$}}%
  \NPLabel{7}{0}{\SMSWhite{\scriptsize$#1_4$}}%
  \NPLabel{9}{0}{\SMSWhite{\scriptsize$#1_5$}}%
  \NPLabelL{10.35}{2}{\scriptsize$D(G')$}%
  \NPLabelL{10.35}{0}{\scriptsize$D(G)$}%
  \ifnum#2=0
    \NPLabel{2}{-.82}{\scriptsize$A_{v_1}^{(t_1)}$}%
    \NPLabel{4}{-.82}{\scriptsize$A_{v_2}^{(t_2)}$}%
    \NPLabel{6}{-.82}{\scriptsize$A_{v_3}^{(t_3)}$}%
    \NPLabel{8}{-.82}{\scriptsize$A_{v_4}^{(t_4)}$}%
    \NPLabel{1}{.86}{\scriptsize$p_1$}%
    \NPLabel{3}{.86}{\scriptsize$p_2$}%
    \NPLabel{5}{.86}{\scriptsize$p_3$}%
    \NPLabel{7}{.86}{\scriptsize$p_4$}%
    \NPLabel{9}{.86}{\scriptsize$p_5$}%
  \else
    \NPLabel{1}{.86}{\small$F_1$}%
    \NPLabel{3}{.86}{\small$F_2$}%
    \NPLabel{5}{.86}{\small$F_3$}%
    \NPLabel{7}{.86}{\small$F_4$}%
    \NPLabel{9}{.86}{\small$F_5$}%
  \fi
\end{picture}\endgroup}
\newcommand{\SMSRestrictionStrip}{%
\begingroup\setlength{\unitlength}{.59cm}%
\begin{picture}(13.4,2.25)(-.35,-1.55)
  \linethickness{.65pt}%
  \NPLine{0}{0}{10}{0}%
  \multiput(2,-.7)(2,0){4}{\line(0,1){.7}}%
  \NPLabel{1}{0}{\SMSWhite{\scriptsize$g_1$}}%
  \NPLabel{3}{0}{\SMSWhite{\scriptsize$g_2$}}%
  \NPLabel{5}{0}{\SMSWhite{\scriptsize$g_3$}}%
  \NPLabel{7}{0}{\SMSWhite{\scriptsize$g_4$}}%
  \NPLabel{9}{0}{\SMSWhite{\scriptsize$g_5$}}%
  \NPLabel{2}{-1.13}{\scriptsize$A_{v_1}^{(t_1)}$}%
  \NPLabel{4}{-1.13}{\scriptsize$A_{v_2}^{(t_2)}$}%
  \NPLabel{6}{-1.13}{\scriptsize$A_{v_3}^{(t_3)}$}%
  \NPLabel{8}{-1.13}{\scriptsize$A_{v_4}^{(t_4)}$}%
  \NPLabelL{10.35}{0}{\scriptsize$D(G)$}%
\end{picture}\endgroup}
\newcommand{\SMSPlaquetteStrip}{\SMSCorrectionRow{h}{1}}

\newcommand{\SMSUnfoldedCell}{%
\begingroup\setlength{\unitlength}{0.56cm}%
\begin{picture}(3.55,2.6)(-0.58,-0.55)
\linethickness{.55pt}\NPLine{-0.3}{0}{2.7}{0}%
\linethickness{.55pt}\NPLine{-0.3}{1.5}{2.7}{1.5}%
\linethickness{.55pt}\NPLine{0}{-0.38}{0}{1.9}%
\linethickness{.55pt}\NPLine{2.4}{-0.38}{2.4}{1.9}%
\NPLabel{1.2}{1.5}{\SMSWhite{\scriptsize $g'_i$}}%
\NPLabel{0}{0.75}{\SMSWhite{\scriptsize $q'_{i-1}$}}%
\NPLabel{2.4}{0.75}{\SMSWhite{\scriptsize $q'_i$}}%
\NPLabel{1.2}{0}{\SMSWhite{\scriptsize $h_i$}}%
\end{picture}\endgroup}
\newcommand{\SMSFoldedCell}{%
\begingroup\setlength{\unitlength}{0.56cm}%
\begin{picture}(3.55,3.35)(-0.58,-1.25)
\linethickness{.55pt}\NPLine{-0.3}{0}{2.7}{0}%
\linethickness{.55pt}\NPLine{-0.3}{1.5}{2.7}{1.5}%
\linethickness{.55pt}\NPLine{0}{0}{0}{1.9}%
\linethickness{.55pt}\NPLine{2.4}{0}{2.4}{1.9}%
\NPLabel{1.2}{1.5}{\SMSWhite{\scriptsize $g'_i$}}%
\NPLabel{0}{0.75}{\SMSWhite{\scriptsize $q'_{i-1}$}}%
\NPLabel{2.4}{0.75}{\SMSWhite{\scriptsize $q'_i$}}%
\NPLabel{1.2}{0}{\SMSWhite{\scriptsize $h'_i$}}%
\linethickness{.55pt}\NPLine{-0.3}{-0.72}{2.7}{-0.72}%
\NPLabel{1.2}{-0.72}{\SMSWhite{\scriptsize $h_i$}}%
\linethickness{.55pt}\NPLine{0}{-1.1}{0}{-0.72}%
\put(0,-0.6){\circle*{.025}}%
\put(0,-0.48){\circle*{.025}}%
\put(0,-0.36){\circle*{.025}}%
\put(0,-0.24){\circle*{.025}}%
\put(0,-0.12){\circle*{.025}}%
\linethickness{.55pt}\NPLine{2.4}{-1.1}{2.4}{-0.72}%
\put(2.4,-0.6){\circle*{.025}}%
\put(2.4,-0.48){\circle*{.025}}%
\put(2.4,-0.36){\circle*{.025}}%
\put(2.4,-0.24){\circle*{.025}}%
\put(2.4,-0.12){\circle*{.025}}%
\end{picture}\endgroup}
\newcommand{\SMSOriginal}{%
\begingroup\setlength{\unitlength}{0.71cm}%
\begin{picture}(6.6,8.0)(-0.3,-0.7)
\linethickness{.55pt}\NPLine{0}{0}{6}{0}%
\linethickness{.55pt}\NPLine{0}{1}{6}{1}%
\linethickness{.55pt}\NPLine{0}{2}{6}{2}%
\linethickness{.55pt}\NPLine{0}{3}{6}{3}%
\linethickness{.55pt}\NPLine{1}{0}{1}{3}%
\linethickness{.55pt}\NPLine{2}{0}{2}{3}%
\linethickness{.55pt}\NPLine{3}{0}{3}{3}%
\linethickness{.55pt}\NPLine{4}{0}{4}{3}%
\linethickness{.55pt}\NPLine{5}{0}{5}{3}%
\linethickness{.55pt}\NPLine{0}{4}{6}{4}%
\linethickness{.55pt}\NPLine{0}{5}{6}{5}%
\linethickness{.55pt}\NPLine{0}{6}{6}{6}%
\linethickness{.55pt}\NPLine{0}{7}{6}{7}%
\linethickness{.55pt}\NPLine{1}{4}{1}{7}%
\linethickness{.55pt}\NPLine{2}{4}{2}{7}%
\linethickness{.55pt}\NPLine{3}{4}{3}{7}%
\linethickness{.55pt}\NPLine{4}{4}{4}{7}%
\linethickness{.55pt}\NPLine{5}{4}{5}{7}%
{\color{paperprimary}\linethickness{3.3pt}\NPLine{3.57}{2}{4.43}{2}}%
{\color{paperprimary}\linethickness{3.3pt}\NPLine{4}{1.57}{4}{2.43}}%
{\color{papersecondary}\linethickness{1.8pt}\NPLine{1}{1}{2}{1}}%
{\color{papersecondary}\linethickness{1.8pt}\NPLine{2}{1}{2}{2}}%
{\color{papersecondary}\linethickness{1.8pt}\NPLine{2}{2}{1}{2}}%
{\color{papersecondary}\linethickness{1.8pt}\NPLine{1}{2}{1}{1}}%
\NPLabel{4.78}{2.4}{\SMSWhite{\color{paperprimary}\scriptsize $A_v^{G}$}}%
\NPLabel{2.55}{1.5}{\SMSWhite{\color{papersecondary}\scriptsize $B_p^e$}}%
{\color{paperprimary}\linethickness{3.3pt}\NPLine{3.57}{6}{4.43}{6}}%
{\color{paperprimary}\linethickness{3.3pt}\NPLine{4}{5.57}{4}{6.43}}%
{\color{papersecondary}\linethickness{1.8pt}\NPLine{1}{5}{2}{5}}%
{\color{papersecondary}\linethickness{1.8pt}\NPLine{2}{5}{2}{6}}%
{\color{papersecondary}\linethickness{1.8pt}\NPLine{2}{6}{1}{6}}%
{\color{papersecondary}\linethickness{1.8pt}\NPLine{1}{6}{1}{5}}%
\NPLabel{4.78}{6.4}{\SMSWhite{\color{paperprimary}\scriptsize $A_v^{G'}$}}%
\NPLabel{2.55}{5.5}{\SMSWhite{\color{papersecondary}\scriptsize $B_p^e$}}%
\NPLabel{3}{7.45}{\SMSWhite{\scriptsize $D(G')$}}%
\NPLabel{3}{-0.45}{\SMSWhite{\scriptsize $D(G)$}}%
{\color{paperprimary}\linethickness{3.3pt}\NPLine{0.57}{4}{1.43}{4}}%
{\color{paperprimary}\linethickness{3.3pt}\NPLine{1}{4}{1}{4.43}}%
{\color{paperprimary}\linethickness{3.3pt}\NPLine{3.57}{3}{4.43}{3}}%
{\color{paperprimary}\linethickness{3.3pt}\NPLine{4}{2.57}{4}{3}}%
\NPLabel{1.0}{3.72}{\SMSWhite{\color{paperprimary}\scriptsize $A_{v,\partial}^{G'}$}}%
\NPLabel{4.0}{3.28}{\SMSWhite{\color{paperprimary}\scriptsize $A_{v,\partial}^{G}$}}%
\end{picture}\endgroup}
\newcommand{\SMSRetained}{%
\begingroup\setlength{\unitlength}{0.71cm}%
\begin{picture}(6.6,8.0)(-0.3,-0.7)
\linethickness{.55pt}\NPLine{0}{0}{6}{0}%
\linethickness{.55pt}\NPLine{0}{1}{6}{1}%
\linethickness{.55pt}\NPLine{0}{2}{6}{2}%
\linethickness{.55pt}\NPLine{0}{3}{6}{3}%
\linethickness{.55pt}\NPLine{1}{0}{1}{3}%
\linethickness{.55pt}\NPLine{2}{0}{2}{3}%
\linethickness{.55pt}\NPLine{3}{0}{3}{3}%
\linethickness{.55pt}\NPLine{4}{0}{4}{3}%
\linethickness{.55pt}\NPLine{5}{0}{5}{3}%
\linethickness{.55pt}\NPLine{0}{4}{6}{4}%
\linethickness{.55pt}\NPLine{0}{5}{6}{5}%
\linethickness{.55pt}\NPLine{0}{6}{6}{6}%
\linethickness{.55pt}\NPLine{0}{7}{6}{7}%
\linethickness{.55pt}\NPLine{1}{4}{1}{7}%
\linethickness{.55pt}\NPLine{2}{4}{2}{7}%
\linethickness{.55pt}\NPLine{3}{4}{3}{7}%
\linethickness{.55pt}\NPLine{4}{4}{4}{7}%
\linethickness{.55pt}\NPLine{5}{4}{5}{7}%
{\color{paperprimary}\linethickness{3.3pt}\NPLine{3.57}{2}{4.43}{2}}%
{\color{paperprimary}\linethickness{3.3pt}\NPLine{4}{1.57}{4}{2.43}}%
{\color{papersecondary}\linethickness{1.8pt}\NPLine{1}{1}{2}{1}}%
{\color{papersecondary}\linethickness{1.8pt}\NPLine{2}{1}{2}{2}}%
{\color{papersecondary}\linethickness{1.8pt}\NPLine{2}{2}{1}{2}}%
{\color{papersecondary}\linethickness{1.8pt}\NPLine{1}{2}{1}{1}}%
\NPLabel{4.78}{2.4}{\SMSWhite{\color{paperprimary}\scriptsize $A_v^{G}$}}%
\NPLabel{2.55}{1.5}{\SMSWhite{\color{papersecondary}\scriptsize $B_p^e$}}%
{\color{paperprimary}\linethickness{3.3pt}\NPLine{3.57}{6}{4.43}{6}}%
{\color{paperprimary}\linethickness{3.3pt}\NPLine{4}{5.57}{4}{6.43}}%
{\color{papersecondary}\linethickness{1.8pt}\NPLine{1}{5}{2}{5}}%
{\color{papersecondary}\linethickness{1.8pt}\NPLine{2}{5}{2}{6}}%
{\color{papersecondary}\linethickness{1.8pt}\NPLine{2}{6}{1}{6}}%
{\color{papersecondary}\linethickness{1.8pt}\NPLine{1}{6}{1}{5}}%
\NPLabel{4.78}{6.4}{\SMSWhite{\color{paperprimary}\scriptsize $A_v^{G'}$}}%
\NPLabel{2.55}{5.5}{\SMSWhite{\color{papersecondary}\scriptsize $B_p^e$}}%
\NPLabel{3}{7.45}{\SMSWhite{\scriptsize $D(G')$}}%
\NPLabel{3}{-0.45}{\SMSWhite{\scriptsize $D(G)$}}%
{\color{paperprimary}\linethickness{3.3pt}\NPLine{0.57}{4}{1.43}{4}}%
{\color{paperprimary}\linethickness{3.3pt}\NPLine{1}{4}{1}{4.43}}%
{\color{paperprimary}\linethickness{3.3pt}\NPLine{3.57}{3}{4.43}{3}}%
{\color{paperprimary}\linethickness{3.3pt}\NPLine{4}{2.57}{4}{3}}%
\NPLabel{1.0}{3.72}{\SMSWhite{\color{paperprimary}\scriptsize $A_{v,\partial}^{G'}$}}%
\NPLabel{4.0}{3.28}{\SMSWhite{\color{paperprimary}\scriptsize $A_{v,\partial}^{H}$}}%
\end{picture}\endgroup}
\newcommand{\SMSUnfolded}{%
\begingroup\setlength{\unitlength}{0.71cm}%
\begin{picture}(6.6,8.0)(-0.3,-0.7)
\linethickness{.55pt}\NPLine{0}{0}{6}{0}%
\linethickness{.55pt}\NPLine{0}{1}{6}{1}%
\linethickness{.55pt}\NPLine{0}{2}{6}{2}%
\linethickness{.55pt}\NPLine{0}{3}{6}{3}%
\linethickness{.55pt}\NPLine{1}{0}{1}{3}%
\linethickness{.55pt}\NPLine{2}{0}{2}{3}%
\linethickness{.55pt}\NPLine{3}{0}{3}{3}%
\linethickness{.55pt}\NPLine{4}{0}{4}{3}%
\linethickness{.55pt}\NPLine{5}{0}{5}{3}%
\linethickness{.55pt}\NPLine{0}{4}{6}{4}%
\linethickness{.55pt}\NPLine{0}{5}{6}{5}%
\linethickness{.55pt}\NPLine{0}{6}{6}{6}%
\linethickness{.55pt}\NPLine{0}{7}{6}{7}%
\linethickness{.55pt}\NPLine{1}{4}{1}{7}%
\linethickness{.55pt}\NPLine{2}{4}{2}{7}%
\linethickness{.55pt}\NPLine{3}{4}{3}{7}%
\linethickness{.55pt}\NPLine{4}{4}{4}{7}%
\linethickness{.55pt}\NPLine{5}{4}{5}{7}%
{\color{paperprimary}\linethickness{3.3pt}\NPLine{3.57}{2}{4.43}{2}}%
{\color{paperprimary}\linethickness{3.3pt}\NPLine{4}{1.57}{4}{2.43}}%
{\color{papersecondary}\linethickness{1.8pt}\NPLine{1}{1}{2}{1}}%
{\color{papersecondary}\linethickness{1.8pt}\NPLine{2}{1}{2}{2}}%
{\color{papersecondary}\linethickness{1.8pt}\NPLine{2}{2}{1}{2}}%
{\color{papersecondary}\linethickness{1.8pt}\NPLine{1}{2}{1}{1}}%
\NPLabel{4.78}{2.4}{\SMSWhite{\color{paperprimary}\scriptsize $A_v^{G}$}}%
\NPLabel{2.55}{1.5}{\SMSWhite{\color{papersecondary}\scriptsize $B_p^e$}}%
{\color{paperprimary}\linethickness{3.3pt}\NPLine{3.57}{6}{4.43}{6}}%
{\color{paperprimary}\linethickness{3.3pt}\NPLine{4}{5.57}{4}{6.43}}%
{\color{papersecondary}\linethickness{1.8pt}\NPLine{1}{5}{2}{5}}%
{\color{papersecondary}\linethickness{1.8pt}\NPLine{2}{5}{2}{6}}%
{\color{papersecondary}\linethickness{1.8pt}\NPLine{2}{6}{1}{6}}%
{\color{papersecondary}\linethickness{1.8pt}\NPLine{1}{6}{1}{5}}%
\NPLabel{4.78}{6.4}{\SMSWhite{\color{paperprimary}\scriptsize $A_v^{G'}$}}%
\NPLabel{2.55}{5.5}{\SMSWhite{\color{papersecondary}\scriptsize $B_p^e$}}%
\NPLabel{3}{7.45}{\SMSWhite{\scriptsize $D(G')$}}%
\NPLabel{3}{-0.45}{\SMSWhite{\scriptsize $D(G)$}}%
{\color{paperprimary}\linethickness{3.3pt}\NPLine{0.57}{4}{1.43}{4}}%
{\color{paperprimary}\linethickness{3.3pt}\NPLine{1}{3.57}{1}{4.43}}%
\NPLabel{1}{4.65}{\SMSWhite{\color{paperprimary}\scriptsize $A_v^{G'}$}}%
{\color{paperprimary}\linethickness{.55pt}\NPLine{1}{3}{1}{4}}%
\NPLatticeArrow{1}{3.5}{90}%
{\color{paperprimary}\linethickness{.55pt}\NPLine{2}{3}{2}{4}}%
\NPLatticeArrow{2}{3.5}{90}%
{\color{paperprimary}\linethickness{.55pt}\NPLine{3}{3}{3}{4}}%
\NPLatticeArrow{3}{3.5}{90}%
{\color{paperprimary}\linethickness{.55pt}\NPLine{4}{3}{4}{4}}%
\NPLatticeArrow{4}{3.5}{90}%
{\color{paperprimary}\linethickness{.55pt}\NPLine{5}{3}{5}{4}}%
\NPLatticeArrow{5}{3.5}{90}%
{\color{paperprimary}\linethickness{3.3pt}\NPLine{0.57}{3}{1.43}{3}}%
{\color{paperprimary}\linethickness{3.3pt}\NPLine{1}{2.57}{1}{3.43}}%
{\color{papersecondary}\linethickness{1.8pt}\NPLine{3}{3}{4}{3}}%
{\color{papersecondary}\linethickness{1.8pt}\NPLine{4}{3}{4}{4}}%
{\color{papersecondary}\linethickness{1.8pt}\NPLine{4}{4}{3}{4}}%
{\color{papersecondary}\linethickness{1.8pt}\NPLine{3}{4}{3}{3}}%
\NPLabel{1.15}{2.34}{\SMSWhite{\color{paperprimary}\scriptsize $A_{v,\mathrm{interface}}^K$}}%
\NPLabel{3.5}{4.5}{\SMSWhite{\color{papersecondary}\scriptsize $B_{p,\mathrm{interface}}^e$}}%
\end{picture}\endgroup}
\newcommand{\SMSFolded}{%
\begingroup\setlength{\unitlength}{0.71cm}%
\begin{picture}(6.6,8.4)(-0.3,-0.7)
\linethickness{.55pt}\NPLine{0}{0}{6}{0}%
\linethickness{.55pt}\NPLine{0}{1}{6}{1}%
\linethickness{.55pt}\NPLine{0}{2}{6}{2}%
\linethickness{.55pt}\NPLine{0}{3}{6}{3}%
\linethickness{.55pt}\NPLine{1}{0}{1}{3}%
\linethickness{.55pt}\NPLine{2}{0}{2}{3}%
\linethickness{.55pt}\NPLine{3}{0}{3}{3}%
\linethickness{.55pt}\NPLine{4}{0}{4}{3}%
\linethickness{.55pt}\NPLine{5}{0}{5}{3}%
\linethickness{.55pt}\NPLine{0}{4.4}{6}{4.4}%
\linethickness{.55pt}\NPLine{0}{5.4}{6}{5.4}%
\linethickness{.55pt}\NPLine{0}{6.4}{6}{6.4}%
\linethickness{.55pt}\NPLine{0}{7.4}{6}{7.4}%
\linethickness{.55pt}\NPLine{1}{4.4}{1}{7.4}%
\linethickness{.55pt}\NPLine{2}{4.4}{2}{7.4}%
\linethickness{.55pt}\NPLine{3}{4.4}{3}{7.4}%
\linethickness{.55pt}\NPLine{4}{4.4}{4}{7.4}%
\linethickness{.55pt}\NPLine{5}{4.4}{5}{7.4}%
{\color{paperprimary}\linethickness{3.3pt}\NPLine{3.57}{2}{4.43}{2}}%
{\color{paperprimary}\linethickness{3.3pt}\NPLine{4}{1.57}{4}{2.43}}%
{\color{papersecondary}\linethickness{1.8pt}\NPLine{1}{1}{2}{1}}%
{\color{papersecondary}\linethickness{1.8pt}\NPLine{2}{1}{2}{2}}%
{\color{papersecondary}\linethickness{1.8pt}\NPLine{2}{2}{1}{2}}%
{\color{papersecondary}\linethickness{1.8pt}\NPLine{1}{2}{1}{1}}%
\NPLabel{4.78}{2.4}{\SMSWhite{\color{paperprimary}\scriptsize $A_v^{G}$}}%
\NPLabel{2.55}{1.5}{\SMSWhite{\color{papersecondary}\scriptsize $B_p^e$}}%
{\color{paperprimary}\linethickness{3.3pt}\NPLine{3.57}{6.4}{4.43}{6.4}}%
{\color{paperprimary}\linethickness{3.3pt}\NPLine{4}{5.97}{4}{6.83}}%
{\color{papersecondary}\linethickness{1.8pt}\NPLine{1}{5.4}{2}{5.4}}%
{\color{papersecondary}\linethickness{1.8pt}\NPLine{2}{5.4}{2}{6.4}}%
{\color{papersecondary}\linethickness{1.8pt}\NPLine{2}{6.4}{1}{6.4}}%
{\color{papersecondary}\linethickness{1.8pt}\NPLine{1}{6.4}{1}{5.4}}%
\NPLabel{4.78}{6.8}{\SMSWhite{\color{paperprimary}\scriptsize $A_v^{G'}$}}%
\NPLabel{2.55}{5.9}{\SMSWhite{\color{papersecondary}\scriptsize $B_p^e$}}%
\NPLabel{3}{7.85}{\SMSWhite{\scriptsize $D(G')$}}%
\NPLabel{3}{-0.45}{\SMSWhite{\scriptsize $D(G)$}}%
{\color{paperprimary}\linethickness{3.3pt}\NPLine{0.57}{4.4}{1.43}{4.4}}%
{\color{paperprimary}\linethickness{3.3pt}\NPLine{1}{3.97}{1}{4.83}}%
\NPLabel{1}{5.05}{\SMSWhite{\color{paperprimary}\scriptsize $A_v^{G'}$}}%
{\color{paperprimary}\linethickness{.55pt}\NPLine{1}{3.4}{1}{4.4}}%
\NPLatticeArrow{1}{3.9}{90}%
{\color{paperprimary}\linethickness{.55pt}\NPLine{2}{3.4}{2}{4.4}}%
\NPLatticeArrow{2}{3.9}{90}%
{\color{paperprimary}\linethickness{.55pt}\NPLine{3}{3.4}{3}{4.4}}%
\NPLatticeArrow{3}{3.9}{90}%
{\color{paperprimary}\linethickness{.55pt}\NPLine{4}{3.4}{4}{4.4}}%
\NPLatticeArrow{4}{3.9}{90}%
{\color{paperprimary}\linethickness{.55pt}\NPLine{5}{3.4}{5}{4.4}}%
\NPLatticeArrow{5}{3.9}{90}%
{\color{paperprimary}\linethickness{.55pt}\NPLine{0}{3.4}{1}{3.4}}%
\NPLatticeArrow{0.5}{3.4}{0}%
{\color{paperprimary}\linethickness{.55pt}\NPLine{1}{3.4}{2}{3.4}}%
\NPLatticeArrow{1.5}{3.4}{0}%
{\color{paperprimary}\linethickness{.55pt}\NPLine{2}{3.4}{3}{3.4}}%
\NPLatticeArrow{2.5}{3.4}{0}%
{\color{paperprimary}\linethickness{.55pt}\NPLine{3}{3.4}{4}{3.4}}%
\NPLatticeArrow{3.5}{3.4}{0}%
{\color{paperprimary}\linethickness{.55pt}\NPLine{4}{3.4}{5}{3.4}}%
\NPLatticeArrow{4.5}{3.4}{0}%
{\color{paperprimary}\linethickness{.55pt}\NPLine{5}{3.4}{6}{3.4}}%
\NPLatticeArrow{5.5}{3.4}{0}%
\put(1,3.1){\circle*{.035}}%
\put(1,3.2){\circle*{.035}}%
\put(1,3.3){\circle*{.035}}%
\put(2,3.1){\circle*{.035}}%
\put(2,3.2){\circle*{.035}}%
\put(2,3.3){\circle*{.035}}%
\put(3,3.1){\circle*{.035}}%
\put(3,3.2){\circle*{.035}}%
\put(3,3.3){\circle*{.035}}%
\put(4,3.1){\circle*{.035}}%
\put(4,3.2){\circle*{.035}}%
\put(4,3.3){\circle*{.035}}%
\put(5,3.1){\circle*{.035}}%
\put(5,3.2){\circle*{.035}}%
\put(5,3.3){\circle*{.035}}%
{\color{paperprimary}\linethickness{3.3pt}\NPLine{0.57}{3}{1.43}{3}}%
{\color{paperprimary}\linethickness{3.3pt}\NPLine{1}{2.57}{1}{3}}%
{\color{paperprimary}\linethickness{3.3pt}\NPLine{0.57}{3.4}{1.43}{3.4}}%
{\color{paperprimary}\linethickness{3.3pt}\NPLine{1}{3.4}{1}{3.83}}%
{\color{papersecondary}\linethickness{1.8pt}\NPLine{3}{3.4}{4}{3.4}}%
{\color{papersecondary}\linethickness{1.8pt}\NPLine{4}{3.4}{4}{4.4}}%
{\color{papersecondary}\linethickness{1.8pt}\NPLine{4}{4.4}{3}{4.4}}%
{\color{papersecondary}\linethickness{1.8pt}\NPLine{3}{4.4}{3}{3.4}}%
\NPLabel{1.15}{2.34}{\SMSWhite{\color{paperprimary}\scriptsize $A_{v,\mathrm{folded}}^K$}}%
\NPLabel{3.5}{4.9}{\SMSWhite{\color{papersecondary}\scriptsize $B_{p,\mathrm{folded}}^e$}}%
\end{picture}\endgroup}
\newcommand{\SMSFoldRelation}{%
\begingroup\setlength{\unitlength}{.68cm}%
\begin{picture}(20.6,8.8)(-.6,-1.2)
  \linethickness{.65pt}%
  \multiput(0,0)(0,1){4}{\line(1,0){4}}%
  \multiput(1,0)(1,0){3}{\line(0,1){3}}%
  \multiput(0,4)(0,1){4}{\line(1,0){4}}%
  \multiput(1,4)(1,0){3}{\line(0,1){3}}%
  {\color{paperprimary}\linethickness{.85pt}%
    \NPLine{1}{3}{1}{4}\NPLine{2}{3}{2}{4}\NPLine{3}{3}{3}{4}}%
  \NPLabelL{4.45}{5.50}{$D(G')$}%
  \NPLabelL{4.45}{1.50}{$D(G)$}%
  \NPLabel{2}{-.82}{\small (a)}%
  \NPLabel{7.6}{3.50}{$\longleftrightarrow$}%
  \put(10.1,.35){\begin{picture}(0,0)
    {\color{black!65}%
      \multiput(2.3,2.46)(0,.12){30}{\circle*{.044}}%
      \multiput(3.4,2.46)(0,.12){30}{\circle*{.044}}%
      \multiput(4.5,2.46)(0,.12){30}{\circle*{.044}}}%
    \multiput(0,.70)(.4,.55){4}{\line(1,0){4.4}}%
    \NPLine{1.1}{.70}{2.3}{2.35}%
    \NPLine{2.2}{.70}{3.4}{2.35}%
    \NPLine{3.3}{.70}{4.5}{2.35}%
    \multiput(0,4.35)(.4,.55){3}{\line(1,0){4.4}}%
    \NPLine{1.1}{4.35}{1.9}{5.45}%
    \NPLine{2.2}{4.35}{3.0}{5.45}%
    \NPLine{3.3}{4.35}{4.1}{5.45}%
    {\color{paperprimary}\linethickness{.85pt}%
      \NPLine{1.2}{6.0}{5.6}{6.0}%
      \NPLine{1.9}{5.45}{2.3}{6.0}%
      \NPLine{3.0}{5.45}{3.4}{6.0}%
      \NPLine{4.1}{5.45}{4.5}{6.0}}%
    \NPLabelL{5.95}{5.05}{$D(G')$}%
    \NPLabelL{5.95}{1.55}{$D(G)$}%
  \end{picture}}%
  \NPLabel{12.9}{-.82}{\small (b)}%
\end{picture}\endgroup}

\newcommand{\NativeGeneralizedMult}[1]{%
  \begingroup\setlength{\unitlength}{.70cm}%
  \begin{picture}(4.4,3.4)(-2.2,-1.65)
    \linethickness{.82pt}%
    \NPLine{-1.48}{.48}{0}{0}\NPArrow{-.74}{.24}{-18}%
    \NPLine{-.90}{-1.18}{0}{0}\NPArrow{-.45}{-.59}{53}%
    \NPLine{0}{0}{.90}{-1.18}\NPArrow{.45}{-.59}{-53}%
    \NPLine{0}{0}{1.48}{.48}\NPArrow{.74}{.24}{18}%
    \NPLine{0}{1.38}{0}{0}\NPArrow{0}{.69}{-90}%
    \NPMult{0}{0}\ifnum#1=1\NPCross{0}{0}\fi
    \NPLabelR{-1.67}{.53}{\scriptsize$h$}%
    \NPLabelR{-1.03}{-1.33}{\scriptsize$g'_2$}%
    \NPLabelL{1.03}{-1.33}{\scriptsize$g'_3$}%
    \NPLabelL{1.67}{.53}{\scriptsize$g'_4$}%
    \NPLabel{0}{1.63}{\scriptsize$g'_5$}%
  \end{picture}\endgroup
}
\newcommand{\NativeGeneralizedCR}{%
  \begingroup\setlength{\unitlength}{.88cm}%
  \begin{picture}(3.45,3.3)(-.60,-1.70)
    \linethickness{.82pt}%
    \NPLine{0}{-1.05}{0}{1.05}\NPCopy{0}{0}%
    \NPLine{0}{0}{1.95}{0}\NPArrow{.98}{0}{180}%
    \NPLine{1.95}{-1.05}{1.95}{1.05}%
    \NPArrow{1.95}{-.55}{90}\NPArrow{1.95}{.55}{90}%
    \NPMult{1.95}{0}\NPCross{1.95}{0}%
    \NPLabel{0}{-1.45}{\scriptsize$\ket{h}$}\NPLabel{1.95}{-1.45}{\scriptsize$\ket{N'}_A$}%
    \NPLabel{0}{1.38}{\scriptsize$x$}\NPLabel{1.95}{1.38}{\scriptsize$z'$}%
    \NPLabel{.98}{.32}{\scriptsize$h$}%
  \end{picture}\endgroup
}

\begin{document}
\raggedbottom

\title{Hybrid Lattice Surgery between Smooth Boundaries}

\author{Grace Ding$^{1,2}$}
\author{Yanzhu Chen$^{1,2}$}
\author{Sheng-Jie Huang$^{1,2}$}
\email{shuang5@fsu.edu}
\affiliation{$^1$ Department of Physics, Florida State University, Tallahassee, FL 32306, USA}
\affiliation{$^2$ FSU Quantum Initiative, Florida State University, Tallahassee, FL 32306, USA}

\begin{abstract}
\noindent
We develop smooth hybrid lattice surgery for group surface codes, enabling joint logical measurements through merge and split operations along smooth boundaries. The code patches may correspond to identical or different finite groups, including non-Abelian groups. Our central result is that smooth merge and split realize a logical $K$-coset measurement, where $K$ is the subgroup defining the interface. This generalizes the joint logical parity measurement of ordinary surface codes. Using a tensor-network representation, we derive explicit measurement circuits and develop procedures for correcting nonidentity measurement outcomes. We illustrate the framework through a generalized controlled multiplication gate between group surface codes and protocols for preparing the $T$, $CZ$, $CCX$, and $CS$ magic states, producing resources encoded in one or more $\mathbb{Z}_2$ surface code patches. Our results show how smooth hybrid lattice surgery broadens the range of logical operations available for group surface codes.
\end{abstract}

\maketitle

\tableofcontents

\section{Introduction}
Fault-tolerant quantum computation requires both quantum error-correcting (QEC) codes that protect quantum information against noise and a universal set of logical operations that can be implemented with manageable resource overhead. The surface code is a promising platform because of its structural simplicity, local checks, and resilience to physical errors~\cite{krinner2022,ra2024,wang2011}. Its regular two-dimensional geometry supports implementations with limited connectivity, motivating the development of logical operations compatible with this architecture.

Lattice surgery provides a way to couple surface code logical qubits through local operations along patch boundaries~\cite{horsman2012,fowler2012,Litinski2019gameofsurfacecodes,Cowtan2025vok}. By merging and splitting code patches, it implements joint logical measurements that serve as building blocks for logical gates. These primitives support Clifford operations, while universality requires additional non-Clifford resources~\cite{fowler2012,boykin1999,bravyiKitaev2005}. Magic-state distillation and code switching provide widely studied routes to such resources~\cite{campbell2017,beverland2021,daguerre2025}. Both methods require substantial resources. These costs motivate exploring other topological codes and operations at their interfaces to expand the available methods for manipulating encoded information.

Recent work has explored non-Abelian QEC codes as resources for non-Clifford computation. Proposed approaches include gauging and code transformations~\cite{Williamson2026gauging,huangChen2025,davydova2025,hsin2024,Bauer2026color,Bauer2025lowoverheadnon,Zhu2026nqldpc,Christos2026nqldpc,Zhu2026sheaf,Chang2026qldpcgauging}, sliding code patches~\cite{brown2020,davydova2025,Sajith2026,manjunath2026}, transversal or constant-depth logical gates~\cite{kobayashi2025,Hsin2026automorphism,Warman2026transversal,manjunath2026}, and rough hybrid lattice surgery~\cite{huang2025Hybrid}. Within this broader direction, group surface codes provide a promising setting for extending surface code computation. They generalize the ordinary surface code to finite groups while retaining geometrically local checks on planar patches~\cite{kitaev2003}. Recent constructions show how suitable group choices and interfaces with ordinary surface codes enable non-Clifford gates and resource-state preparation~\cite{huang2025Hybrid,davydova2025,manjunath2026}. Numerical simulations of a related non-Abelian twisted quantum-double code also find error-correction thresholds comparable to those of the toric code under phenomenological noise~\cite{Magdalena2026decoding}. These developments motivate expanding the set of logical operations available for group surface codes. Lattice surgery offers a natural route to this goal by coupling code patches and manipulating encoded information through local measurements along their boundaries.

In this work, we develop smooth hybrid lattice surgery for group surface codes. We introduce merge and split operations between the smooth boundaries of surface codes based on different groups and derive their logical action. Our central result is that smooth merge and split realize a logical $K$-coset measurement, where $K$ is the subgroup defining the interface between the two patches. Different measurement outcomes project onto different logical coset subspaces while preserving coherence among logical basis states within the selected coset. For two $\mathbb{Z}_2$ surface codes, this operation reduces to the familiar joint logical parity measurement. Smooth hybrid lattice surgery generalizes the logical parity measurements to a family of joint logical measurements determined by the interface subgroup.

Using the tensor-network representation, we derive explicit measurement circuits for smooth merge and split. We also develop procedures for correcting nonidentity outcomes of the boundary and interface plaquette measurements. We illustrate the framework through generalized controlled multiplication between different group surface codes and protocols for preparing the $T$, $CZ$, $CCX$, and $CS$ magic states. These examples demonstrate how smooth hybrid lattice surgery expands the available methods for implementing logical operations in group surface codes and preparing non-Clifford resources in one or more $\mathbb{Z}_2$ surface code patches. 

The paper is organized as follows. Section~\ref{sec:review_QD} reviews quantum-double models, their boundary conditions, and the logical operators of group surface codes. Section~\ref{sec:smooth-surgery} introduces smooth merge and split, specifies the bulk and interface projectors, and derives their logical action for identity measurement outcomes. Section~\ref{sec:smooth-correction} uses a tensor-network representation to construct the measurement circuits and discusses parallel procedures for correcting nonidentity outcomes of boundary and interface plaquette measurements. Section~\ref{sec:protocols} applies the framework to generalized controlled multiplication and magic-state preparation. Section~\ref{sec:discussion} discusses future directions, including protocol optimization, fault tolerance, and resource costs. The appendices provide supporting constructions and derivations. Appendix~\ref{sec:review-rough-merge-split} reviews rough merge and split, and Appendix~\ref{sec:smooth-merge-split-folded} presents an equivalent description of smooth surgery using the folded picture. Appendix~\ref{sec:row-algebra} derives the logical operation for identity and nonidentity final plaquette outcomes, while Appendix~\ref{sec:regular-tensors} includes the tensor-network formalism and the measurement circuits. Appendix~\ref{sec:parallel-correction} gives sequential correction procedures for general finite-group interface data and identifies conditions that permit parallel corrections. Finally, Appendix~\ref{sec:logical_recovery_smooth} discusses logical recovery for different $K$-coset measurement outcomes and its application to the protocols in Section~\ref{sec:protocols}.

\section{Review}
\label{sec:review_QD}
A $G$ surface code is a Kitaev's quantum double model defined on a two-dimensional patch with
smooth and rough boundaries~\cite{kitaev2003,beigi2011,Cowtan2025vok,huang2025Hybrid,manjunath2026} as shown in Fig.~\ref{fig:review-code-patch}. This section
briefly reviews the model with boundaries to set the stage for our construction of smooth merge and split operations in
Sec.~\ref{sec:smooth-surgery}.

\subsection{Quantum Double Model}
\label{sec:review-quantum-double-patch}

\begin{figure}[!t]
\pdfdest name {float-01-start} xyz
\centering
\begingroup\setlength{\unitlength}{0.4mm}
\begin{picture}(100,78)(-14,-9)
  \linethickness{0.45pt}
  \multiput(0,0)(0,18){4}{\line(1,0){72}}
  \multiput(18,0)(18,0){3}{\line(0,1){54}}
  \multiput(9,0)(18,0){4}{\makebox(0,0){\LatticeMidArrow}}
  \multiput(9,18)(18,0){4}{\makebox(0,0){\LatticeMidArrow}}
  \multiput(9,36)(18,0){4}{\makebox(0,0){\LatticeMidArrow}}
  \multiput(9,54)(18,0){4}{\makebox(0,0){\LatticeMidArrow}}
  \multiput(18,9)(18,0){3}{\makebox(0,0){\rotatebox{90}{\LatticeMidArrow}}}
  \multiput(18,27)(18,0){3}{\makebox(0,0){\rotatebox{90}{\LatticeMidArrow}}}
  \multiput(18,45)(18,0){3}{\makebox(0,0){\rotatebox{90}{\LatticeMidArrow}}}
  \put(36,63){\makebox(0,0){\scriptsize Smooth}}
  \put(36,-7){\makebox(0,0){\scriptsize Smooth}}
  \put(-8,27){\makebox(0,0){\rotatebox{90}{\scriptsize Rough}}}
  \put(80,27){\makebox(0,0){\rotatebox{90}{\scriptsize Rough}}}
\end{picture}
\endgroup
\caption{Orientation and boundary convention for a \(D(G)\) code patch.
The vertical boundaries are rough, and the horizontal boundaries are smooth.}
\label{fig:review-code-patch}
\pdfdest name {float-01-end} xyz
\end{figure}
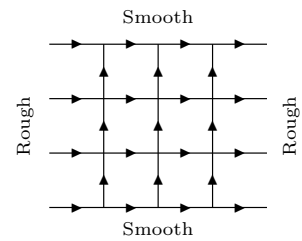

Let \(G\) be any finite group. The quantum double model is defined on an
oriented square lattice, with a \(\lvert G\rvert\)-dimensional local Hilbert
space assigned to each edge. This space is spanned by the orthonormal basis
states \(\{\ket{g}:g\in G\}\), so each edge carries a qudit whose basis
states are indexed by the elements of \(G\). Reversing an edge orientation
maps \(g\) to \(g^{-1}\). The toric code corresponds to
\(G=\mathbb Z_2\), for which each edge hosts a two-level spin.
The boundary and edge-orientation conventions are shown in
Fig.~\ref{fig:review-code-patch}.

For \(a,g\in G\), define left and right multiplication by
\begin{equation}
  L^a\ket{g}=\ket{ag},
  \qquad
  R^a\ket{g}=\ket{ga^{-1}}.
  \label{eq:review-left-right}
\end{equation}
An edge directed away from a vertex is acted on by \(L^a\), whereas an edge
directed into the vertex is acted on by \(R^a\).

The group transformation at vertex \(v\) and the vertex stabilizer are defined as
\begin{align}
  A_v^{(a)}
  &=\prod_{v\to e}L_e^a
    \prod_{e\to v}R_e^a,
  \label{eq:review-vertex-action}\\
  A_v&=\frac{1}{\lvert G\rvert}\sum_{a\in G}A_v^{(a)}.
  \label{eq:review-vertex-projector}
\end{align}
Here \(v\to e\) and \(e\to v\) denote edges directed away from and into
\(v\), respectively. On a vertex in the oriented square lattice,
\begin{equation}
  A_v^{(a)}=
  \vcenter{\hbox{%
  \begingroup\setlength{\unitlength}{0.34mm}%
  \begin{picture}(80,54)(-40,-27)
    \linethickness{0.45pt}
    \put(-36,0){\line(1,0){72}}
    \put(0,-24){\line(0,1){48}}
    \NPLatticeArrow{-18}{0}{0}
    \NPLatticeArrow{18}{0}{0}
    \NPLatticeArrow{0}{-12}{90}
    \NPLatticeArrow{0}{12}{90}
    \put(5,5){\makebox(0,0)[l]{\scriptsize\(v\)}}
    \put(-20,8){\makebox(0,0){\scriptsize\(R^a\)}}
    \put(20,8){\makebox(0,0){\scriptsize\(L^a\)}}
    \put(10,-14){\makebox(0,0){\scriptsize\(R^a\)}}
    \put(10,14){\makebox(0,0){\scriptsize\(L^a\)}}
  \end{picture}\endgroup}}
  .
  \label{eq:review-vertex-graphical}
\end{equation}

Label the four edges of a plaquette as shown below. The projector
onto flux $h\in G$ acts as
\begin{equation}
\begin{aligned}
  &\vcenter{\hbox{%
  \begingroup\setlength{\unitlength}{0.32mm}%
  \begin{picture}(76,68)(-38,-34)
    \linethickness{0.45pt}
    \put(-22,-22){\line(1,0){44}}
    \put(-22,22){\line(1,0){44}}
    \put(-22,-22){\line(0,1){44}}
    \put(22,-22){\line(0,1){44}}
    \NPLatticeArrow{0}{-22}{0}\NPLatticeArrow{0}{22}{0}
    \NPLatticeArrow{-22}{0}{90}\NPLatticeArrow{22}{0}{90}
    \put(0,-31){\makebox(0,0){\scriptsize\(g_1\)}}
    \put(-33,0){\makebox(0,0){\scriptsize\(g_2\)}}
    \put(0,31){\makebox(0,0){\scriptsize\(g_3\)}}
    \put(33,0){\makebox(0,0){\scriptsize\(g_4\)}}
    \put(0,0){\makebox(0,0){\scriptsize\(p\)}}
  \end{picture}\endgroup}}
  \quad\Longrightarrow\quad B_p^{(h)}\ket{g_1,g_2,g_3,g_4}
  \\[2pt]
  &\qquad=\delta_{g_1^{-1}g_2g_3g_4^{-1},\,h}
  \ket{g_1,g_2,g_3,g_4}.
\end{aligned}
\label{eq:review-plaquette-graphical}
\end{equation}
The plaquette stabilizer is $B_p=B_p^{(e)}$, where $e\in G$ is
the identity.

It can be shown that \(A_v\) and
\(B_p\) are Hermitian projectors and
\begin{equation}
  [A_v,A_{v'}]=[B_p,B_{p'}]=[A_v,B_p]=0.
  \label{eq:review-commuting-projectors}
\end{equation}

The bulk Hamiltonian and code space are therefore
\begin{align}
  H_G&=-\sum_v A_v-\sum_p B_p,
  \label{eq:review-bulk-hamiltonian}\\
  \mathcal H_C^{\mathrm{bulk}}
  &=\left\{\ket{\psi}\in\bigotimes_e\mathbb C[G]:
  \begin{aligned}
    A_v\ket{\psi}&=\ket{\psi}\quad\forall v,\\
    B_p\ket{\psi}&=\ket{\psi}\quad\forall p
  \end{aligned}\right\}.
  \label{eq:review-bulk-code-space}
\end{align}
Here \(v\), \(p\), and \(e\) label bulk vertices, plaquettes, and oriented
edges, respectively. \(\mathcal H_C^{\mathrm{bulk}}\) is the common
\(+1\) eigenspace of projectors \(A_v\) and \(B_p\).

\subsection{Boundaries and $G$ surface codes}

A gapped boundary is specified by a subgroup \(K\leq G\) and a cohomology
class in \(H^2(K,U(1))\)~\cite{beigi2011}. The commuting boundary projectors are
\begin{equation}
  \widetilde A_v^{K,\phi}
  =\frac{1}{\lvert K\rvert}\sum_{k\in K}
    \widetilde A_v^{(k,\phi)},
  \qquad
  B_p^K=\sum_{k\in K}B_p^{(k)}.
  \label{eq:review-boundary-projectors}
\end{equation}
The boundary gauge operator \(\widetilde A_v^{(k,\phi)}\) is obtained
by restricting \(A_v^{(k)}\) to \(k\in K\) and including the cocycle
phases determined by \(\phi\). 
For a vertex on the right smooth boundary, its action is
\begin{equation}
\widetilde A_v^{(k,\phi)}
\vcenter{\hbox{%
\begingroup\setlength{\unitlength}{0.25mm}%
\begin{picture}(70,64)(-36,-32)
  \linethickness{0.45pt}
  \put(-28,0){\line(1,0){28}}
  \put(0,-24){\line(0,1){48}}
  \NPLatticeArrow{-14}{0}{0}
  \NPLatticeArrow{0}{-12}{90}
  \NPLatticeArrow{0}{12}{90}
  \put(-16,17){\makebox(0,0){\scriptsize\(\ket z\)}}
  \put(20,20){\makebox(0,0){\scriptsize\(\ket y\)}}
  \put(20,-20){\makebox(0,0){\scriptsize\(\ket x\)}}
\end{picture}\endgroup}}
=\phi(x,k^{-1})\phi(k,y)
\vcenter{\hbox{%
\begingroup\setlength{\unitlength}{0.25mm}%
\begin{picture}(86,64)(-43,-32)
  \linethickness{0.45pt}
  \put(-34,0){\line(1,0){34}}
  \put(0,-24){\line(0,1){48}}
  \NPLatticeArrow{-17}{0}{0}
  \NPLatticeArrow{0}{-12}{90}
  \NPLatticeArrow{0}{12}{90}
  \put(-20,18){\makebox(0,0){\scriptsize\(\ket{zk^{-1}}\)}}
  \put(27,20){\makebox(0,0){\scriptsize\(\ket{ky}\)}}
  \put(30,-20){\makebox(0,0){\scriptsize\(\ket{xk^{-1}}\)}}
\end{picture}\endgroup}}.
  \label{eq:review-boundary-vertex-graphical}
\end{equation}
The plaquette projector $B_p^{(h)}$ selects states whose oriented
boundary product equals $h$, so $B_p^K$ restricts this product to $K$.
On the left rough boundary, the identity-flux term is
\begingroup
\newcommand{\BoundaryPlaquetteGraph}{%
  \vcenter{\hbox{%
    \setlength{\unitlength}{0.30mm}%
    \begin{picture}(76,68)(-38,-34)
      \linethickness{0.45pt}
      \put(-22,-22){\line(1,0){44}}
      \put(-22,22){\line(1,0){44}}
      \put(22,-22){\line(0,1){44}}
      \NPLatticeArrow{0}{-22}{0}\NPLatticeArrow{0}{22}{0}
      \NPLatticeArrow{22}{0}{90}
      \put(0,-31){\makebox(0,0){\scriptsize\(g_1\)}}
      \put(0,31){\makebox(0,0){\scriptsize\(g_3\)}}
      \put(33,0){\makebox(0,0){\scriptsize\(g_4\)}}
      \put(0,0){\makebox(0,0){\scriptsize\(p\)}}
    \end{picture}%
  }}%
}
\begin{equation}
  B_p\BoundaryPlaquetteGraph
  =\delta_{g_1^{-1}g_3g_4^{-1},\,e}\BoundaryPlaquetteGraph.
  \label{eq:review-boundary-plaquette-graphical}
\end{equation}
\endgroup

With boundary terms included, the commuting-projector Hamiltonian is
\begin{equation}
  \widetilde H_G
  =-\sum_{v\in\mathrm{bulk}}A_v
   -\sum_{p\in\mathrm{bulk}}B_p
   -\sum_{(v,p)\in\partial}
    \left(\widetilde A_v^{K,\phi}+B_p^K\right).
  \label{eq:review-boundary-hamiltonian}
\end{equation}
Its code space is
\begin{equation}
\begin{aligned}
  \mathcal H_C=\{\ket{\psi}:\;&
  A_v\ket{\psi}=B_p\ket{\psi}=\ket{\psi}
  &&\text{in the bulk},\\
  &\widetilde A_v^{K,\phi}\ket{\psi}
   =B_p^K\ket{\psi}=\ket{\psi}
  &&\text{on }\partial\}.
\end{aligned}
  \label{eq:review-boundary-code-space}
\end{equation}

The canonical boundary choices used in this paper are
\begin{equation}
\begin{aligned}
  \text{smooth}: \quad (K,\phi) &= (G,1),\\
  \text{rough}: \quad (K,\phi) &= (\{e\},1).
\end{aligned}
  \label{eq:review-boundary-choice}
\end{equation}
We define $G$ surface codes using the smooth and rough boundaries arranged as in Fig.~\ref{fig:review-code-patch}. For these choices, any boundary term that reduces to the identity is omitted
from \(\widetilde H_G\), while the remaining truncated stabilizers enforce
the local boundary constraint. We sometimes refer to a $G$ surface code as a $D(G)$ patch, where $D(G)$ denotes the quantum double of $G$.

The boundary condition determines which stabilizer violation can terminate at
the edge of the patch. A rough boundary retains the truncated plaquette
constraint but relaxes the vertex constraint, so a vertex defect, or electric
charge, is absorbed when it reaches that boundary. A smooth boundary instead
retains the truncated vertex constraint and relaxes the plaquette constraint,
so a plaquette defect, or magnetic flux, can be absorbed there.

\subsection{Logical Operators}

\begin{figure}[!t]
\pdfdest name {float-03-start} xyz
\centering
\begingroup\setlength{\unitlength}{.40mm}
\begin{picture}(190,78)(0,0)
  \put(41,72){\makebox(0,0){\scriptsize(a)}}
  \put(144,72){\makebox(0,0){\scriptsize(b)}}
  \linethickness{.4pt}
  \multiput(5,6)(0,18){4}{\line(1,0){72}}
  \multiput(23,6)(18,0){3}{\line(0,1){54}}
  \multiput(108,6)(0,18){4}{\line(1,0){72}}
  \multiput(126,6)(18,0){3}{\line(0,1){54}}
  {\color{paperprimary}\linethickness{1.1pt}
   \multiput(5,6)(0,18){4}{\line(1,0){18}}
   \multiput(162,6)(0,18){4}{\line(1,0){18}}}
  \multiput(14,11)(0,18){4}{%
    \makebox(0,0){\scriptsize\color{paperprimary}\(L^a\)}}
  \multiput(171,11)(0,18){4}{%
    \makebox(0,0){\scriptsize\color{paperprimary}\(R^a\)}}
\end{picture}
\endgroup
\caption{The definition of the logical
(a) left multiplication \(\overline L^{\,a}\) on the left boundary or
(b) right multiplication \(\overline R^{\,a}\) on the right boundary.}
\label{fig:review-boundary-logicals}
\pdfdest name {float-03-end} xyz
\end{figure}
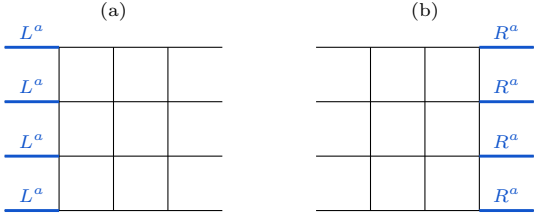

One set of the logical operators acting on the left and right rough boundaries are
\begin{equation}
  \overline L^{\,a}=\prod_{e\in\partial_L}L_e^a,
  \qquad
  \overline R^{\,a}=\prod_{e\in\partial_R}R_e^a.
  \label{eq:review-logical-left-right}
\end{equation}
Here \(a\in G\), \(\partial_L\) and \(\partial_R\) denote the edge sets on the left and right rough boundaries, and \(L_e^a\) and \(R_e^a\) act on edge \(e\) by left and right multiplication by \(a\), respectively, as illustrated in Figs.~\ref{fig:review-boundary-logicals}(a) and \ref{fig:review-boundary-logicals}(b). These operators commute with \(\widetilde H_G\) and obey
\begin{equation}
  \overline L^{\,a}\overline L^{\,b}=\overline L^{\,ab},
  \quad
  \overline R^{\,a}\overline R^{\,b}=\overline R^{\,ab},
  \quad
  [\overline L^{\,a},\overline R^{\,b}]=0.
  \label{eq:review-logical-algebra}
\end{equation}

A convenient logical group basis is obtained from a normalized fiducial
state,
\begin{equation}
  \ket{\Phi_e}=\mathcal N
  \left(\prod_v A_v\right)\bigotimes_{e'}\ket{e}_{e'},
  \qquad
  \ket{\Phi_g}=\overline L^{\,g}\ket{\Phi_e}.
  \label{eq:review-logical-basis}
\end{equation}
The logical multiplication operators permute this basis as
\begin{equation}
  \overline L^{\,a}\ket{\Phi_g}=\ket{\Phi_{ag}},
  \qquad
  \overline R^{\,a}\ket{\Phi_g}=\ket{\Phi_{ga^{-1}}}.
  \label{eq:review-logical-permutations}
\end{equation}

A physical \(G\)-qudit may be measured in the group basis
\(\{\ket g:g\in G\}\) or in the basis defined by the irreducible representations of \(G\). A projective measurement in either basis returns
the corresponding group or representation label.

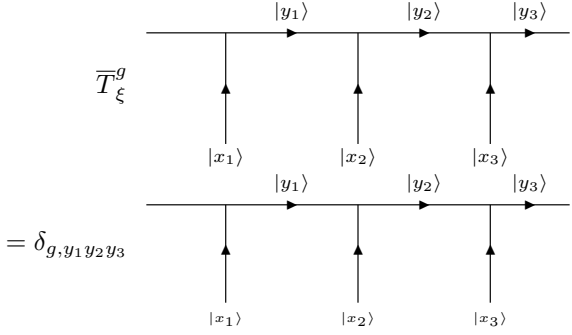
\begin{figure}[!t]
\pdfdest name {float-02-start} xyz
\centering
\begingroup\setlength{\unitlength}{0.35mm}
\begin{picture}(190,128)(0,-14)
  \linethickness{0.45pt}
  \put(20,100){\line(1,0){160}}
  \put(50,58){\line(0,1){42}}\put(100,58){\line(0,1){42}}
  \put(150,58){\line(0,1){42}}
  \NPLatticeArrow{75}{100}{0}\NPLatticeArrow{125}{100}{0}
  \NPLatticeArrow{165}{100}{0}
  \NPLatticeArrow{50}{79}{90}\NPLatticeArrow{100}{79}{90}
  \NPLatticeArrow{150}{79}{90}
  \put(13,80){\makebox(0,0)[r]{\normalsize\(\overline{T}_\xi^{g}\)}}
  \put(75,108){\makebox(0,0){\scriptsize\(\ket{y_1}\)}}
  \put(125,108){\makebox(0,0){\scriptsize\(\ket{y_2}\)}}
  \put(165,108){\makebox(0,0){\scriptsize\(\ket{y_3}\)}}
  \put(50,53){\makebox(0,0){\scriptsize\(\ket{x_1}\)}}
  \put(100,53){\makebox(0,0){\scriptsize\(\ket{x_2}\)}}
  \put(150,53){\makebox(0,0){\scriptsize\(\ket{x_3}\)}}

  \put(20,35){\line(1,0){160}}
  \put(50,-2){\line(0,1){37}}\put(100,-2){\line(0,1){37}}
  \put(150,-2){\line(0,1){37}}
  \NPLatticeArrow{75}{35}{0}\NPLatticeArrow{125}{35}{0}
  \NPLatticeArrow{165}{35}{0}
  \NPLatticeArrow{50}{18}{90}\NPLatticeArrow{100}{18}{90}
  \NPLatticeArrow{150}{18}{90}
  \put(13,20){\makebox(0,0)[r]{\normalsize\(=\delta_{g,y_1y_2y_3}\)}}
  \put(75,43){\makebox(0,0){\scriptsize\(\ket{y_1}\)}}
  \put(125,43){\makebox(0,0){\scriptsize\(\ket{y_2}\)}}
  \put(165,43){\makebox(0,0){\scriptsize\(\ket{y_3}\)}}
  \put(50,-8){\makebox(0,0){\tiny\(\ket{x_1}\)}}
  \put(100,-8){\makebox(0,0){\tiny\(\ket{x_2}\)}}
  \put(150,-8){\makebox(0,0){\tiny\(\ket{x_3}\)}}
\end{picture}
\endgroup
\caption{The definition of the logical operator \(\overline{T}_\xi^{g}\), where $\xi$ is a path along the top edges.}
\label{fig:review-ribbon}
\pdfdest name {float-02-end} xyz
\end{figure}

The other set of logical operators is defined as 
\begin{equation}
  \overline{T}_{\xi}^{h}\ket{\Phi_g}
  =\delta_{h,g}\ket{\Phi_g},
  \label{eq:review-horizontal-ribbon-measurement}
\end{equation}
as shown in Fig.~\ref{fig:review-ribbon}.
These logical operators can be used to distinguish
the logical group element basis states.

\section{Smooth lattice surgery}
\label{sec:smooth-surgery}

Here we introduce lattice surgery between the smooth boundaries of the $G$ surface code and the $G'$ surface code. We start by discussing how to perform a smooth merge between the two codes. A smooth merge describes a procedure where the smooth boundaries of the $G$ and the $G'$ surface code patches are brought together and merged into one hybrid code patch. 

We will then discuss the smooth split operation. A smooth split describes the procedure where a hybrid code patch is returned to the two separate code patches $D(G)$ and $D(G')$ with smooth boundaries in between by measuring the ancilla qudits introduced in the merge. We construct the smooth merge and split between the two patches (Sec.~\ref{sec:sms-procedure}) and derive the logical operation (Sec.~\ref{sec:sms-state-algebra}). We then
describe the stabilizers before, during, and after the procedure, including the required modifications (Sec.~\ref{sec:sms-stabilizers}), and give a tensor-network description of the measurement circuits used in Secs.~\ref{sec:sms-procedure} and~\ref{sec:sms-stabilizers} (Sec.~\ref{sec:tensor-network-measurements}). The procedure in Sec.~\ref{sec:sms-procedure} assumes identity outcomes
for all measurements. Corrections for nonidentity boundary and plaquette
outcomes are developed in Sec.~\ref{sec:smooth-correction} and
Appendix~\ref{sec:parallel-correction}.

\begin{figure*}[!t]
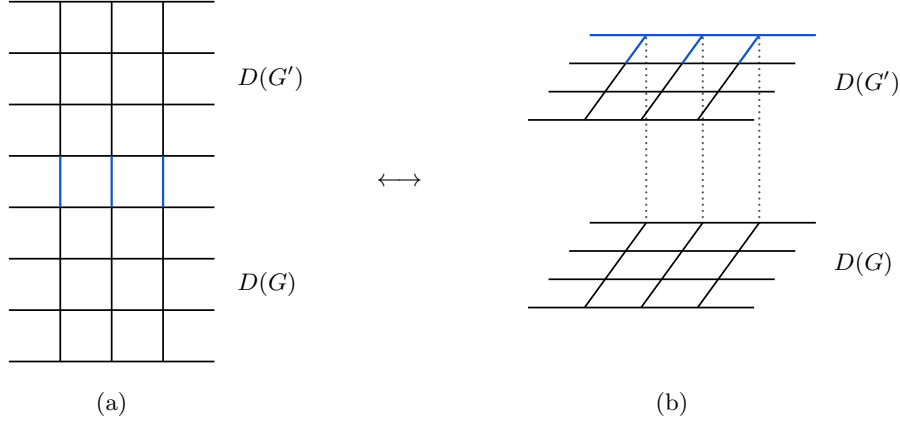

\pdfdest name {float-05-start} xyz
\centering\small
\SMSFoldRelation
\caption{The interface between two code patches $D(G)$
and $D(G')$ viewed as a boundary of a $D(G \times G')$ patch.}
\label{fig:Av-folded-unfolded}
\pdfdest name {float-05-end} xyz
\end{figure*}

\subsection{Smooth Merge and Split}
\label{sec:sms-procedure}

Consider two planar patches $D(G)$ and $D(G')$ with rough boundaries
on the left and right and facing smooth boundaries. If we fold the $D(G')$ code patch over the $D(G)$ code patch, the interface now becomes a boundary of a $D(G \times G')$ code patch as illustrated in Fig.~\ref{fig:Av-folded-unfolded}. The interface can thus be specified by a subgroup $K \leq G \times G'$ and a second cohomology class $\phi \in H^{2}(K,U(1))$. We restrict to the trivial cohomology class, so the interface is specified by \(K\). Here we define the subgroup $K$ as follows. Let
\begin{equation}
\begin{gathered}
H\leq G,\quad H'\leq G',\quad
N\trianglelefteq H,\quad N'\trianglelefteq H',\\
\varphi:H/N\xrightarrow{\sim}H'/N',\\
K \equiv \{(h,h')\in H\times H':N'h' = \varphi(Nh)\}.
\end{gathered}
\label{eq:sms-group-data}
\end{equation}
We write the quotient label $\varphi(Nh)$ as a right coset $N'h'$. 

\begin{figure}[!t]
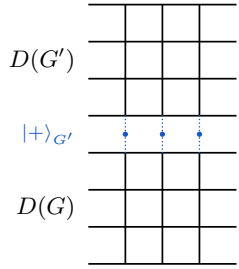

\pdfdest name {float-06-start} xyz
\centering
\begin{minipage}[b]{.48\columnwidth}\centering
\NativeUnfoldedAncilla\\[-1pt]
\end{minipage}\hfill
\caption{Ancilla initialization for smooth merge and split.}
\label{fig:unfolded-ancilla}
\pdfdest name {float-06-end} xyz
\end{figure}
The smooth merge is specified by the subgroup $K$. We now describe a smooth merge--split procedure that uses fewer ancillas than the folded construction based on the boundary terms of Ref.~\cite{beigi2011}. Appendix~\ref{sec:smooth-merge-split-folded} describes the folded construction and shows that both procedures yield the same logical operation.

\begin{enumerate}
\item \emph{Initialize.} Prepare every vertical interface ancilla
(Fig.~\ref{fig:unfolded-ancilla}) in
\begin{equation}
\ket+_{G'}=\frac{1}{\sqrt{|G'|}}\sum_{q'\in G'}\ket{q'}.
\label{eq:sms-Gprime-ancilla}
\end{equation}

\item \emph{Boundary projection.} Measure
\begin{equation}
P_H=\sum_{h\in H}\ket h\bra h
\label{eq:sms-PH}
\end{equation}
on each facing boundary edge of $D(G)$. After the projection, denote the labels of these edges by $h_i\in H$.

\item \emph{Smooth merge.} Enforce the interface plaquette stabilizers one plaquette at a time, from left to right. The interface stabilizer is
defined by
\begin{equation}
\begin{gathered}
B_{p_i,\mathrm{interface}}^e\quad\vcenter{\hbox{\SMSUnfoldedCell}}\\[-2pt]
={}\delta_{N'q'_{i-1}g'_i(q'_i)^{-1},\,\varphi(Nh_i)}
\quad\vcenter{\hbox{\SMSUnfoldedCell}}.
\end{gathered}
\label{eq:Bp-unfolded-graphical}
\end{equation}
The plaquette stabilizers for the leftmost and rightmost plaquettes are truncated because each of these plaquettes has only three edges. The $G$ and $G'$ surface codes are then merged into a hybrid code patch.

\item \emph{Smooth split.} Measure the added ancillas in the representation
basis of $G'$ and retain the trivial-representation outcomes. The merged code is then split into the $G$ and $G'$ surface codes.
\end{enumerate}

In Sec.~\ref{sec:smooth-correction}, we explain how to use measurements to implement the boundary projection and the smooth merge. Here we assume identity measurement outcomes throughout this procedure. In Appendix.~\ref{sec:smooth-merge-split-folded}, we develop an equivalent smooth merge and split operations based on the folded picture and show that the two descriptions have the same logical effect.

\subsection{Bulk and Interface Projectors}
\label{sec:sms-stabilizers}

\begin{figure*}[!t]
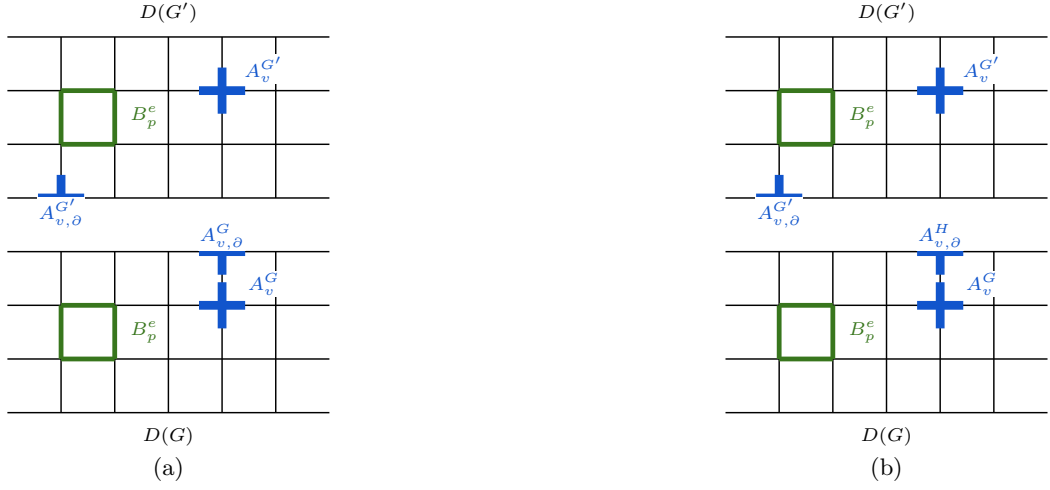

\pdfdest name {float-09-start} xyz
\centering
\begin{minipage}[b]{.47\textwidth}\centering
\SMSOriginal\\[-1pt]
\small (a) 
\end{minipage}\hfill
\begin{minipage}[b]{.47\textwidth}\centering
\SMSRetained\\[-1pt]
\small (b) 
\end{minipage}
\caption{Bulk and interface projectors of the separated patches.
(a) Before and (b) after smooth merge and split.}
\label{fig:sms-before-after}
\pdfdest name {float-09-end} xyz
\end{figure*}

In this section, we show the bulk and the interface projectors explicitly. As shown in Fig.~\ref{fig:sms-before-after}, the bulk projectors are $A_v^G$, $A_v^{G'}$, and $B_p^e$. Before merging, the facing smooth boundaries carry
$A_{v,\partial}^G$ and $A_{v,\partial}^{G'}$.
Both splits leave $A_{v,\partial}^H$ on the restricted $D(G)$ boundary
and restore $A_{v,\partial}^{G'}$ on the $D(G')$ boundary.

\begin{figure*}[!t]
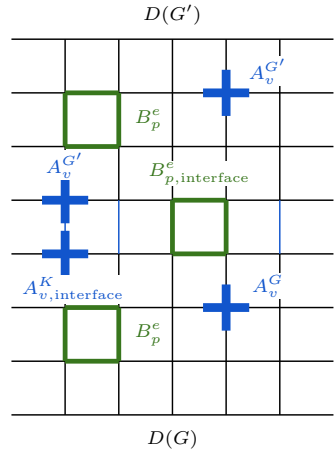

\pdfdest name {float-10-start} xyz
\centering
\renewcommand{\NPLatticeArrow}[3]{}
\begin{minipage}[b]{.47\textwidth}\centering
\SMSUnfolded\\[-1pt] 
\end{minipage}\hfill
\caption{Bulk and interface projectors of the hybrid code patch after the smooth merge. The added edges are blue.}
\label{fig:sms-interface}
\label{fig:unfolded-stabilizer-support}
\pdfdest name {float-10-end} xyz
\end{figure*}

During merging, the upper boundary vertices carry $A_v^{G'}$
including the added vertical edge as shown in Fig.~\ref{fig:unfolded-stabilizer-support}. The lower interface action is
\begin{equation}
\begin{aligned}
&A_{v,\mathrm{interface}}^K
\vcenter{\hbox{\resizebox{.39\columnwidth}{!}{\SMSAvInput}}}\\[3pt]
&\quad=\frac{1}{|K|}\sum_{(h,h')\in K}
\vcenter{\hbox{\resizebox{.60\columnwidth}{!}{\SMSAvOutput}}}.
\end{aligned}
\label{eq:sms-Av-unfolded}
\end{equation}
Here $|K|=|H|\,|N'|$.

\subsection{Identity Logical Operation}
\label{sec:sms-state-algebra}

Now we derive the logical operation, assuming identity measurement outcomes. After performing the boundary projection, let the disjoint patch states be $\ket{\Psi_1}_G$ and
$\ket{\Psi_2}_{G'}$. With $n_e=n-1$ vertical ancillas, the joint state after the smooth merge is
\begin{equation}
\begin{aligned}
\ket{\Psi_{1,2}}_{\mathrm{joint}}
={}&\left(\prod_{i=1}^nB_{p_i,\mathrm{interface}}^e\right)
\ket{\Psi_1}_G\ket{\Psi_2}_{G'}\\
&\quad\otimes\bigotimes_{l=1}^{n-1}\ket+_{G'}.
\end{aligned}
\label{eq:joint-hybrid-state}
\end{equation}
Define the diagonal constraint operator by
\begin{equation}
\begin{aligned}
&\mathcal C(\{q'_l\})\ket{\boldsymbol h,\boldsymbol g'}\\
&\quad=\left[\prod_{i=1}^n
\delta_{N'q'_{i-1}g'_i(q'_i)^{-1},\,\varphi(Nh_i)}\right]
\ket{\boldsymbol h,\boldsymbol g'}.
\end{aligned}
\label{eq:sms-state-constraints}
\end{equation}
Then
\begin{equation}
\begin{aligned}
\ket{\Psi_{1,2}}_{\mathrm{joint}}
={}&\frac1{|G'|^{n_e/2}}
\sum_{q'_1,\ldots,q'_{n-1}\in G'}\mathcal C(\{q'_l\})\\
&\quad\times\ket{\Psi_1}_G\ket{\Psi_2}_{G'}
\bigotimes_{l=1}^{n-1}\ket{q'_l}.
\end{aligned}
\label{eq:sms-joint-expanded}
\end{equation}
The representation basis used for the split is
\begin{equation}
\ket{\rho_l,a_l,b_l}
=\sqrt{\frac{d_{\rho_l}}{|G'|}}
\sum_{r_l\in G'}D^{\rho_l}_{a_lb_l}(r_l)\ket{r_l},
\label{eq:sms-interface-fourier-basis}
\end{equation}
where $\rho_l$ labels an irrep of the group $G'$, $d_{\rho_l}$ is the dimension of the representation, $D^{\rho_l}_{a_lb_l}$ is a map from $G'$ to $d_R \times d_R$ matrices, and $a_l,b_l$ label a matrix element.

Projecting every added ancilla onto its trivial representation and
removing it leaves
\begin{equation}
\begin{aligned}
\ket{\Psi_{1,2}(\mathrm{trivial})}_{\mathrm{disjoint}}
={}&\frac1{|G'|^{n_e}}
\sum_{q'_1,\ldots,q'_{n-1}\in G'}\mathcal C(\{q'_l\})\\
&\quad\times\ket{\Psi_1}_G\ket{\Psi_2}_{G'}.
\end{aligned}
\label{eq:sms-disjoint-trivial}
\end{equation}

Write $h=h_1\cdots h_n$ and $h'=g'_1\cdots g'_n$ for the ordered
boundary products. Summing over the ancilla labels contracts the
local coset deltas from left to right and gives
\begin{equation}
\begin{aligned}
&\left[\sum_{q'_1,\ldots,q'_{n-1}\in G'}\mathcal C(\{q'_l\})\right]
\ket{\boldsymbol h,\boldsymbol g'}\\
&\qquad=|N'|^{n_e}\delta_{N'h',\,\varphi(Nh)}
\ket{\boldsymbol h,\boldsymbol g'}.
\end{aligned}
\label{eq:sms-summed-constraint}
\end{equation}
Appendix~\ref{sec:row-algebra} evaluates this contraction explicitly
on the five-plaquette row of Fig.~\ref{fig:sms-B}. The smooth merge and split procedure therefore implements the logical projection
\begin{equation}
\overline{\mathsf{P}}_K
=\sum_{h\in H}\sum_{h'\in H'}
\delta_{N'h',\,\varphi(Nh)}\ket{h,h'}\bra{h,h'}.
\label{eq:sms-logical-projector}
\end{equation}
Its action on logical group-basis states is
\begin{equation}
\overline{\mathsf{P}}_K \ket{j,k}
=\begin{cases}
\delta_{N'k,\,\varphi(Nj)}\ket{j,k},&j\in H,\ k\in H',\\
0,&\text{otherwise}.
\end{cases}
\label{eq:sms-logical-action}
\end{equation}
The logical projector projects the product of the $G$ and $G'$ surface-code spaces onto the subspace associated with the subgroup $K$.

\section{Smooth merge and split through measurements}
\label{sec:smooth-correction}

In Sec.~\ref{sec:smooth-surgery}, we assumed identity outcomes for measurements of the boundary projector $P_H$ and the plaquette operators $B_p^e$. In this section, we explain how to perform the smooth merge and split using measurements and appropriate corrections. We use the tensor network representation to derive the measurement circuits, explain how to correct nonidentity outcomes, and derive the resulting logical measurement.

\subsection{Measurement Circuits and Tensor Network Representation}
\label{sec:tensor-network-measurements}

We use the copy tensors and a generalization of the multiplication tensors in Ref.~\cite{manjunath2026} to represent the interface plaquette terms and their measurement circuits. We discuss these tensors further in Appendix~\ref{sec:regular-tensors}.

\subsubsection{Generalized multiplication tensor}
\label{sec:generalized-tensor}

The interface requires multiplication tensors with one label $h\in H$
and the remaining labels in $G'$. For each $h$, fix $h'\in H'$ such
that $N'h'=\varphi(Nh)$. The $H$ leg contributes $h'$ when its arrow
points inward and $(h')^{-1}$ when it points outward. An open circle
multiplies counterclockwise, and a crossed circle multiplies clockwise:
\begin{equation}
\begin{gathered}
\vcenter{\hbox{\NativeGeneralizedMult{0}}}\\[-2pt]
={}\delta_{h'g'_2(g'_3)^{-1}(g'_4)^{-1}g'_5,\,e},
\end{gathered}
\label{eq:generalized-multiplication-graphical}
\end{equation}
\begin{equation}
\begin{gathered}
\vcenter{\hbox{\NativeGeneralizedMult{1}}}\\[-2pt]
={}\delta_{g'_5(g'_4)^{-1}(g'_3)^{-1}g'_2h',\,e}.
\end{gathered}
\label{eq:generalized-multiplication-clockwise}
\end{equation}
Combining a copy tensor in Eq.~\ref{eq:copy-tensor} with the generalized clockwise multiplication
tensor constructs a controlled multiplication gate. As an example, consider an ancilla $A$ initialized in the subgroup state
\begin{equation}
\ket{N'}_{A} \equiv \frac{1}{\sqrt{|N'|}}\sum_{n'\in N'}\ket{n'}_{A}.
\end{equation}
The gate acts on
$\ket{h}\ket{N'}_A$ as
\begin{equation}
\widetilde{\mathrm{CR}}\ket{h}\ket{N'}_A
=\vcenter{\hbox{\NativeGeneralizedCR}}.
\label{eq:generalized-CR-graphical}
\end{equation}
The copy tensor fixes $x=h$. For each $n'\in N'$ in the ancilla
sum, the clockwise tensor gives
\begin{equation}
(z')^{-1}n'(h')^{-1}=e,
\qquad z'=n'(h')^{-1}.
\label{eq:generalized-CR-labels}
\end{equation}
Summing the output components yields
\begin{equation}
\begin{aligned}
&\widetilde{\mathrm{CR}}\ket{h}\ket{N'}_A\\
&\quad=\frac{1}{\sqrt{|N'|}}\sum_{n'\in N'}
\ket{h}\ket{n'(h')^{-1}}_A\\
&\quad=\ket{h}\ket{\varphi(Nh)^{-1}}_A.
\end{aligned}
\label{eq:generalized-CR-action}
\end{equation}

\subsubsection{Plaquette operator}
\label{sec:tensor-Bp}

The plaquette projectors have a compact tensor network representation
\begin{equation}
\begin{aligned}
B_{p_i,\mathrm{interface}}^e
&=\vcenter{\hbox{\InterfacePlaquetteCompact{h_i}{q'_{i-1}}{g'_i}{q'_i}}},\\[3pt]
\end{aligned}
\label{eq:Bp-generalized-compact}
\end{equation}

Using $\widetilde{\mathrm{CR}}_{h_i,A}$ from
Sec.~\ref{sec:generalized-tensor} and the controlled gates of
Appendix~\ref{sec:regular-tensors}, initialize the readout ancilla in
$\ket{N'}_A=\ket+_{N'}$. Assuming the identity measurement outcome, we derive the following measurement circuit:
\begin{equation}
\begin{gathered}
B_{p_i,\mathrm{interface}}^e
={}_A\!\bra{N'}U_i\ket{N'}_A\\[3pt]
=\vcenter{\hbox{\CosetPlaquetteNetwork{0}}},
\end{gathered}
\label{eq:Bp-generalized-decomposed}
\end{equation}
where
\begin{equation}
U_i=\mathrm{CR}_{q'_i,A}\,
\mathrm{C}\overline{\mathrm R}_{g'_i,A}\,
\mathrm{C}\overline{\mathrm R}_{q'_{i-1},A}\,\widetilde{\mathrm{CR}}_{h_i,A}.
\label{eq:sms-Bp-unitary}
\end{equation}

Reading the gates
from bottom to top gives
\begin{equation}
\begin{aligned}
\ket{N'}_A
&\xrightarrow{\widetilde{\mathrm{CR}}_{h_i,A}}\ket{\varphi(Nh_i)^{-1}}_A\\
&\xrightarrow{\mathrm{C}\overline{\mathrm R}_{q'_{i-1},A}}
\ket{\varphi(Nh_i)^{-1}q'_{i-1}}_A\\
&\xrightarrow{\mathrm{C}\overline{\mathrm R}_{g'_i,A}}
\ket{\varphi(Nh_i)^{-1}q'_{i-1}g'_i}_A\\
&\xrightarrow{\mathrm{CR}_{q'_i,A}}
\ket{\varphi(Nh_i)^{-1}q'_{i-1}g'_i(q'_i)^{-1}}_A.
\end{aligned}
\label{eq:sms-Bp-ancilla-steps}
\end{equation}
The final projection onto $\ket{N'}_A$ yields
\begin{equation}
\begin{aligned}
&{}_A\!\braket{N'}{\varphi(Nh_i)^{-1}q'_{i-1}g'_i(q'_i)^{-1}}_A\\
&=\delta_{\varphi(Nh_i)^{-1}q'_{i-1}g'_i(q'_i)^{-1},\,N'}\\
&=\delta_{N'q'_{i-1}g'_i(q'_i)^{-1},\,\varphi(Nh_i)}.
\end{aligned}
\label{eq:sms-Bp-internal-contraction}
\end{equation}

To convert the tensor network into a
measurement, we replace the projection with a measurement of the readout ancilla in the group basis. For the measurement outcome $N'm'_i$, the measurement projects the readout ancilla onto $\ket{N'm'_i}_A$:
\begin{equation}
\mathcal M_{p_i,\mathrm{interface}}(N'm'_i)
=\vcenter{\hbox{\CosetPlaquetteNetwork{1}}}.
\label{eq:Bp-generalized-measurement}
\end{equation}
The bold leg denotes the measurement, which acts conditionally on the outcome as follows:
\begin{equation}
\begin{aligned}
&\mathcal M_{p_i,\mathrm{interface}}(N'm'_i)
\ket{h_i,q'_{i-1},g'_i,q'_i}\\
&\quad=\delta_{\varphi(Nh_i)^{-1}q'_{i-1}g'_i(q'_i)^{-1},\,N'm'_i}\\
&\qquad\times\ket{h_i,q'_{i-1},g'_i,q'_i}.
\end{aligned}
\label{eq:sms-Bp-measurement-contraction}
\end{equation}
For the measurement outcome $N'$, this reduces to the $B_{p_i}^e$ measurement.

\subsection{Parallel Correction for Nonidentity Measurement Outcomes}

Here, we describe procedures for correcting nonidentity measurement outcomes. Parallel boundary correction applies when $H$ is subnormal in $G$. Parallel interface plaquette correction applies when $G'/Z_0$ is nilpotent for some subgroup $Z_0\leq N'$ that is normal in $G'$. Both conditions hold, in particular, when the patch groups $G$ and $G'$ are nilpotent. We formulate these procedures using general group labels and illustrate them with the $D_4$ interfaces used in our protocols. Appendix~\ref{sec:parallel-schemes} extends these procedures to more general interface data and derives sufficient conditions for parallel correction. These conditions hold, in particular, when both patch groups $G$ and $G'$ are nilpotent. For arbitrary finite-group interface data, corrections can instead be applied sequentially, one edge or plaquette at a time, as described in Appendix~\ref{sec:sequential-schemes}.

\subsubsection{Parallel correction for boundary projections}
\label{sec:correction-PH-example}

Here we explain how to correct nonidentity measurement outcomes of the boundary $P_H$ measurement in parallel when $H$ is subnormal in $G$. In each round, we perform simultaneous measurements and then apply corrections based on their outcomes. All smooth interfaces in the protocols of Sec.~\ref{sec:protocols} satisfy these conditions. We illustrate this procedure below with examples from these protocols.

We use five-plaquette rows to illustrate the corrections; the
constructions extend straightforwardly to a row of $n$ plaquettes.

As an example, consider $G=D_4$ and $H=\langle r^2,rs\rangle\cong\mathbb Z_2\times\mathbb Z_2$,
as in the second smooth interface of the $CS$ magic state protocol
(Sec.~\ref{sec:cs-magic-state}), where
\begin{equation}
D_4=\langle r,s\mid r^4=s^2=e,\ srs=r^{-1}\rangle.
\label{eq:correction-D4}
\end{equation}
The subgroup $H$ is normal in $D_4$.

The projector $P_H$ can be implemented by a coset measurement. Note that we can partition a group into right cosets of $H$. For each coset we define a projector
\begin{equation}
    Q_{r_i}=\sum_{h\in H}\ket{hr_i}\bra{hr_i},
\end{equation}
where $r_i$ is the representative for each right coset $Hr_i$. It can be shown that the set of projectors $Q_{r_i}$ forms a projective measurement. For the outcome $e$, we have $P_H = Q_{e}$.

\begin{figure}[!t]
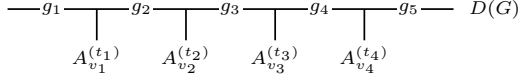

\pdfdest name {float-07-start} xyz
\centering
\SMSRestrictionStrip
\caption{Illustrate the coset measurement and the correction on the $D(G)$ boundary. The correction operators are the boundary vertex terms $A_{v_i}^{(t_i)}$. The set of group elements $\{t_{i}\}$ that is used in the corrections is determined by the coset measurements.}
\label{fig:sms-P}
\label{fig:parallel-PH}
\pdfdest name {float-07-end} xyz
\end{figure}

We can implementing the coset measurement with an ancilla. Let $R$ contain one representative of each right coset of $H$, with $e$ representing the identity coset. The ancilla has the Hilbert space
\begin{equation}
\mathcal H_A=\operatorname{span}{\ket{r_i}_A\in R}.
\end{equation}
Prepare an ancilla in $\ket{e}_{A}$ and apply the unitary
\begin{equation}
    U_{\mathrm{coset}}=\sum_{r_i}Q_{r_i}\otimes L_A^{r_i},
\end{equation}
we have
\begin{equation}
    U_{\mathrm{coset}}\ket{hr_i}\ket e_A=\ket{hr_i}\ket{r_i}_A.
\label{eq:sms-coset}
\end{equation}
Here the register $A$ records that representative. On the boundary of Fig.~\ref{fig:sms-P}, simultaneously
measure the right cosets of the first four labels using
Eq.~\eqref{eq:sms-coset}, obtaining $g_i\in Hr_i$.

To perform the parallel correction, we choose correction factors $t_i$ through their quotient labels:
\begin{equation}
\begin{aligned}
Ht_1&=Hr_1,& Ht_2&=Hr_1r_2,\\
Ht_3&=Hr_1r_2r_3,& Ht_4&=Hr_1r_2r_3r_4.
\end{aligned}
\label{eq:correction-PH-factors}
\end{equation}
We then apply the boundary vertex terms $A_{v_i}^{(t_i)}$ of the original $D(G)$ patch simultaneously to perform the correction. The resulting boundary labels are
\begin{equation}
\begin{aligned}
h_1&=g_1t_1^{-1},& h_2&=t_1g_2t_2^{-1},\\
h_3&=t_2g_3t_3^{-1},& h_4&=t_3g_4t_4^{-1},\\
\widetilde g_5&=t_4g_5.
\end{aligned}
\label{eq:correction-PH-row}
\end{equation}
The coset multiplication in $D_4/H$ gives
\begin{equation}
\begin{aligned}
Hh_i&=(Ht_{i-1})(Hr_i)(Ht_i)^{-1}=H,\\[-2pt]
&\hspace{38mm}i=1,2,3,4.
\end{aligned}
\label{eq:correction-PH-proof}
\end{equation}
Thus $h_1,\ldots,h_4\in H$. Note that, at adjacent vertices, one correction acts on the shared edge by left multiplication, while the other acts by right multiplication. These corrections commute because $a(gb^{-1})=(ag)b^{-1}$. 

The product $g_1g_2g_3g_4g_5$ labels the logical basis state. After performing the corrections, we have
\begin{equation}
g_1g_2g_3g_4g_5 \rightarrow h_1h_2h_3h_4\widetilde g_5.
\label{eq:correction-PH-product}
\end{equation}
Finally, we measure the coset on the fifth edge and postselect on the outcome with coset label $e$:
\begin{equation}
    h_1h_2h_3h_4\widetilde g_5 \rightarrow h_1h_2h_3h_4h_5.
\end{equation}
This completes the $P_H$ measurement procedure. After the $P_H$ measurement with corrections and postselection, the logical space of the $D(G)$ patch reduces to $|H|$. Only the final edge measurement requires postselection, so the success probability is finite. We note that for the smooth merge with $H=G$, we do not need to measure $P_H$.

\subsubsection{Parallel correction of the interface plaquette measurements}
\label{sec:correction-Bp-example}

\begin{figure}[!t]
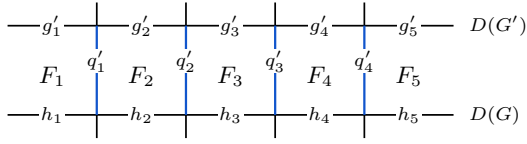

\pdfdest name {float-08-start} xyz
\centering
\SMSPlaquetteStrip
\caption{Five-plaquette interface for plaquette correction.}
\label{fig:sms-B}
\label{fig:parallel-Bp}
\pdfdest name {float-08-end} xyz
\end{figure}

We move on to discuss the parallel correction of the interface plaquette measurements. We can apply parallel corrections when $G'/Z_0$ is nilpotent for some subgroup $Z_0\leq N'$ that is normal in $G'$. As an illustrating example, we consider
$G'=D_4$ and $N'=\{e\}$, as in the first smooth interface
of the $T$ protocol (Sec.~\ref{sec:smooth magic protocol}). The
quotient label $\varphi(Nh_i)$ is then a group element. On each plaquette of the row in Fig.~\ref{fig:sms-B}, we define
\begin{equation}
F_i=\varphi(Nh_i)^{-1}q'_{i-1}g'_i(q'_i)^{-1} \in G',
\qquad i=1,\ldots,5,
\label{eq:sms-plaquette-word}
\end{equation}
with $q'_0=q'_5=e$. The interface plaquette constraint is satisfied precisely when $F_i=e$.

We begin with simultaneous measurements of the interface plaquette terms on the first four plaquettes. Nonidentity outcomes are corrected by acting in parallel on the vertical ancillas $q'_1,\ldots,q'_4$, while the fifth plaquette is reserved for the final logical readout. The measurement and correction procedure uses the center $Z=Z(D_4)=\langle r^2\rangle$ and the fact that $D_4/Z\cong\Z_2\times\Z_2$ is Abelian. These properties allow us to perform the correction in two rounds, as described below.

\begin{enumerate}
\item \emph{Correct modulo $Z$.} 
First, we simultaneously perform the coset measurement for the interface plaquette terms. Each measurement resolves the plaquette $F_i$ modulo $Z$, without distinguishing group elements within the same coset. Denote the outcome on the plaquette $i$ for $i=1,2,3,4$ by $C_i\in G'/Z$, and choose
a representative $m'_i\in G'$ such that $C_i=Zm'_i$. Thus the
measurement establishes $ZF_i=Zm'_i$.

Using these representatives, we choose the correction factors $\{t_i\}$ satisfying
\begin{equation}
\begin{gathered}
t_0=e,\quad Zt_1=Zm'_1,\quad Zt_2=Zm'_1m'_2,\\
Zt_3=Zm'_1m'_2m'_3,\quad Zt_4=Zm'_1m'_2m'_3m'_4.
\end{gathered}
\label{eq:correction-Bp-factors}
\end{equation}
Apply $L^{t_i}_{q'_i}$ to the four ancillas in parallel:
\begin{equation}
q'_i\longmapsto t_iq'_i,\qquad i=1,2,3,4.
\label{eq:correction-Bp-update}
\end{equation}
Since all elements in $G'$ commute modulo $Z$, the updated words satisfy
\begin{equation}
\begin{aligned}
ZF_i^{\mathrm{new}}
&=Zt_{i-1}F_it_i^{-1}\\
&=Zt_{i-1}m'_it_i^{-1}=Z,
\end{aligned}
\label{eq:correction-Bp-coarse-proof}
\end{equation}
where $Zt_i=Zt_{i-1}m'_i$. Hence $F_i^{\mathrm{new}}\in Z$ for $i=1,\ldots,4$: the first correction round leaves only central elements, which are removed in the second round.

\item \emph{Correct to identity.} 
We now simultaneously measure
these remaining central elements. Denote the outcome on
plaquette $i$ by $z_i\in Z$, so that $F_i=z'_i$.

For this round, set $t_0=e$ and choose new correction factors
\begin{equation}
t_i=z'_1\cdots z'_i,
\qquad i=1,\ldots,4.
\label{eq:correction-Bp-central-factors}
\end{equation}
Apply Eq.~\eqref{eq:correction-Bp-update} to the current
ancillas in parallel. All correction factors $t_i$ now belong
to $Z$ and therefore commute with every element of $D_4$. Consequently,
\begin{equation}
F_i^{\mathrm{new}}=t_{i-1}F_it_i^{-1}
=t_{i-1}z'_it_i^{-1}=e.
\label{eq:correction-Bp-fine-proof}
\end{equation}
\end{enumerate}
Thus the first four interface plaquette constraints are
satisfied. Note that the corrections act only on the added vertical ancillas, so the
ordered logical products $h=h_1\cdots h_5$ and $h'=g'_1\cdots g'_5$
are unchanged. 

After correcting the first four plaquettes, we measure the
remaining plaquette $F_5$ in the group element basis and denote its outcome by $m'\in G'$. This outcome imposes the logical constraint
\begin{equation}
F_5=m'
\quad\Longleftrightarrow\quad
h'=\varphi(Nh)m',
\label{eq:correction-Bp-logical-test}
\end{equation}
as derived in Appendix~\ref{sec:row-algebra}. 

\subsubsection{Logical operations}
Combining the correction and postselection procedure for $P_H$ with the correction procedure for the interface plaquette terms in Appendix~\ref{sec:parallel-correction}, we conclude that the logical operation implemented by the smooth merge and split specified by the subgroup $K$ defined in Eq.~\ref{eq:sms-group-data} is given by 
\begin{equation}
\overline{\mathsf{P}}_K(m') \ket{j,k}
=\begin{cases}
\delta_{N'k,\,\varphi(Nj)m'} \ket{j,k},&j\in H,\\
0,&\text{otherwise}.
\end{cases}
\label{eq:sms-logical-measurement-postselect}
\end{equation}
We define the right coset of \(K\) as 
\begin{equation}
K(e,m') \equiv \{(h,h')\in H\times G':N'h' = \varphi(Nh)m'\}.
\label{eq:kcoset}
\end{equation}
Here $m'\in G'$ labels the right coset $K(e,m')$. The projector in Eq.~\eqref{eq:sms-logical-measurement-postselect} projects onto the
subspace spanned by logical basis states whose labels
$(j,k)$ belong to the right coset $K(e,m')$ defined
in Eq.~\eqref{eq:kcoset}. We call this operation a
\emph{postselected logical $K$-coset measurement}. The
postselection consists of retaining the $P_H=1$ outcome
on the final horizontal edge qudit in the $P_H$ measurement
procedure. This postselection has a finite success probability.

When $H=G$, the projector $P_H$ is the identity, so its
measurement and the associated postselection are unnecessary.
The smooth merge and split then implements a logical
$K$-coset measurement on the full logical space, with
\begin{equation}
\overline{\mathsf{P}}_K(m')\ket{j,k}
=
\delta_{N'k,\,\varphi(Nj)m'}\ket{j,k},
\quad j\in G,\quad k\in G'.
\label{eq:sms-logical-measurement}
\end{equation}
When the meaning is clear, we omit the overline from the notation for logical operations.

\section{Applications of smooth hybrid lattice surgery}
\label{sec:protocols}

In this section, we discuss some applications of the smooth merge-and-split operation between different group surface codes. We demonstrate that, by combining with other logical operations in the group surface codes including rough merge-and-split and transversal gates, various logical operations can be realized. This includes a generalization of the controlled multiplication gates between $G$ and $G'$ surface codes, and magic state generation protocols for the $T$, $CZ$, $CCX$, and $CS$ magic states in copies of $\Z_2$ surface codes. 

Here we describe the protocols for the desired logical measurement outcome, corresponding to the identity coset. For this outcome, smooth merge and split implement the logical projector $\mathsf P_K$ in Eq.~\eqref{eq:sms-logical-action}. When the logical measurement yields a nonidentity outcome, we apply a corresponding logical recovery operation. Appendix~\ref{sec:logical_recovery_smooth} discuss these recovery procedures for all protocols in this section.

\subsection{Generalized Controlled Multiplication Gate}
\label{sec:controlled-multiplication}

To perform the generalized controlled multiplication on a $D(G)$ patch in the state
$\ket{j}_C$ and a $D(G')$ patch in the state $\ket{k}_T$, where
$\ket{j}$ is the control qudit and $\ket{k}$ is the target qudit, we
use the patch layout shown in Fig.~\ref{fig:int-target-control} and
proceed as follows:

\begingroup
\begin{figure}[!t]
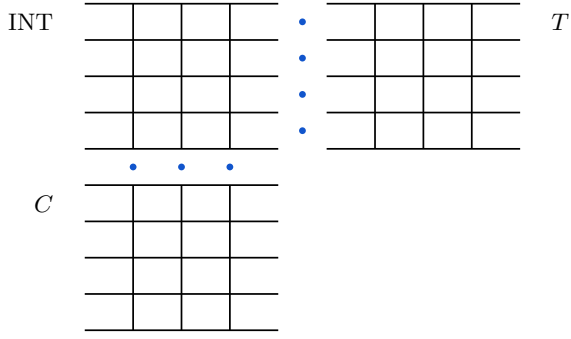

\pdfdest name {float-11-start} xyz
\centering
\NativeCNOTLayout
\caption{Intermediate $(\mathrm{INT})$ $D(G')$, target $(T)$ $D(G')$, and control $(C)$ $D(G)$
patches for the generalized controlled multiplication gate. Blue dots mark added ancillas.}
\label{fig:int-target-control}
\pdfdest name {float-11-end} xyz
\end{figure}
\endgroup

\begin{enumerate}
    \item First, initialize an intermediate $D(G')$ patch in the state
    \begin{equation}
        \ket{+}_{\mathrm{INT}}
        =
        \ket{+}_{G'}.
    \end{equation}

    \item Perform a smooth merge and split between $\ket{j}_C$ and
    $\ket{+}_{\mathrm{INT}}$. This gives
\begin{equation}
    \begin{aligned}
        \mathsf P_{K,C\mathrm{INT}}\ket j_C\ket+_{\mathrm{INT}}
        &\propto\sum_{t\in H'}\delta_{N't,\,\varphi(Nj)}
        \ket j_C\ket t_{\mathrm{INT}}.
    \end{aligned}
    \end{equation}
\item Perform a rough merge and split between the intermediate code
    patch and the $D(G')$ patch in the state $\ket{k}_T$. The resulting
    state is
    \begin{equation}
        \sum_{t\in H'}\sum_{h'\in H'}
        \delta_{N't,\,\varphi(Nj)}
        \ket{j}_{C}\ket{t{h'}^{-1}}_{\mathrm{INT}}
        \ket{h'k}_{T}.
    \end{equation}

    \item Finally, measure the intermediate code patch in the
    group-element basis and denote the measurement outcome by $m$.
    Since $m=t(h')^{-1}$ gives $h'=m^{-1}t$, apply left
    multiplication by $m$ to the target patch.
\end{enumerate}

The final state is
\begin{equation}
    \sum_{t\in H'}
    \delta_{N't,\,\varphi(Nj)}
    \ket{j}_{G}\ket{tk}_{G'},
\end{equation}
For $N'=\{e\}$, this gives the controlled multiplication
\begin{equation}
\ket{j}_G\ket{k}_{G'}
\longmapsto
\ket{j}_G\ket{\varphi(Nj)k}_{G'}.
\end{equation}
This map is unitary on the control subspace $j\in H$ because
$\varphi(Nj)$ specifies a unique element of $H'$. For nontrivial
$N'$, the preceding sum is a noninvertible coset projection followed
by the controlled multiplication.

\subsection{\texorpdfstring{\(T\) magic state and \(T\) gate}{T-magic state and T gate}}
\label{sec:smooth magic protocol}
Here we discuss how to use smooth hybrid lattice surgery to generate $T$ magic state or implement $T$ gate in the $Z_2$ surface code.

\begingroup
\begin{figure}[!t]
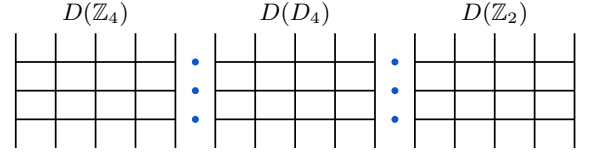


\pdfdest name {float-12-start} xyz
\centering

\NativeThreePatchLayout
\caption{Layout of the
\(D(\mathbb{Z}_4)\), \(D(D_4)\), and \(D(\mathbb{Z}_2)\) 
patches for generating the $T$-magic state or implementing the $T$ gate.
Blue dots mark added ancillas.}
\label{fig:smooth-three-patch}
\pdfdest name {float-12-end} xyz
\end{figure}
\endgroup

We will use three $D(G)$ patches with the following groups
\begin{equation}
\begin{aligned}
\mathbb{Z}_4^{(a)}
&=\langle a\mid a^4=e\rangle,\\
D_4
&=\langle r,s\mid r^4=s^2=e,\;srs=r^{-1}\rangle,\\
\mathbb{Z}_2^{(m)}
&=\langle m\mid m^2=e\rangle.
\end{aligned}
\end{equation}
and define
\begin{equation}
\eta=e^{i\pi/4}.
\end{equation}
The three code patches and the interface ancillas used in this
construction are arranged as in
Fig.~\ref{fig:smooth-three-patch}. 

To generate a $T$-magic state on a $D(\mathbb{Z}_2)$ code patch, we
proceed as follows:

\begin{enumerate}
    \item Begin with the $\Z_4$ stabilizer state
    \begin{equation}
        \ket{\phi_0}_{\mathbb{Z}_4}
        =
        \frac{1}{2}
        \left(
            \ket{e}
            +\eta\ket{a}
            -\ket{a^2}
            +\eta\ket{a^3}
        \right)
    \end{equation}
    on the $D(\mathbb{Z}_4)$ patch and the trivial-irrep state
    \begin{equation}
    \begin{aligned}
        \ket{+}_{D_4}
        &={}
        \frac{1}{\sqrt{8}}
        \Bigl(
            \ket{e}+\ket{r}+\ket{r^2}+\ket{r^3}
            \\[1pt]
        &\qquad
            +\ket{s}+\ket{rs}+\ket{r^2s}+\ket{r^3s}
        \Bigr)
    \end{aligned}
    \end{equation}
    on the $D(D_4)$ patch.

\item To perform a smooth merge and split, choose
    \begin{equation}
    \begin{aligned}
        G=H&=\mathbb{Z}_4^{(a)}, & N&=\{e\},\\
        G'&=D_4, & H'&=\langle r\rangle, & N'&=\{e\}.
    \end{aligned}
    \end{equation}
    with the isomorphism
    \begin{equation}
        \varphi(a)=r.
    \end{equation}
The corresponding diagonal subgroup is
    \begin{equation}
        K_1^{\mathrm{diag}}
        =
        \left\{
            (e,e),(a,r),(a^2,r^2),(a^3,r^3)
        \right\}.
    \end{equation}
The smooth merge and split gives
    \begin{align}
        \ket{\phi_1}
        &\propto\mathsf P_{K_1^{\mathrm{diag}}}
        \ket{\phi_0}_{\mathbb{Z}_4}\ket{+}_{D_4}
        \nonumber\\
        &\propto
        \ket{e}\ket{e}
        +\eta\ket{a}\ket{r}
        \nonumber\\[-1pt]
        &\quad
        -\ket{a^2}\ket{r^2}
        +\eta\ket{a^3}\ket{r^3}.
    \end{align}
\item Measure the $D(\mathbb{Z}_4)$ patch in the eigenbasis of the
    generalized Pauli operator
    \begin{equation}
    \begin{aligned}
        X_{\mathbb{Z}_4}
        &={}
        \sum_{j=0}^{3}
        \ket{a^{j+1}}\!\bra{a^j},
    \end{aligned}
    \end{equation}
    where the exponent is understood modulo $4$. The
    $+1$ measurement outcome gives
    \begin{equation}
        \ket{\phi_2}_{D_4}
        =
        \frac{1}{2}
        \left(
            \ket{e}
            +\eta\ket{r}
            -\ket{r^2}
            +\eta\ket{r^3}
        \right).
    \end{equation}
    For other outcomes, we perform a logical recovery operation to return to this state.

\item Next, measure the right-multiplication operator $R^s$. For the $+1$ outcome, it means applying the logical projector
    \begin{equation}
    \begin{aligned}
        P_{R^s=+1}
        &={}
        \frac{1}{2}\left(I+R^s\right),
        \\[2pt]
        R^s\ket{g}&=\ket{gs}.
    \end{aligned}
    \end{equation}
The resulting unnormalized state is
    \begin{align}
        \ket{\phi_3}_{D_4}
        &=
        \frac{1}{2}\left(I+R^s\right)\ket{\phi_2}_{D_4}
        \nonumber\\
        &=
        \frac{1}{4}
        \Bigl(
            \ket{e}
            +\eta\ket{r}
            -\ket{r^2}
            +\eta\ket{r^3}
        \Bigr)
        \nonumber\\[-1pt]
        &\quad+
        \frac{1}{4}
        \Bigl(
            \ket{s}
            +\eta\ket{rs}
            -\ket{r^2s}
            +\eta\ket{r^3s}
        \Bigr).
    \end{align}
\item Initialize a $D(\mathbb{Z}_2)$ patch in
    \begin{equation}
        \ket{+}_{\mathbb{Z}_2}
        =
        \frac{1}{\sqrt{2}}
        \left(\ket{e}+\ket{m}\right).
    \end{equation}
Perform a smooth merge and split between
    $\ket{\phi_3}_{D_4}$ and $\ket{+}_{\mathbb{Z}_2}$ with
\begin{equation}
    \begin{aligned}
        G&=D_4, & H&=\langle rs\rangle, & N&=\{e\},\\
        G'=H'&=\mathbb{Z}_2^{(m)}, & N'&=\{e\}.
    \end{aligned}
    \end{equation}
    and
    \begin{equation}
    \begin{gathered}
        \varphi(rs)=m,\\
        K_2^{\mathrm{diag}}=\{(e,e),(rs,m)\}.
    \end{gathered}
    \end{equation}
    Thus,
    \begin{align}
        \ket{\phi_4}
        &\propto\mathsf P_{K_2^{\mathrm{diag}}}
        \ket{\phi_3}_{D_4}\ket{+}_{\mathbb{Z}_2}
        \nonumber\\
        &\propto
        \ket{e}_{D_4}\ket{e}_{\mathbb{Z}_2}
        +\eta\ket{rs}_{D_4}\ket{m}_{\mathbb{Z}_2}.
    \end{align}
\item Finally, measure the $D(D_4)$ patch in the Fourier basis and, for the trivial-irrep outcome, we produce the $T$ magic state:
    \begin{align}
        \ket{\phi_5}
        &\propto{}_{D_4}\!\bra{+}\ket{\phi_4}
        \nonumber\\
        &\propto{}_{D_4}\!\bra{+}
        \left(
            \ket{e}_{D_4}\ket{e}_{\mathbb{Z}_2}
            +\eta\ket{rs}_{D_4}\ket{m}_{\mathbb{Z}_2}
        \right)
        \nonumber\\
        &\propto
        \ket{e}_{\mathbb{Z}_2}
        +\eta\ket{m}_{\mathbb{Z}_2}.
    \end{align}
\end{enumerate}
For other outcomes, logical recovery can be applied and recover the $T$ magic state.

Normalizing $\ket{\phi_5}$ gives the $T$-magic state
\begin{equation}
    \ket{T}_{\mathbb{Z}_2}
    =
    \frac{1}{\sqrt{2}}
    \left(
        \ket{e}
        +e^{i\pi/4}\ket{m}
    \right).
\end{equation}

This protocol actually implements the $T$ gate. To see this, we replace $\ket{+}_{\mathbb{Z}_2}$ in the second smooth merge and split by an arbitrary state
\begin{equation}
    \ket{\psi}_{\mathbb{Z}_2}
    =
    \alpha\ket{e}+\beta\ket{m}.
\end{equation}
The smooth merge and split then gives
\begin{equation}
\begin{aligned}
    \mathsf P_{K_2^{\mathrm{diag}}}\ket{\phi_3}_{D_4}\ket{\psi}_{\mathbb Z_2}
    &\propto\alpha\ket{e}_{D_4}\ket{e}_{\mathbb Z_2}\\
    &\quad+\eta\beta\ket{rs}_{D_4}\ket{m}_{\mathbb Z_2}.
\end{aligned}
\end{equation}
For the trivial-irrep measurement outcome on the $D(D_4)$ patch,
\begin{align}
    &{}_{D_4}\!\bra{+}
    \left(
        \alpha\ket{e}_{D_4}\ket{e}_{\mathbb{Z}_2}
        +
        \eta\beta\ket{rs}_{D_4}\ket{m}_{\mathbb{Z}_2}
    \right)
    \nonumber\\
    &\propto
    \alpha\ket{e}_{\mathbb{Z}_2}
    +
    e^{i\pi/4}\beta\ket{m}_{\mathbb{Z}_2}
    =
    T\ket{\psi}_{\mathbb{Z}_2}.
\end{align}
The preceding construction produces a single-qubit non-Clifford resource on a \(D(\mathbb Z_2)\) patch.

\subsection{$CZ$ Magic State}
Here we show that the $CZ$ magic state \texorpdfstring{\(\ket{CZ}\)}{M} can be generated by applying a non-Clifford transversal gate in the \(D_4\) patch and teleporting the state to two $D(Z_2)$ patches. Using the encoding \(a=s\), \(b=r^2\), and \(c=rs\), each
element of \(D_4\) has the unique normal form
\(a^\alpha b^\beta c^\gamma\). We encode the logical basis of the
\(D(D_4)\) patch as
\begin{equation}
    \ket{a^\alpha b^\beta c^\gamma}_L
    :=
    \ket{\alpha,\beta,\gamma}_L,
    \qquad
    \alpha,\beta,\gamma\in\mathbb{Z}_2.
\end{equation}
via smooth merge and split. 

First, initialize the $D(D_4)$ patch in
\begin{align}
    \ket{+0+}_{D_4}
    &=
    \frac{1}{2}
    \left(
        \ket{000}+\ket{001}+\ket{100}+\ket{101}
    \right)
    \nonumber\\
    &=
    \frac{1}{2}
    \left(
        \ket{e}+\ket{c}+\ket{a}+\ket{ac}
    \right).
\end{align}

Following Ref.~\cite{manjunath2026}, apply the transversal $\operatorname{SWAP}_{13}\operatorname{CCX}_{132}$ gate. This produces
\begin{align}
    \ket{\tau}_{D_4}
    &=
    \operatorname{SWAP}_{13}\operatorname{CCX}_{132}
    \ket{+0+}_{D_4}
    \nonumber\\
    &=
    \frac{1}{2}
    \left(
        \ket{000}+\ket{100}+\ket{001}+\ket{111}
    \right)
    \nonumber\\
    &=
    \frac{1}{2}
    \left(
        \ket{e}+\ket{a}+\ket{c}+\ket{abc}
    \right)
    \nonumber\\
    &=
    \frac{1}{2}
    \left(
        \ket{e}+\ket{a}+\ket{c}+\ket{ca}
    \right).
\end{align}

Next, initialize two $D(\mathbb{Z}_2)$ patches, labeled $A$ and $C$,
in 
\begin{equation}
    \ket{++}_{AC}
    =
    \frac{1}{2}
    \sum_{x,y\in\mathbb{Z}_2}\ket{xy}_{AC}.
\end{equation}

We first perform a smooth merge-and-split between the $D(D_4)$ patch and patch
$A$, choose
\begin{equation}
    G=H=D_4,\qquad
    N_A=\langle b,c\rangle,\qquad
    H/N_A\cong\mathbb{Z}_2,
\end{equation}
\begin{equation}
    G_A'=H_A'=\mathbb{Z}_2^{(A)},\qquad N_A'=\{0\},
\end{equation}
with
\begin{equation}
    \varphi_A
    \left(
        a^\alpha b^\beta c^\gamma N_A
    \right)
    =
    \alpha.
\end{equation}
The corresponding diagonal subgroup is
\begin{equation}
    K_A^{\mathrm{diag}}
    =
    \left\{
        \left(a^\alpha b^\beta c^\gamma,\alpha\right)
        \,\middle|\,
        \alpha,\beta,\gamma\in\mathbb{Z}_2
    \right\}.
\end{equation}

Then, for the smooth merge and split between the $D(D_4)$ patch and patch
$C$, choose
\begin{equation}
    G=H=D_4,\qquad
    N_C=\langle a,b\rangle,\qquad
    H/N_C\cong\mathbb{Z}_2,
\end{equation}
\begin{equation}
    G_C'=H_C'=\mathbb{Z}_2^{(C)},\qquad N_C'=\{0\},
\end{equation}
with
\begin{equation}
    \varphi_C
    \left(
        a^\alpha b^\beta c^\gamma N_C
    \right)
    =
    \gamma.
\end{equation}
Thus,
\begin{equation}
    K_C^{\mathrm{diag}}
    =
    \left\{
        \left(a^\alpha b^\beta c^\gamma,\gamma\right)
        \,\middle|\,
        \alpha,\beta,\gamma\in\mathbb{Z}_2
    \right\}.
\end{equation}

Since $ca=abc$, the input amplitudes can be written as
\begin{equation}
\begin{aligned}
\ket\tau_{D_4}
&=\frac12\bigl(\ket e+\ket c+\ket a+\ket{abc}\bigr)\\
&=\frac12\sum_{\alpha,\gamma\in\mathbb Z_2}
\ket{a^\alpha b^{\alpha\gamma}c^\gamma}.
\end{aligned}
\label{eq:tau-input-amplitudes}
\end{equation}
Applying the two smooth projectors gives
\begin{equation}
\begin{aligned}
\ket{\Phi_{\mathrm{smooth}}}
&\propto\mathsf P_{K_A^{\mathrm{diag}}}
\mathsf P_{K_C^{\mathrm{diag}}}\ket\tau_{D_4}\ket{++}_{AC}\\
&\propto\sum_{\alpha,\gamma,x,y\in\mathbb Z_2}
\delta_{\alpha,x}\delta_{\gamma,y}
\ket{a^\alpha b^{\alpha\gamma}c^\gamma}\ket{xy}\\
&=\sum_{\alpha,\gamma\in\mathbb Z_2}
\ket{a^\alpha b^{\alpha\gamma}c^\gamma}\ket{\alpha\gamma}\\
&=\ket e\ket{00}+\ket c\ket{01}
+\ket a\ket{10}+\ket{ca}\ket{11}.
\end{aligned}
\label{eq:smooth-three-term-state}
\end{equation}
Finally, measure the $D(D_4)$ patch onto $\ket{+0+}_{D_4}$. Since
$\ket{+0+}_{D_4}$ contains the basis states $\ket{e}$, $\ket{c}$,
$\ket{a}$, and $\ket{ac}$, with zero coefficient for $\ket{ca}$, we obtain
\begin{equation}
\begin{aligned}
    {}_{D_4}\!\braket{+0+}{\Phi_{\mathrm{smooth}}}
    &\propto{}
    \frac{1}{\sqrt{3}}
    \left(
        \ket{00}+\ket{01}+\ket{10}
    \right)_{AC}
    \\[2pt]
    &={}
    \ket{CZ}_{AC}.
\end{aligned}
\end{equation}

\subsection{$CCX$ Magic State}
Here we show that the $CCX$ magic state \texorpdfstring{\(\ket{CCX}\)}{M} can be generated similarly. First, initialize the $D(D_4)$ patch in the state $\ket{+0+}_{D_4}$. Then, applying the transversal gate
$\operatorname{SWAP}_{13}\operatorname{CCX}_{132}$ to
$\ket{+0+}_{D_4}$ prepares
\begin{align}
    \ket{\mathrm{CCX}}_{D_4}
    &:=
    \operatorname{SWAP}_{13}\operatorname{CCX}_{132}
    \ket{+0+}_{D_4}
    \nonumber\\
    &=
    \frac{1}{2}
    \left(
        \ket{e}+\ket{a}+\ket{c}+\ket{ca}
    \right),
\end{align}

To teleport the $CCX$ magic state to three copies of $D(\mathbb{Z}_2)$ patches, we initialize the three $D(\mathbb{Z}_2)$ patches in $\ket{+++}_{ABC}$ so that the ititial state is
\begin{equation}
    \ket{\Phi_0}
    =
    \ket{\mathrm{CCX}}_{D_4}\ket{+++}_{ABC}.
\end{equation}

First, perform a smooth merge and split between the $D(D_4)$ patch
and patch $A$, with
\begin{equation}
\begin{aligned}
&
    (G,G',H,H',N,N',\varphi_A)
    \\[2pt]
&\quad={}
    \bigl(
        D_4,\mathbb{Z}_2^{(A)},D_4,\mathbb{Z}_2^{(A)},
    \\[-1pt]
&\qquad
        \langle a,b\rangle,\{0\},\varphi_A
    \bigr).
\end{aligned}
\end{equation}
where
\begin{equation}
    \varphi_A
    \left(
        a^\alpha b^\beta c^\gamma N
    \right)
    =
    N'+\gamma.
\end{equation}
Thus,
\begin{equation}
    K_A^{\mathrm{diag}}
    =
    \left\{
        \left(a^\alpha b^\beta c^\gamma,\gamma\right)
        \,\middle|\,
        \alpha,\beta,\gamma\in\mathbb{Z}_2
    \right\}.
\end{equation}
The resulting state is
\begin{equation}
\begin{aligned}
    \ket{\Phi_1}
    &\propto\mathsf P_{K_A^{\mathrm{diag}}}\ket{\Phi_0}\\
    &\propto{}
    \ket{e}\ket{0++}
    +\ket{c}\ket{1++}
    \\[1pt]
    &\quad{}
    +\ket{a}\ket{0++}
    +\ket{ca}\ket{1++}.
\end{aligned}
\end{equation}
Next, perform a smooth merge and split with patch $C$, using
\begin{equation}
\begin{aligned}
&
    (G,G',H,H',N,N',\varphi_C)
    \\[2pt]
&\quad={}
    \bigl(
        D_4,\mathbb{Z}_2^{(C)},D_4,\mathbb{Z}_2^{(C)},
    \\[-1pt]
&\qquad
        \langle ac\rangle,\{0\},\varphi_C
    \bigr).
\end{aligned}
\end{equation}
where
\begin{equation}
    \langle ac\rangle=\{e,ac,b,ca\}
\end{equation}
and
\begin{equation}
    \varphi_C
    \left(
        a^\alpha b^\beta c^\gamma N
    \right)
    =
    N'+\alpha+\gamma.
\end{equation}
The corresponding diagonal subgroup is
\begin{equation}
    K_C^{\mathrm{diag}}
    =
    \left\{
        \left(
            a^\alpha b^\beta c^\gamma,
            \alpha+\gamma
        \right)
        \,\middle|\,
        \alpha,\beta,\gamma\in\mathbb{Z}_2
    \right\}.
\end{equation}
We obtain
\begin{equation}
\begin{aligned}
    \ket{\Phi_2}
    &\propto\mathsf P_{K_C^{\mathrm{diag}}}\ket{\Phi_1}\\
    &\propto{}
    \ket{e}\ket{0+0}
    +\ket{c}\ket{1+1}
    \\[1pt]
    &\quad{}
    +\ket{a}\ket{0+1}
    +\ket{ca}\ket{1+0}.
\end{aligned}
\end{equation}
Apply the projector
\begin{equation}
    Q_{bc}
    =
    \frac{1}{2}
    \left(
        I+\overline{R}^{\,bc}
    \right)
\end{equation}
to the $D(D_4)$ patch. This gives
\begin{align}
    \ket{\Phi_3}
    &=
    Q_{bc}\ket{\Phi_2}
    \nonumber\\
    &\propto
    \left(\ket{e}+\ket{bc}\right)\ket{0+0}
    \nonumber\\
    &\quad+
    \left(\ket{c}+\ket{b}\right)\ket{1+1}
    \nonumber\\
    &\quad+
    \left(\ket{a}+\ket{ca}\right)\ket{0+1}
    \nonumber\\
    &\quad+
    \left(\ket{ca}+\ket{a}\right)\ket{1+0}.
\end{align}

Finally, perform a smooth merge and split with patch $B$, using
\begin{equation}
\begin{aligned}
&
    (G,G',H,H',N,N',\varphi_B)
    \\[2pt]
&\quad={}
    \bigl(
        D_4,\mathbb{Z}_2^{(B)},D_4,\mathbb{Z}_2^{(B)},
    \\[-1pt]
&\qquad
        \langle a,b\rangle,\{0\},\varphi_B
    \bigr).
\end{aligned}
\end{equation}
where
\begin{equation}
    \varphi_B
    \left(
        a^\alpha b^\beta c^\gamma N
    \right)
    =
    N'+\gamma.
\end{equation}
The resulting state is
\begin{align}
    \ket{\Phi_4}
    &\propto\mathsf P_{K_B^{\mathrm{diag}}}\ket{\Phi_3}\\
    &\propto{}
    \ket{e}\ket{000}
    +\ket{bc}\ket{010}
    \nonumber\\
    &\quad+
    \ket{c}\ket{111}
    +\ket{b}\ket{101}
    \nonumber\\
    &\quad+
    \ket{a}\ket{001}
    +\ket{ca}\ket{011}
    \nonumber\\
    &\quad+
    \ket{ca}\ket{110}
    +\ket{a}\ket{100}.
\end{align}
Measure the $D(D_4)$ patch on to
\begin{equation}
    \ket{+0+}_{D_4}
    =
    \frac{1}{2}
    \left(
        \ket{e}+\ket{c}+\ket{a}+\ket{ac}
    \right).
\end{equation}
Only the components proportional to $\ket{e}$, $\ket{c}$, and
$\ket{a}$ survive. Hence,
\begin{align}
    \ket{\Phi_5}
    &\propto
    {}_{D_4}\!\braket{+0+}{\Phi_4}
    \nonumber\\
    &\propto
    \ket{000}+\ket{001}+\ket{100}+\ket{111}.
\end{align}
After normalization,
\begin{equation}
\begin{aligned}
    \ket{\Phi_5}
    &={}
    \frac{1}{2}
    \left(
        \ket{000}+\ket{001}+\ket{100}+\ket{111}
    \right)
    \\[2pt]
    &={}
    \ket{\mathrm{CCX}}_{ABC}.
\end{aligned}
\end{equation}

\subsection{$CS$ Magic State}
\label{sec:cs-magic-state}
Finally, we discuss how to generate the $CS$ magic state. Let
\begin{equation}
\begin{split}
    \mathbb{Z}_2^2=\{e,x,y,xy\},
    \qquad
    &\ket{e}=\ket{00},\quad
    \ket{x}=\ket{01},\quad
    \\
    &\ket{y}=\ket{10},\quad
    \ket{xy}=\ket{11}.
\end{split}
\end{equation}
The $CS$ magic state is generated as follows:
\begin{enumerate}
    \item Initialize a $D(\mathbb{Z}_2)$ patch in $\ket{+}$ and apply the logical $S$ gate
    \begin{equation}
        S\ket{+}_{\mathbb{Z}_2}
        =
        \frac{1}{\sqrt{2}}
        \left(
            \ket{0}+i\ket{1}
        \right),
    \end{equation}
    and a $D(D_4)$ patch in the trivial-irrep state
    \begin{equation}
        \ket{+}_{D_4}
        =
        \frac{1}{\sqrt{8}}
        \sum_{g\in D_4}\ket{g}.
    \end{equation}
    Perform a smooth merge and split with
    \begin{equation}
    \begin{aligned}
        G&=H=\mathbb{Z}_2, & N&=\{0\},\\
        G'&=D_4, & H'&=\langle s\rangle,
        & N'&=\{e\},
    \end{aligned}
    \end{equation}
    with isomorphism defined as
    \begin{equation}
        \varphi_1(0)=e,
        \qquad
        \varphi_1(1)=s.
    \end{equation}
    The corresponding diagonal subgroup is
    \begin{equation}
        K_1^{\mathrm{diag}}
        =
        \{(0,e),(1,s)\}.
    \end{equation}
For the identity outcome, we apply the logical projector
    \begin{equation}
    \begin{aligned}
    &\mathsf P_{K_1^{\mathrm{diag}}}
    \bigl(S\ket+_{\mathbb Z_2}\bigr)\ket+_{D_4}\\
    &\qquad\propto\ket0\ket e+i\ket1\ket s.
    \end{aligned}
    \end{equation}
    Measuring the $D(\mathbb{Z}_2)$ patch onto $\ket{+}$, the remaining state is
\begin{equation}
        \ket{\psi_1}_{D_4}
        =
        \frac{1}{\sqrt{2}}
        \left(
            \ket{e}+i\ket{s}
        \right).
    \end{equation}

    \item On the $D(D_4)$ patch, sequentially measure $L^{rs}$ and
    $L^s$, for the $+1$ outcome in each case, we apply the following logical projectors:
    \begin{equation}
    \begin{aligned}
        P_{L^{rs}=+1}
        &=
        \frac{1}{2}\left(I+L^{rs}\right),\\
        P_{L^s=+1}
        &=
        \frac{1}{2}\left(I+L^s\right),
        \qquad
        L^h\ket{g}=\ket{hg}.
    \end{aligned}
    \end{equation}
    After normalization, the resulting state is
    \begin{align}
        \ket{\psi_2}_{D_4}
        &=
        \frac{1}{2\sqrt{2}}
        \Bigl[
            (1+i)\ket{e}
            +i\ket{r}
            +\ket{r^3}
            \nonumber\\[-1pt]
        &\qquad\qquad
            +(1+i)\ket{s}
            +\ket{rs}
            +i\ket{r^3s}
        \Bigr].
    \end{align}

    \item Initialize a $D(\mathbb{Z}_2^2)$ patch in
    \begin{equation}
        \ket{++}_{\mathbb{Z}_2^2}
        =
        \frac{1}{2}
        \left(
            \ket{e}+\ket{x}+\ket{y}+\ket{xy}
        \right).
    \end{equation}
    Perform a second smooth merge-and-split with
    \begin{equation}
    \begin{aligned}
        G&=D_4,
        & H&=\langle r^2,rs\rangle,
        & N&=\{e\},\\
        G'&=H'=\mathbb{Z}_2^2,
        && N'=\{e\},
    \end{aligned}
    \end{equation}
    with isomorphism defined as
    \begin{equation}
    \begin{aligned}
        \varphi_2(e)&=e,
        &\varphi_2(r^2)&=x,\\
        \varphi_2(rs)&=xy,
        &\varphi_2(r^3s)&=y.
    \end{aligned}
    \end{equation}
    and
    \begin{equation}
        K_2^{\mathrm{diag}}
        =
        \{(e,e),(r^2,x),(rs,xy),(r^3s,y)\}.
    \end{equation}
Applying the logical projector for the second merge-and-split gives
    \begin{equation}
    \begin{aligned}
    &\mathsf P_{K_2^{\mathrm{diag}}}\ket{\psi_2}_{D_4}
    \ket+_{\mathbb Z_2^2}\\
    &\qquad\propto(1+i)\ket e\ket e
    +i\ket{r^3s}\ket y+\ket{rs}\ket{xy}.
    \end{aligned}
    \end{equation}
    
Then, measure the $D(D_4)$ patch in the Fourier basis. For the trivial-irrep outcome. The $D(\mathbb{Z}_2^2)$ patch is left in
\begin{equation}
        \ket{\psi_3}
        =
        \frac{1}{2}
        \left[
            (1+i)\ket{00}
            +i\ket{10}
            +\ket{11}
        \right].
    \end{equation}
This state is Clifford equivalent to the $CS$ magic state. To see this explicitly, we apply the logical Hadamard gate
    $H_2=I\otimes H$. This gives
    \begin{align}
        H_2\ket{\psi_3}
        &=
        \frac{1+i}{2\sqrt{2}}
        \left(
            \ket{00}+\ket{01}+\ket{10}+i\ket{11}
        \right)
        \nonumber\\
        &=
        e^{i\pi/4}\ket{\mathrm{CS}},
    \end{align}
    where
    \begin{equation}
        \ket{\mathrm{CS}}
        :=
        \mathrm{CS}\ket{++}
        =
        \frac{1}{2}
        \left(
            \ket{00}+\ket{01}+\ket{10}+i\ket{11}
        \right).
    \end{equation}
\end{enumerate}

\section{Discussion}
\label{sec:discussion}
We have developed smooth hybrid lattice surgery for group surface codes, extending the rough-boundary construction of Ref.~\cite{huang2025Hybrid}. Our central result is that smooth merge and split realize a logical $K$-coset measurement, with the interface subgroup determining the measured subspaces and the operation preserving coherence within the selected subspace. We provide explicit measurement circuits and procedures for correcting nonidentity measurement outcomes. Combining these operations with rough surgery and other logical operations yields generalized controlled multiplication, extending the surface-code CNOT construction of Ref.~\cite{horsman2012}, and protocols for preparing the $T$, $CZ$, $CCX$, and $CS$ magic states. The resulting resources reside in one or more $\mathbb{Z}_2$ surface-code patches, while other group surface codes serve as auxiliary patches.

Fault-tolerant implementations of these protocols can build on just-in-time decoding for non-Abelian codes with $D(D_4)$ topological order, together with suitable decoders for the Abelian patches~\cite{bombin2018,brown2020,scruby2022,davydova2025,Magdalena2026decoding}. Applying them to smooth hybrid lattice surgery requires incorporating the boundary and interface that change during merge and split. Numerical simulations are needed to determine the thresholds and logical error rates of the resulting protocols, which we left as future work.

The protocols presented here are not expected to exhaust the possibilities of hybrid lattice surgery. Different choices of group surface codes, interface subgroups, and sequences of rough and smooth surgery may implement the same logical operation with lower space--time cost. Optimizing these choices is therefore an important direction for future work. A natural direction is to extend the smooth hybrid lattice surgery protocols to larger dihedral groups and investigate the preparation of resource states for gates at higher levels of the Clifford hierarchy. The corresponding rough-surgery constructions provide a starting point~\cite{huang2025Hybrid}. Such extensions could enable direct preparation of resources suited to the non-Clifford operations required by a given quantum algorithm. Preparing these states directly from stabilizer inputs could avoid intermediate conversions and reduce the resource overhead of implementing the algorithm.

A further direction is to develop the topological quantum field theory (TQFT) description of smooth hybrid lattice surgery. Our previous work related rough hybrid lattice surgery to gauging boundary symmetries and identifying anyon data across interfaces~\cite{huang2025Hybrid}. Further developing the corresponding description of smooth surgery and connecting it to the logical coset measurements derived here could clarify how interface choices and sequences of surgery operations determine the resulting logical action. This perspective could guide both the optimization of the present protocols and the construction of protocols for additional non-Clifford resources.

\begin{acknowledgments}
While completing this manuscript, we became aware of the work of Li et al.~\cite{Li2026irrep}, which presents smooth merge and split operations between irrep surface codes and ordinary surface codes for joint logical measurements.
We used Anthropic’s Claude Opus 5.5 and ChatGPT Astra to assist with checking some computations and to improve the presentation of parts of this work. This material is based upon work supported by the U.S. Department of Energy, Office of Science, Advanced Scientific Computing Research under Award No. DE-SC0026209. GD and SJ-H are supported by startup funds at Florida State University.

\end{acknowledgments}

\appendix

\section{Review Rough Merge and Split}
\label{sec:review-rough-merge-split}

This subsection reviews our previous rough merge and split construction
from Ref.~\cite{huang2025Hybrid}. We give the definitions and logical action
needed for the novel protocols in Sec.~\ref{sec:protocols}; the
full derivation and implementation details are provided in the original
work.
Consider \(D(G)\) and \(D(G')\) patches whose facing boundaries are rough.
After folding the \(D(G')\) patch, the interface is a boundary of
\(D(G\times G')\). The interface is specified by
\begin{equation}
\begin{gathered}
  K\leq G,\qquad N\trianglelefteq K,\qquad
  p:K\longrightarrow G',\\
  \ker p=N,\qquad p(K)\cong K/N\leq G',\\
  K^{\mathrm{diag}}
  =\{(h,p(h)):h\in K\}\leq G\times G'.
\end{gathered}
  \label{eq:review-diagonal-subgroup}
\end{equation}
Here \(K\) is a subgroup of \(G\), \(N\) is a normal subgroup of \(K\),
\(p:K\to G'\) is a homomorphism with kernel \(N\), and
\(K^{\mathrm{diag}}\) is its graph in \(G\times G'\). 
A cocycle on \(K^{\mathrm{diag}}\) may also be included; the protocols
reviewed here use the trivial cocycle.

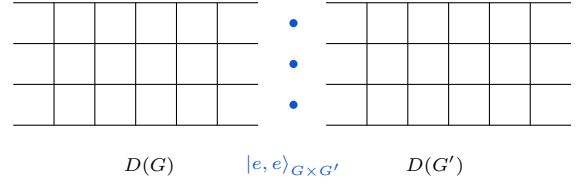
\begin{figure}[!t]
\pdfdest name {float-04-start} xyz
\centering
\begingroup\setlength{\unitlength}{0.36mm}
\begin{picture}(215,78)(0,-4)
  \linethickness{0.4pt}
  \multiput(5,25)(0,15){4}{\line(1,0){90}}
  \multiput(120,25)(0,15){4}{\line(1,0){90}}
  \multiput(20,25)(15,0){5}{\line(0,1){45}}
  \multiput(135,25)(15,0){5}{\line(0,1){45}}
  {\color{paperprimary}\multiput(108,32.5)(0,15){3}{\circle*{3}}}
  \put(55,10){\makebox(0,0){\scriptsize\(D(G)\)}}
  \put(160,10){\makebox(0,0){\scriptsize\(D(G')\)}}
  \put(108,10){\makebox(0,0){\scriptsize\color{paperprimary}
    \(\ket{e,e}_{G\times G'}\)}}
\end{picture}
\endgroup
\caption{Rough-boundary interface with ancillas initialized in
$\ket{e,e}_{G\times G'}$.}
\label{fig:review-rough-merge-split}
\pdfdest name {float-04-end} xyz
\end{figure}

The rough-merge layout is shown in Fig.~\ref{fig:review-rough-merge-split}.
The merge introduces a \((G\times G')\) ancilla qudit on each interface edge,
initialized in \(\ket{e,e}\), and then enforces
\begin{equation}
  \widetilde A_v^{K^{\mathrm{diag}}}
  =\frac{1}{\lvert K\rvert}\sum_{h\in K}
    A_{v,G}^{(h)}\otimes A_{v,G'}^{(p(h))}.
  \label{eq:review-interface-vertex}
\end{equation}
Here \(A_{v,G}^{(h)}\) and \(A_{v,G'}^{(p(h))}\) act on the
\(G\)- and \(G'\)-edges, respectively.
The orientations determine whether each local action is implemented by a
controlled \(L\) or \(R\). Measuring these vertex projectors sequentially
along the interface from bottom to top realizes the merge operation. The split reverses the code deformation by measuring out the ancilla qudits in the group basis.

For identity outcomes on all interface qudits, this merge and split procedure
gives
\begin{equation}
\begin{aligned}
\ket{\Psi_{1,2}(\mathrm{id},\ldots,\mathrm{id})}_{\mathrm{disjoint}}
={}&\frac{1}{\lvert K\rvert^{n_v}}\sum_{k\in K}
\left(\prod_{v'=1}^{n_v}R_{v'}^k\ket{\Psi_1}_G\right)\\
&\otimes
\left(\prod_{v''=1}^{n_v}L_{v''}^{p(k)}\ket{\Psi_2}_{G'}\right).
\end{aligned}
  \label{eq:review-rough-final-operation}
\end{equation}
Here \(\ket{\Psi_1}_G\) and \(\ket{\Psi_2}_{G'}\) are the input code
states, and \(n_v\) is the number of interface vertices. The indices
\(v'\) and \(v''\) label the corresponding boundary vertices of the
\(D(G)\) and \(D(G')\) patches. At each such vertex, \(R_{v'}^k\) acts on
the adjacent \(G\)-boundary edge and \(L_{v''}^{p(k)}\) acts on the
adjacent \(G'\)-boundary edge. Their products are the logical operators
\(\overline R_G^{k}\) and \(\overline L_{G'}^{p(k)}\), respectively.
Thus the rough merge followed by rough split projects onto the logical
subspace stabilized by
\(\overline R_G^{k}\otimes\overline L_{G'}^{p(k)}\) for \(k\in K\).
Nonidentity split outcomes give known boundary defects or logical
byproducts; they can be corrected or absorbed into subsequent stabilizer
definitions as described in Ref.~\cite{huang2025Hybrid}.

\section{Smooth Merge and Split from the folded picture}
\label{sec:smooth-merge-split-folded}

\begin{figure}[!t]
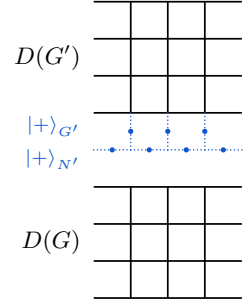

\pdfdest name {float-06-start} xyz
\centering
\begin{minipage}[b]{.48\columnwidth}\centering
\NativeFoldedAncilla\\[-1pt]
\end{minipage}
\caption{Ancilla initialization for smooth merge and split from the folded picture.}
\label{fig:sms-ancilla-geometries}
\label{fig:folded-ancilla}
\pdfdest name {float-06-end} xyz
\end{figure}

In this appendix, we introduce another smooth merge and split protocol. If we fold along the horizontal line containing the horizontal ancillas (Fig.~\ref{fig:sms-ancilla-geometries}), the resulting boundary terms match those in Ref.~\cite{beigi2011}. This standard form lets us identify the subgroup associated with the interface, making the protocol conceptually simpler. However, the protocol requires more ancillas than the one in Sec.~\ref{sec:sms-state-algebra}, making it less practical. We show that both protocols implement the same logical operation.

\begin{enumerate}
\item \emph{Initialize.} Add horizontal and vertical
$G'$ ancillas (Fig.~\ref{fig:sms-ancilla-geometries}). Prepare the
vertical ancillas in $\ket+_{G'}$, the horizontal ancillas in
\begin{equation}
\ket+_{N'}=\frac{1}{\sqrt{|N'|}}\sum_{n'\in N'}\ket{n'}.
\label{eq:sms-Nprime-ancilla}
\end{equation}
\item \emph{Boundary projection.} Project all edge states on
the facing boundary of $D(G)$ to the subgroup $H$, using
Eq.~\eqref{eq:sms-PH} as in the unfolded picture. Couple each
boundary edge to its horizontal ancilla by
\begin{equation}
U\ket h\ket+_{N'}
=\frac{1}{\sqrt{|N'|}}\sum_{h'\in H'}
\delta_{N'h',\,\varphi(Nh)}\ket{h,h'}.
\label{eq:sms-pair-state}
\end{equation}
This is implemented by controlled left multiplication with a fixed
representative of $\varphi(Nh)$; its action on $\ket+_{N'}$ is independent
of that choice. Disjoint pairs are coupled simultaneously.

\item \emph{Smooth merge.} Enforce the interface plaquette stabilizers
one plaquette at a time, from left to right. The interface stabilizer is
defined by
\begin{equation}
\begin{gathered}
B_{p_i,\mathrm{folded}}^e\quad\vcenter{\hbox{\SMSFoldedCell}}\\[-2pt]
={}\delta_{N'q'_{i-1}g'_i(q'_i)^{-1},\,N'h'_i}
\quad\vcenter{\hbox{\SMSFoldedCell}}.
\end{gathered}
\label{eq:Bp-folded-graphical}
\end{equation}

\item \emph{Split.} Measure both sets of added ancillas in the
representation basis of $G'$ and retain the trivial-representation
outcomes.
\end{enumerate}

\begin{figure*}[!t]
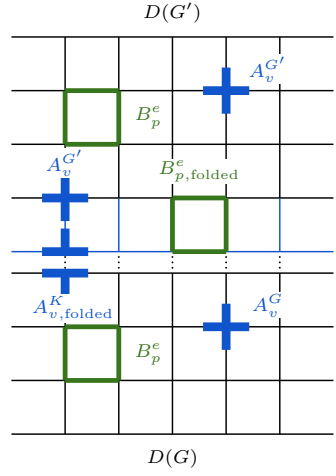

\pdfdest name {float-10-start} xyz
\centering
\renewcommand{\NPLatticeArrow}[3]{}
\begin{minipage}[b]{.47\textwidth}\centering
\SMSFolded\\[-1pt]
\end{minipage}
\caption{Bulk and interface projectors of the hybrid code patch. The added edges are shown in blue.}
\label{fig:sms-interface}
\label{fig:folded-stabilizer-support}
\pdfdest name {float-10-end} xyz
\end{figure*}

In the folded picture, the interface vertex projector is
\begin{equation}
A_{v,\mathrm{folded}}^K
=\frac{1}{|K|}\sum_{(h,h')\in K}
A_{v,\mathrm{folded}}^{(h,h')},
\end{equation}
with local action
\begin{equation}
\begin{aligned}
&A_{v,\mathrm{folded}}^{(h,h')}
\vcenter{\hbox{\resizebox{.44\columnwidth}{!}{\NativeAvUnfoldedInput}}}\\[3pt]
&\qquad={}\vcenter{\hbox{\resizebox{.70\columnwidth}{!}{\NativeAvUnfoldedOutput}}}.
\end{aligned}
\label{eq:Av-unfolded}
\end{equation}
The bulk and the interface projectors are shown in Fig.~\ref{fig:folded-stabilizer-support}.

We can now compare the protocol from the folded picture with the one in Sec.~\ref{sec:sms-state-algebra}. On the paired subspace,
$N'h'_i=\varphi(Nh_i)$, so
\begin{equation}
\delta_{N'q'_{i-1}g'_i(q'_i)^{-1},\,N'h'_i}
=\delta_{N'q'_{i-1}g'_i(q'_i)^{-1},\,\varphi(Nh_i)}.
\label{eq:sms-local-equivalence}
\end{equation}
Each horizontal ancilla contributes the same split overlap
$\sqrt{|N'|/|G'|}$, independently of $h_i$.
Thus, the folded picture produces the same logical operator $\mathsf{P}_K$ up to an overall scalar factor.

\section{Derivation of the Logical Operation}
\label{sec:row-algebra}

We evaluate the ancilla sum on the five-plaquette row of
Fig.~\ref{fig:sms-B} by contracting the coset deltas from left to right,
enforcing one plaquette constraint at each step. The group data are
those of Eq.~\eqref{eq:sms-group-data}, with $h_i\in H$ and
$g'_i,q'_i\in G'$. Write $h=h_1h_2h_3h_4h_5$ and
$h'=g'_1g'_2g'_3g'_4g'_5$.

\subsection{Identity Measurement Outcomes}
\label{sec:row-identity}

With $q'_0=q'_5=e$, the five local constraints are
\begin{equation}
\begin{aligned}
N'g'_1(q'_1)^{-1}&=\varphi(Nh_1),\\
N'q'_1g'_2(q'_2)^{-1}&=\varphi(Nh_2),\\
N'q'_2g'_3(q'_3)^{-1}&=\varphi(Nh_3),\\
N'q'_3g'_4(q'_4)^{-1}&=\varphi(Nh_4),\\
N'q'_4g'_5&=\varphi(Nh_5).
\end{aligned}
\label{eq:row-five-constraints}
\end{equation}
The first delta restricts $q'_1$ to
$N'q'_1=\varphi(Nh_1)^{-1}g'_1$. Substituting this into the
second delta gives
\begin{equation}
\begin{aligned}
&\sum_{q'_1\in G'}
\delta_{N'g'_1(q'_1)^{-1},\,\varphi(Nh_1)}
\delta_{N'q'_1g'_2(q'_2)^{-1},\,\varphi(Nh_2)}\\
&\qquad\propto
\delta_{N'g'_1g'_2(q'_2)^{-1},\,\varphi(Nh_1h_2)}.
\end{aligned}
\label{eq:row-product-two}
\end{equation}
Here $\varphi(Nh_1)\varphi(Nh_2)=\varphi(Nh_1h_2)$ in $H'/N'$.
Contracting the next two deltas in order yields
\begin{align}
&\sum_{q'_1,q'_2\in G'}\prod_{i=1}^3
\delta_{N'q'_{i-1}g'_i(q'_i)^{-1},\,\varphi(Nh_i)}\nonumber\\
&\qquad\propto
\delta_{N'g'_1g'_2g'_3(q'_3)^{-1},\,\varphi(Nh_1h_2h_3)},
\label{eq:row-product-three}\\[2pt]
&\sum_{q'_1,q'_2,q'_3\in G'}\prod_{i=1}^4
\delta_{N'q'_{i-1}g'_i(q'_i)^{-1},\,\varphi(Nh_i)}\nonumber\\
&\qquad\propto
\delta_{N'g'_1g'_2g'_3g'_4(q'_4)^{-1},\,\varphi(Nh_1h_2h_3h_4)}.
\label{eq:row-product-four}
\end{align}
Finally, contracting with the fifth delta and summing over $q'_4$ gives
\begin{equation}
\begin{aligned}
&\sum_{q'_1,q'_2,q'_3,q'_4\in G'}
\prod_{i=1}^5
\delta_{N'q'_{i-1}g'_i(q'_i)^{-1},\,\varphi(Nh_i)}\\
&\qquad\propto\delta_{N'h',\,\varphi(Nh)}.
\end{aligned}
\label{eq:row-five-count}
\end{equation}
Uniform ancilla preparation and trivial-representation readout
therefore give the logical operator $\mathsf P_K$.

\subsection{Other Final Plaquette Outcomes}
\label{sec:row-shifted}

Retain the first four constraints in
Eq.~\eqref{eq:row-five-constraints}, and resolve the final right coset as
\begin{equation}
F_5=\varphi(Nh_5)^{-1}q'_4g'_5=N'm'.
\label{eq:row-final-readout}
\end{equation}
Equation~\eqref{eq:row-product-four} implies
\begin{equation}
N'q'_4=\varphi(Nh_1h_2h_3h_4)^{-1}g'_1g'_2g'_3g'_4.
\label{eq:row-final-ancilla}
\end{equation}
Substituting into the final word gives
\begin{equation}
\begin{aligned}
F_5
&=\varphi(Nh_5)^{-1}
  \varphi(Nh_1h_2h_3h_4)^{-1}h'\\
&=\varphi(Nh_1h_2h_3h_4h_5)^{-1}h'\\
&=\varphi(Nh)^{-1}h'.
\end{aligned}
\label{eq:sms-B-final}
\end{equation}
Multiplying on the left by the quotient coset $\varphi(Nh)$ yields
\begin{equation}
\delta_{\varphi(Nh)^{-1}h',\,N'm'}
=\delta_{N'h',\,\varphi(Nh)m'}.
\label{eq:sms-logical-delta}
\end{equation}
The shifted ancilla sum is consequently
\begin{equation}
\begin{aligned}
&\sum_{q'_1,q'_2,q'_3,q'_4\in G'}
\left[\prod_{i=1}^4
\delta_{N'q'_{i-1}g'_i(q'_i)^{-1},\,\varphi(Nh_i)}\right]\\
&\quad\times
\delta_{\varphi(Nh_5)^{-1}q'_4g'_5,\,N'm'}\\
&\qquad\propto\delta_{N'h',\,\varphi(Nh)m'}.
\end{aligned}
\label{eq:row-shifted-count}
\end{equation}
The final test $F_5=N'm'$ thus measures the logical relation
$N'h'=\varphi(Nh)m'$ and selects the branch $\mathsf P_K(m')$.
The shift acts on the right of $\varphi(Nh)$, and the retained
labels $h'$ may lie outside $H'$.

In the folded picture, the pair constraint
$N'h'_i=\varphi(Nh_i)$ gives the same deltas. The contractions above
therefore yield the same logical operation, as in
Sec.~\ref{sec:sms-state-algebra}.

\section{Tensor Network Formalism}
\label{sec:regular-tensors}

This appendix collects the ordinary-group tensor networks used as
reference cases in Sec.~\ref{sec:tensor-network-measurements}. 

Following the tensor-network construction of
Ref.~\cite{manjunath2026}, we use a copy tensor and an
oriented multiplication tensor.  Every leg carries a group-basis
index.  The \(n\)-leg copy tensor is
\begin{equation}
\begin{aligned}
\Delta_G(g_1,\ldots,g_n)
&={}
\prod_{j=2}^{n}\delta_{g_1,g_j}
\\[2pt]
&={}
\begin{cases}
1,&g_1=g_2=\cdots=g_n,\\
0,&\text{otherwise}.
\end{cases}
\end{aligned}
\label{eq:copy-tensor}
\end{equation}
It is represented by a solid dot:
\begin{equation}
\NativeCopyTensor
\;=\;
\Delta_G(g_1,\ldots,g_n).
\label{eq:copy-tensor-graphical}
\end{equation}

The \(n\)-leg multiplication tensor uses the counterclockwise order
of the legs. Set \(\epsilon_j=+1\) for an inward-pointing leg and
\(\epsilon_j=-1\) for an outward-pointing leg. Then
\begin{equation}
M_G(g_1,\ldots,g_n)
=\delta_{g_1^{\epsilon_1}g_2^{\epsilon_2}\cdots g_n^{\epsilon_n},\,e_G}.
\label{eq:multiplication-tensor-ccw}
\end{equation}
It is represented by an open circle:
\begin{equation}
\NativeNLegMultTensor
\;=\;
M_G(g_1,\ldots,g_n).
\label{eq:multiplication-tensor-graphical}
\end{equation}
Because the product is constrained to the identity, its starting
point may be shifted cyclically without changing the tensor.

The copy and multiplication tensors give the controlled-left
multiplication gate.  Taking the lower legs as inputs and the upper
legs as outputs, its graphical definition is
\begin{equation}
\mathrm{CL}_{12}
=
\NativeCLGate
.
\label{eq:CL12-graphical}
\end{equation}
Indeed,
\begin{align}
\bra{g',h'}\mathrm{CL}_{12}\ket{g,h}
&=
\sum_{x\in G}
\Delta_G(g',g,x)\,
\delta_{xh(h')^{-1},\,e_G}
\notag\\
&=
\delta_{g',g}\,\delta_{h',gh},
\label{eq:CL12-contraction}
\end{align}
and therefore
\begin{equation}
\mathrm{CL}_{12}\ket{g,h}
=
\ket{g,gh}.
\label{eq:CL12-action}
\end{equation}

The remaining leg orientations give the inverse-left and the two
right-multiplication variants.  We use a crossed open circle to denote
the opposite cyclic ordering of the multiplication tensor:
\begin{align}
\mathrm{C}\overline{\mathrm L}_{12}
&=
\NativeControlled{0}{0}
,
&
\mathrm{C}\overline{\mathrm L}_{12}\ket{g,h}
&=\ket{g,g^{-1}h},
\\
\mathrm{CR}_{12}
&=
\NativeControlled{0}{1}
,
&
\mathrm{CR}_{12}\ket{g,h}
&=\ket{g,hg^{-1}},
\\
\mathrm{C}\overline{\mathrm R}_{12}
&=
\NativeControlled{1}{1}
,
&
\mathrm{C}\overline{\mathrm R}_{12}\ket{g,h}
&=\ket{g,hg}.
\label{eq:controlled-multiplication-variants}
\end{align}

For
arbitrary \(g',h'\in G\),  
\begin{align}
\bra{g',h'}
\mathrm{C}\overline{\mathrm L}_{12}
\ket{g,h}
&=
\sum_{x\in G}
\Delta_G(g',g,x)\,
\delta_{x^{-1}h(h')^{-1},\,e_G}
\notag\\
&=
\delta_{g',g}\,
\delta_{h',\,g^{-1}h}.
\label{eq:CLbar-contraction}
\end{align}
\begin{align}
\bra{g',h'}\mathrm{CR}_{12}\ket{g,h}
&=
\sum_{x\in G}
\Delta_G(g',g,x)\,
\delta_{hx^{-1}(h')^{-1},\,e_G}
\notag\\
&=
\delta_{g',g}\,
\delta_{h',\,hg^{-1}},
\label{eq:CR-contraction}\\[3pt]
\bra{g',h'}
\mathrm{C}\overline{\mathrm R}_{12}
\ket{g,h}
&=
\sum_{x\in G}
\Delta_G(g',g,x)\,
\delta_{hx(h')^{-1},\,e_G}
\notag\\
&=
\delta_{g',g}\,
\delta_{h',\,hg}.
\label{eq:CRbar-contraction}
\end{align}
Thus, in every case the first register is unchanged, while the second
is multiplied by \(g\) or \(g^{-1}\) on the side selected by the
tensor orientation.

These controlled-multiplication identities provide the elementary
contractions used to decompose the stabilizer networks below.

For an ordinary $G'$ plaquette with the edge labels of
Eq.~\eqref{eq:review-plaquette-graphical},
\begin{equation}
\begin{aligned}
B_p^{G'}\ket{g'_1,g'_2,g'_3,g'_4}
&=\delta_{(g'_1)^{-1}g'_2g'_3(g'_4)^{-1},\,e_{G'}}\\[-1pt]
&\quad\times\ket{g'_1,g'_2,g'_3,g'_4}.
\end{aligned}
\label{eq:Bp-regular-action}
\end{equation}
The compact schematic representation is
\begin{equation}
B_p^{G'}
=\vcenter{\hbox{\PlaquetteCompact{g'_1}{g'_2}{g'_3}{g'_4}}}.
\label{eq:Bp-regular-compact}
\end{equation}

The corresponding decomposed tensor network is
\begin{equation}
B_p^{G'}
=\vcenter{\hbox{\RegularPlaquetteNetwork{0}}}.
\label{eq:Bp-regular-decomposed}
\end{equation}
The ancillary endpoints denote $\ket e_A$ and ${}_A\bra e$.
Reading from bottom to top, the four gates are
$\mathrm{CR}_{1A}$,
$\mathrm{C}\overline{\mathrm R}_{2A}$,
$\mathrm{C}\overline{\mathrm R}_{3A}$, and
$\mathrm{CR}_{4A}$. Their product is
\begin{equation}
V_p=\mathrm{CR}_{4A}\,
\mathrm{C}\overline{\mathrm R}_{3A}\,
\mathrm{C}\overline{\mathrm R}_{2A}\,
\mathrm{CR}_{1A}.
\label{eq:Bp-regular-unitary}
\end{equation}
The physical labels control the gates and remain unchanged. The
ancilla evolves as
\begin{equation}
\begin{aligned}
\ket e_A
&\xrightarrow{\mathrm{CR}_{1A}}\ket{(g'_1)^{-1}}_A\\
&\xrightarrow{\mathrm{C}\overline{\mathrm R}_{2A}}
\ket{(g'_1)^{-1}g'_2}_A\\
&\xrightarrow{\mathrm{C}\overline{\mathrm R}_{3A}}
\ket{(g'_1)^{-1}g'_2g'_3}_A\\
&\xrightarrow{\mathrm{CR}_{4A}}
\ket{(g'_1)^{-1}g'_2g'_3(g'_4)^{-1}}_A.
\end{aligned}
\label{eq:Bp-regular-steps}
\end{equation}
Projecting the final ancilla onto $\ket e_A$ gives
\begin{equation}
\begin{aligned}
&{}_A\bra e V_p\ket e_A\ket{\boldsymbol g}\\
&\quad=\delta_{(g'_1)^{-1}g'_2g'_3(g'_4)^{-1},\,e_{G'}}
\ket{\boldsymbol g},
\end{aligned}
\label{eq:Bp-regular-contraction}
\end{equation}
where $\boldsymbol g=(g'_1,g'_2,g'_3,g'_4)$. Thus the network
implements $B_p^{G'}$.

Replacing the final projection by a group-basis measurement of the
ancilla gives
\begin{equation}
\mathcal M_p^{G'}(m'_p)
=\vcenter{\hbox{\RegularPlaquetteNetwork{1}}}.
\label{eq:Bp-regular-measurement}
\end{equation}
Its conditional action is
\begin{equation}
\begin{aligned}
\mathcal M_p^{G'}(m'_p)\ket{\boldsymbol g}
&={}_A\bra{m'_p}V_p\ket e_A\ket{\boldsymbol g}\\
&=\delta_{(g'_1)^{-1}g'_2g'_3(g'_4)^{-1},\,m'_p}
\ket{\boldsymbol g}.
\end{aligned}
\label{eq:Bp-regular-measured-action}
\end{equation}

The corresponding vertex construction uses one group element to
control the actions on all four incident edges.

The regular vertex projector is
\begin{equation}
A_v^G
=
\frac{1}{|G|}
\sum_{a\in G}
R_{\mathrm L}^{a}L_{\mathrm R}^{a}
R_{\mathrm B}^{a}L_{\mathrm T}^{a}.
\label{eq:Av-regular}
\end{equation}

the compact network is
\begin{equation}
A_v^G
=
\NativeVertexCompact{%
  \NPLabelR{-1.72}{.30}{\scriptsize$g'_{\mathrm L}$}%
  \NPLabelL{1.72}{-.30}{\scriptsize$g'_{\mathrm R}$}%
  \NPLabel{-.86}{-1.66}{\scriptsize$g'_{\mathrm B}$}%
  \NPLabel{.86}{1.66}{\scriptsize$g'_{\mathrm T}$}}
.
\label{eq:Av-regular-compact}
\end{equation}

It decomposes as
\begin{equation}
A_v^G
=
\NativeVertexCircuit{%
  \NPLabel{-1.90}{2.98}{\scriptsize$g'_{\mathrm L}$}%
  \NPLabel{-.95}{2.98}{\scriptsize$g'_{\mathrm B}$}%
  \NPLabel{.95}{2.98}{\scriptsize$g'_{\mathrm R}$}%
  \NPLabel{1.90}{2.98}{\scriptsize$g'_{\mathrm T}$}}
.
\label{eq:Av-regular-decomposed}
\end{equation}

where
\begin{equation}
\begin{aligned}
&
\NativeOneLegKet{.72cm}
=\ket{+}_G,
\qquad
\NativeOneLegBra{.72cm}
=\bra{+}_G,
\\[3pt]
&
\ket{+}_G
=
\frac{1}{\sqrt{|G|}}\sum_{a\in G}\ket{a}.
\end{aligned}
\label{eq:Av-regular-one-leg}
\end{equation}
Replacing \(G\) by \(G'\) in
Eqs.~\eqref{eq:Av-regular}--\eqref{eq:Av-regular-one-leg} defines
\(A_v^{G'}\) and its measurement circuit.

\section{General Correction Procedures}
\label{sec:parallel-correction}

\subsection{Parallel Correction}
\label{sec:parallel-schemes}

We generalize the parallel corrections of Sec.~\ref{sec:smooth-correction}
and identify the group conditions under which they apply.

\subsubsection{Boundary restriction}
\label{sec:parallel-boundary}
\label{sec:parallel-PH}
\label{sec:correction-PH-two-round}

The second smooth interface of the $T$ protocol
(Sec.~\ref{sec:smooth magic protocol}) uses $H=\langle rs\rangle
\cong\mathbb Z_2$. Although $H$ is not normal in $D_4$, it is normal in
$H_1=\langle r^2,rs\rangle\cong\mathbb Z_2\times\mathbb Z_2$,
which is normal in $D_4$. Apply the procedure of
Sec.~\ref{sec:correction-PH-example} twice: first restrict from
$D_4$ to $H_1$, then from $H_1$ to $H$.

This repeated restriction extends to any \emph{subnormal} subgroup
$H\leq G$, meaning that there is a finite chain
\begin{equation}
H=H_0\trianglelefteq H_1\trianglelefteq\cdots
\trianglelefteq H_d=G.
\label{eq:parallel-H-chain}
\end{equation}
Use one parallel round for each proper inclusion, proceeding from
$H_d$ down to $H_0$. At the start of round $k$, the first four
current labels lie in $H_k$. The round is:
\begin{enumerate}
\item Simultaneously resolve $g_i\in H_{k-1}r_i$ for
$i=1,2,3,4$ by Eq.~\eqref{eq:sms-coset}.
\item Choose $t_i\in H_k$ with
\begin{equation}
H_{k-1}t_i=H_{k-1}r_1\cdots r_i,
\qquad i=1,2,3,4,
\label{eq:parallel-PH-factors}
\end{equation}
and apply $A_{v_i}^{(t_i)}$ simultaneously. With $t_0=e$, the
updated boundary labels are
\begin{equation}
\begin{gathered}
\widetilde g_i=t_{i-1}g_it_i^{-1},\qquad i=1,2,3,4,\\
\widetilde g_5=t_4g_5.
\end{gathered}
\label{eq:parallel-PH-row}
\end{equation}
\end{enumerate}
Because $H_{k-1}\trianglelefteq H_k$, multiplication in
$H_k/H_{k-1}$ gives
\begin{equation}
\begin{aligned}
H_{k-1}\widetilde g_i
&=(H_{k-1}t_{i-1})(H_{k-1}r_i)(H_{k-1}t_i)^{-1}\\
&=H_{k-1}.
\end{aligned}
\label{eq:parallel-PH-proof}
\end{equation}
Each round preserves the ordered logical boundary product by
Eq.~\eqref{eq:correction-PH-product}.
Repeat on the current labels for $k=d-1,\ldots,1$, then test
$P_H$ on edge five. With $h_1,\ldots,h_4\in H$, the product identity gives
\begin{equation}
\widetilde g_5\in H
\quad\Longleftrightarrow\quad g_1g_2g_3g_4g_5\in H.
\label{eq:parallel-PH-final}
\end{equation}
Thus the final test implements the logical restriction to $H$.

\subsubsection{Plaquette measurements}
\label{sec:parallel-Bp}

The plaquette correction of Sec.~\ref{sec:correction-Bp-example}
can be extended through a sequence of central quotients. For general $N'$,
the word $F_i$ in Eq.~\eqref{eq:sms-plaquette-word} is a right coset,
and $B_{p_i}^e=1$ means $F_i=N'$.

Given $G'$ and $N'$, choose $Z_0\leq N'$ with
$Z_0\trianglelefteq G'$. Form $Z_1$ from all elements of $G'$ whose
cosets in $G'/Z_0$ lie in its center. Next, form $Z_2$ from all
elements whose cosets in $G'/Z_1$ lie in its center. Continue by the rule
\begin{equation}
Z_j/Z_{j-1}=Z(G'/Z_{j-1}).
\label{eq:parallel-central-rule}
\end{equation}
Starting from the identity subgroup, this construction defines the
\emph{upper central series}. A group is \emph{nilpotent} if this
series reaches the whole group in finitely many steps. We require
$G'/Z_0$ to be nilpotent for at least one subgroup $Z_0\leq N'$
that is normal in $G'$. This gives
\begin{equation}
Z_0\trianglelefteq Z_1\trianglelefteq\cdots
\trianglelefteq Z_c=G'
\label{eq:parallel-central-chain}
\end{equation}
for finite $c$. The least such $c$ is the nilpotency class of $G'/Z_0$.

For example, take $G'=D_4$ and start with $Z_0=\{e\}$. The two successive constructions are
\begin{equation}
\begin{aligned}
Z_1&=Z(D_4)=\langle r^2\rangle,\\
Z_2/Z_1&=Z(D_4/Z_1)=D_4/Z_1,\qquad Z_2=D_4.
\end{aligned}
\label{eq:parallel-D4-center}
\end{equation}
Here $Z_1$ is the center of $D_4$, and $D_4/Z_1$ is Abelian, so its
center is the entire quotient. Thus $D_4$ is nilpotent of class two;
these are the two layers corrected in Sec.~\ref{sec:correction-Bp-example}.

The useful property is that each element of $Z_j$ has a central coset
in $G'/Z_{j-1}$ and hence commutes with every element of $G'$ modulo
$Z_{j-1}$:
\begin{equation}
Z_{j-1}zx=Z_{j-1}xz,\qquad z\in Z_j,\quad x\in G'.
\label{eq:parallel-central-commutation}
\end{equation}

This construction includes several useful special cases. For normal
$N'$, choosing $Z_0=N'$ requires only that $G'/N'$ be nilpotent.
If $G'$ itself is nilpotent, $Z_0=\{e\}$ works for every allowed $N'$,
including $N'=\{e\}$. In an Abelian group there is only one central
layer, so at most one round is needed.

We construct the chain upward from $Z_0$ to $Z_c$, then correct
downward from $Z_c$ to $Z_0$. For the first four plaquettes of
Fig.~\ref{fig:sms-B}, at the start of round $j$,
$F_i\subseteq N'Z_j$ for $i=1,2,3,4$. This holds automatically
in the first round. Each round consists of:
\begin{enumerate}
\item Simultaneously measure the coarse right cosets
\begin{equation}
N'Z_{j-1}F_i=N'Z_{j-1}m'_i,
\qquad i=1,2,3,4.
\label{eq:parallel-Bp-measure}
\end{equation}
Since $F_i\subseteq N'Z_j$, it has a representative $n'z$ with
$n'\in N'$ and $z\in Z_j$. Then
\begin{equation}
\begin{aligned}
N'Z_{j-1}F_i
&=N'Z_{j-1}n'z\\
&=N'Z_{j-1}z=N'Z_{j-1}m'_i,
\end{aligned}
\label{eq:parallel-Bp-representative}
\end{equation}
so we can choose $m'_i=z\in Z_j$.

\item Set $t_0=e$ and choose $t_i\in Z_j$ satisfying
\begin{equation}
N'Z_{j-1}t_i=N'Z_{j-1}m'_1\cdots m'_i,
\qquad i=1,2,3,4.
\label{eq:parallel-Bp-factors}
\end{equation}
Apply the corrections simultaneously:
\begin{equation}
L^{t_i}_{q'_i}:\quad q'_i\longmapsto t_iq'_i,
\qquad i=1,2,3,4.
\label{eq:parallel-Bp-correction}
\end{equation}
\end{enumerate}

The updated word is
\begin{equation}
F_i^{\mathrm{new}}
=\varphi(Nh_i)^{-1}t_{i-1}q'_{i-1}g'_i(q'_i)^{-1}t_i^{-1}.
\label{eq:parallel-Bp-update}
\end{equation}
Because $t_{i-1}\in Z_j$, it commutes with every factor of the word
modulo $Z_{j-1}$. Therefore
\begin{equation}
\begin{aligned}
N'Z_{j-1}F_i^{\mathrm{new}}
&=N'Z_{j-1}\varphi(Nh_i)^{-1}\\[-2pt]
&\quad\times t_{i-1}q'_{i-1}g'_i(q'_i)^{-1}t_i^{-1}\\
&=N'Z_{j-1}\varphi(Nh_i)^{-1}\\[-2pt]
&\quad\times q'_{i-1}g'_i(q'_i)^{-1}t_{i-1}t_i^{-1}\\
&=N'Z_{j-1}F_it_{i-1}t_i^{-1}\\
&=N'Z_{j-1}m'_it_{i-1}t_i^{-1}\\
&=N'Z_{j-1}.
\end{aligned}
\label{eq:parallel-Bp-proof}
\end{equation}
We successively
restrict the words to the identity cosets of
$N'Z_c,N'Z_{c-1},\ldots,N'Z_0$. Since $Z_0\leq N'$, the last
condition is $N'F_i^{\mathrm{new}}=N'$.
Only vertical ancillas are changed, so both ordered logical boundary
products are preserved. Once $F_1,\ldots,F_4=N'$, the final test
$F_5=N'$ is exactly the logical test $N'h'=\varphi(Nh)$ of
Appendix~\ref{sec:row-algebra}.

Table~\ref{tab:parallel-rounds} lists the numbers of correction rounds
for the $P_H$ restriction and $B_p^e$ measurements in the smooth
magic-state protocols.

\begin{table}[!t]
\pdfdest name {float-15-start} xyz
\caption{Parallel correction rounds for the smooth magic-state
protocols of Sec.~\ref{sec:protocols}. The columns $d$ and $c$ count
boundary-restriction and plaquette rounds, respectively. Subscripts
label the first and second smooth interfaces.}
\label{tab:parallel-rounds}
\centering\small
\setlength{\tabcolsep}{3pt}
\renewcommand{\arraystretch}{1.16}
\begin{tabular}{@{}lcccc@{}}
\hline\hline
Interface & $(G,H)$ & $(G',H')$ & $d$ & $c$\\
\hline
$T_1$ & $(\mathbb Z_4,\mathbb Z_4)$ & $(D_4,\langle r\rangle)$ & $0$ & $2$\\
$T_2$ & $(D_4,\langle rs\rangle)$ & $(\mathbb Z_2,\mathbb Z_2)$ & $2$ & $1$\\
CZ, CCX & $(D_4,D_4)$ & $(\mathbb Z_2,\mathbb Z_2)$ & $0$ & $1$\\
CS$_1$ & $(\mathbb Z_2,\mathbb Z_2)$ & $(D_4,\langle s\rangle)$ & $0$ & $2$\\
CS$_2$ & $(D_4,\langle r^2,rs\rangle)$ & $(\mathbb Z_2^2,\mathbb Z_2^2)$ & $1$ & $1$\\
\hline\hline
\end{tabular}
\pdfdest name {float-15-end} xyz
\end{table}

\subsection{Sequential Correction}
\label{sec:sequential-schemes}

Sequential correction applies to arbitrary finite-group data in
Eq.~\eqref{eq:sms-group-data}, without the subnormality or nilpotence
conditions used above. Measurements and corrections alternate from
left to right.

\subsubsection{Boundary restriction}
\label{sec:sequential-PH}

Use the boundary of Fig.~\ref{fig:sms-P} and set $r_0=e$.
For each edge $i=1,2,3,4$:
\begin{enumerate}
\item Measure the right coset of the current label $r_{i-1}g_i$,
obtaining $Hr_i$.
\item Apply the full vertex transformation $A_{v_i}^{(r_i)}$,
whose boundary action is
\begin{equation}
(r_{i-1}g_i,g_{i+1})
\longmapsto(r_{i-1}g_ir_i^{-1},r_ig_{i+1}).
\label{eq:sms-P-sweep}
\end{equation}
\end{enumerate}
The successive rows are
\begin{equation}
\begin{aligned}
&(h_1,r_1g_2,g_3,g_4,g_5),\\
&(h_1,h_2,r_2g_3,g_4,g_5),\\
&(h_1,h_2,h_3,r_3g_4,g_5),\\
&(h_1,h_2,h_3,h_4,r_4g_5),
\end{aligned}
\label{eq:sequential-PH-row}
\end{equation}
where $h_i=r_{i-1}g_ir_i^{-1}\in H$. The ordered product obeys
\begin{equation}
h_1h_2h_3h_4(r_4g_5)=g_1g_2g_3g_4g_5.
\label{eq:sms-P-final}
\end{equation}
Test $P_H$ on edge five and retain outcome one, writing $h_5=r_4g_5$.
Because the first four labels lie in $H$, this final test is
equivalent to restricting the original logical product to $H$.

\subsubsection{Plaquette measurements}
\label{sec:sequential-Bp}

On Fig.~\ref{fig:sms-B}, first measure $F_1=N'm'_1$ and apply
$L_{q'_1}^{m'_1}$. Then measure $F_2$ using the corrected $q'_1$
and apply $L_{q'_2}^{m'_2}$. Continue one plaquette at a time,
performing each displayed measurement and correction before the next:

\begin{equation}
\begin{aligned}
\varphi(Nh_1)^{-1}g'_1(q'_1)^{-1}&=N'm'_1,
&\widehat q'_1&=m'_1q'_1,\\
\varphi(Nh_2)^{-1}\widehat q'_1g'_2(q'_2)^{-1}&=N'm'_2,
&\widehat q'_2&=m'_2q'_2,\\
\varphi(Nh_3)^{-1}\widehat q'_2g'_3(q'_3)^{-1}&=N'm'_3,
&\widehat q'_3&=m'_3q'_3,\\
\varphi(Nh_4)^{-1}\widehat q'_3g'_4(q'_4)^{-1}&=N'm'_4,
&\widehat q'_4&=m'_4q'_4.
\end{aligned}
\label{eq:sms-B-sweep-general}
\end{equation}
At each step,
\begin{equation}
F_i^{\mathrm{new}}=F_i(m'_i)^{-1}=N'.
\label{eq:sequential-Bp-proof}
\end{equation}
In Fig.~\ref{fig:sms-B}, correcting $F_i$ changes its right ancilla
$q'_i$; the next correction changes $q'_{i+1}$. The fifth plaquette
has no right ancilla to update and provides the logical test.
Both horizontal boundary products are preserved. After the fourth update, read
\begin{equation}
\varphi(Nh_5)^{-1}\widehat q'_4g'_5=N'm'.
\label{eq:sequential-Bp-final}
\end{equation}
By Appendix~\ref{sec:row-shifted}, this final readout tests the
logical relation $N'h'=\varphi(Nh)m'$.

\section{Logical Recovery for the Logical $K$-Coset Measurement}
\label{sec:logical_recovery_smooth}
In this appendix, we discuss how to perform the recovery operation for general measurement outcomes in the logical $K$-coset measurement in smooth hybrid lattice surgery. Given an outcome $m'$, the logical measurement is given by
\begin{equation}
\mathsf{P}_K(m')\ket{j,k}
=
\delta_{N'k,\,\varphi(Nj)m'}\ket{j,k},
\quad j\in H,\quad k\in G'.
\label{eq:sms-logical-measurement-recovery}
\end{equation}

We define the recovery operation as the right multiplication on the $D(G')$ patch
\begin{equation}
  R^{m'}\lvert j,k\rangle=\lvert j,km'^{-1}\rangle.
  \label{eq:recovery-translation}
\end{equation}
Consequently,
\begin{align}
  R^{m'}\mathsf{P}_K(m')\lvert j,k\rangle
  &=\delta_{N'k,\,\varphi(Nj)m'}\lvert j,km'^{-1}\rangle
    \notag\\
  &=\delta_{N'(km'^{-1}),\,\varphi(Nj)}\lvert j,km'^{-1}\rangle
    \notag\\
  &=\mathsf{P}_K R^{m'}\lvert j,k\rangle,
\end{align}
where we have used
\begin{equation}
  N'k=\varphi(Nj)m'
  \quad\Longleftrightarrow\quad
  N'(km'^{-1})=\varphi(Nj),
  \label{eq:recovery-coset-equivalence}
\end{equation}
and $\mathsf{P}_K := \mathsf{P}_K(e)$.
Therefore, we have
\begin{equation}
  R^{m'}\mathsf{P}_K(m')=\mathsf{P}_K R^{m'}.
  \label{eq:recovery-intertwining}
\end{equation}
Equation~\eqref{eq:recovery-intertwining} shows that a sufficient condition
for recovering the identity output on an input $\lvert\Phi\rangle$ is
$R^{m'}\lvert\Phi\rangle=\lvert\Phi\rangle$. Under this condition,
\begin{equation}
  R^{m'} \mathsf{P}_K(m')\lvert\Phi\rangle
  =\mathsf{P}_K R^{m'}\lvert\Phi\rangle
  =\mathsf{P}_K \lvert\Phi\rangle.
  \label{eq:recovery-identity-output}
\end{equation}
In particular, the condition holds when the $D(G')$ patch is in the logical state
\begin{equation}
  \lvert+\rangle_{G'}
  =\frac{1}{\sqrt{|G'|}}\sum_{k\in G'}\lvert k\rangle.
  \label{eq:recovery-uniform-state}
\end{equation}
Every smooth merge–split in the protocols in Sec.~\ref{sec:protocols} satisfies this condition,
except for the second interface of the direct T-gate protocol, which we discuss below.

At the second interface of the direct $T$-gate protocol in Sec.~\ref{sec:smooth magic protocol}, the $D(\Z_2)$ patch carries an arbitrary logical qubit,
\begin{equation}
  \lvert\psi\rangle=\alpha\lvert0\rangle+\beta\lvert1\rangle,
\end{equation}
The condition used above therefore does not hold. Nevertheless, we can recover the identity outcome. If the outcome is $1$, we apply the $S$ gate to recover the identity outcome:
\begin{align}
  S\bigl(\eta\alpha\lvert0\rangle+\beta\lvert1\rangle\bigr)
  &=\eta\alpha\lvert0\rangle+i\beta\lvert1\rangle\notag\\
  &=\eta\bigl(\alpha\lvert0\rangle+\eta\beta\lvert1\rangle\bigr)
    \notag\\
  &=\eta T\lvert\psi\rangle.
  \label{eq:recovery-t-phase}
\end{align}

\bibliography{references}

\end{document}